\documentclass[journal]{IEEEtran}

\renewcommand{\arraystretch}{0.9}
\usepackage{amsfonts}
\usepackage{amssymb}
\usepackage{amsthm}
\usepackage{amsmath,amsfonts,amssymb}
\usepackage[dvips]{graphicx}
\usepackage{subfigure}
\usepackage{verbatim}
\usepackage{setspace}
\usepackage{bm}
\usepackage{algorithmic} %format of the algorithm
\usepackage[ruled,vlined]{algorithm2e}
\usepackage{cite}
\usepackage{balance}

\usepackage{changepage}
\usepackage{pdfpages}
\usepackage{color}
\usepackage{epstopdf}
\usepackage{hyperref}
\usepackage{multirow}
\usepackage{makecell,array}
\usepackage{graphicx}

\newtheorem{assumption}{Assumption}

\newcommand*{\dif}{\mathop{}\!\mathrm{d}}
\allowdisplaybreaks

\newcommand{\figwidth}{8.0}
\IEEEoverridecommandlockouts

\begin{document}
\title{\huge A Tutorial on Movable Antennas for Wireless Networks}
\author{Lipeng Zhu, ~\IEEEmembership{Member,~IEEE,}
	Wenyan Ma,~\IEEEmembership{Graduate Student Member,~IEEE,}
        Weidong Mei,~\IEEEmembership{Member,~IEEE,}
        Yong Zeng,~\IEEEmembership{Fellow,~IEEE,}
        Qingqing Wu,~\IEEEmembership{Senior Member,~IEEE,}
        Boyu Ning,~\IEEEmembership{Member,~IEEE,}
	Zhenyu Xiao,~\IEEEmembership{Senior Member,~IEEE,}
        Xiaodan Shao,~\IEEEmembership{Member,~IEEE,}
	Jun Zhang,
	and Rui Zhang,~\IEEEmembership{Fellow,~IEEE}
	\vspace{-0.5 cm}
        \thanks{L. Zhu and W. Ma are with the Department of Electrical and Computer Engineering, National University of Singapore, Singapore 117583 (e-mail: zhulp@nus.edu.sg, wenyan@u.nus.edu).}
        \thanks{W. Mei and B. Ning are with the National Key Laboratory of Wireless Communications, University of Electronic Science and Technology of China (UESTC), Chengdu 611731, China (e-mail: wmei@uestc.edu.cn, boydning@outlook.com).}
        \thanks{Y. Zeng is with the National Mobile Communications Research Laboratory and Frontiers Science Center for Mobile Information Communication and Security, Southeast University, Nanjing 210096, China, and the Purple Mountain Laboratories, Nanjing 211111, China (e-mail: yong\_zeng@seu.edu.cn).}
        \thanks{Q. Wu is with the Department of Electronic Engineering, Shanghai Jiao Tong University, Shanghai 200240, China (e-mail: qingqingwu@sjtu.edu.cn).}
        \thanks{Z. Xiao is with the School of Electronic and Information Engineering, Beihang University, Beijing 100191, China (e-mail: xiaozy@buaa.edu.cn).}
        \thanks{X. Shao is with the Department of Electrical and Computer Engineering, University of Waterloo, Waterloo, ON N2L 3G1, Canada (e-mail: x6shao@uwaterloo.ca).}
        \thanks{J. Zhang is with the State Key Laboratory of CNS/ATM \& MIIT Key Laboratory of Complex-field Intelligent Sensing, Beijing Institute of Technology, Beijing 100081, China (e-mail: zhjun@bit.edu.cn). He is also with the School of Electronic and Information Engineering, Beihang University, Beijing 100191, China.}
        \thanks{R. Zhang is with the School of Science and Engineering, Shenzhen Research Institute of Big Data, The Chinese University of Hong Kong, Shenzhen, Guangdong 518172, China (e-mail: rzhang@cuhk.edu.cn). He is also with the Department of Electrical and Computer Engineering, National University of Singapore, Singapore 117583 (e-mail: elezhang@nus.edu.sg).}
        % \thanks{Corresponding author: Wenyan Ma and Rui Zhang}
}

\maketitle


\begin{abstract}
    Movable antenna (MA) has been recognized as a promising technology to enhance the performance of wireless communication and sensing by enabling antenna movement. Such a significant paradigm shift from conventional fixed antennas (FAs) to MAs offers tremendous new opportunities towards realizing more versatile, adaptive and efficient next-generation wireless networks such as 6G. In this paper, we provide a comprehensive tutorial on the fundamentals and advancements in the area of MA-empowered wireless networks. First, we overview the historical development and contemporary applications of MA technologies. Next, to characterize the continuous variation in wireless channels with respect to antenna position and/or orientation, we present new field-response channel models tailored for MAs, which are applicable to narrowband and wideband systems as well as far-field and near-field propagation conditions. Subsequently, we review the state-of-the-art architectures for implementing MAs and discuss their practical constraints. A general optimization framework is then formulated to fully exploit the spatial degrees of freedom (DoFs) in antenna movement for performance enhancement in wireless systems. In particular, we delve into two major design issues for MA systems. First, we address the intricate antenna movement optimization problem for various communication and/or sensing systems to maximize the performance gains achievable by MAs. Second, we deal with the challenging channel acquisition issue in MA systems for reconstructing the channel mapping between arbitrary antenna positions inside the transmitter and receiver regions. Moreover, we show existing prototypes developed for MA-aided communication/sensing and the experimental results based on them. Finally, the extension of MA design to other wireless systems and its synergy with other emerging wireless technologies are discussed. We also highlight promising research directions in this area to inspire future investigations.
\end{abstract}
% Note that keywords are not normally used for peer-review papers.
\begin{IEEEkeywords}
	Movable antenna (MA), field-response channel model, MA architecture, sparse array, antenna movement optimization, channel acquisition, 6G.
\end{IEEEkeywords}

\IEEEpeerreviewmaketitle

\section{Introduction} \label{Sec_Intro}
\subsection{Background}
\IEEEPARstart{O}{ver} the past few decades, wireless communication technologies have advanced at an unprecedented pace and profoundly reshaped human life. Among these innovations, multiple-input multiple-output (MIMO) is a pivotal technology that significantly improves the capacity, reliability, and efficiency of wireless systems \cite{Paulraj2004Anover}. As mobile communication networks evolve across generations, two primary technical routes have been adopted to meet the ever-growing demands for higher data rates, i.e., increasing bandwidth in the frequency domain or deploying more antennas in the spatial domain. Due to the scarcity of spectrum, the available frequency bands for wireless communications are limited. Therefore, deploying more antennas at base stations (BSs) has become an essential trend for improving spectrum efficiency in both current and future cellular networks. This evolution has driven the development of MIMO technologies towards more advanced forms, such as massive MIMO \cite{Larsson2014massive, Lu2014Anover} and extremely large-scale MIMO (XL-MIMO) \cite{lu2024nearXL, Wang2024XLMIMO, you2024next}, which leverage hundreds or even thousands of antennas at a single BS to achieve superior spatial multiplexing and beamforming gains. However, the shift towards large-scale antenna arrays also introduces challenging issues, particularly in terms of hardware cost, energy consumption, and signal processing overhead.

In light of this challenge, significant research efforts have been devoted to developing low-cost and energy-efficient antennas. In the past, thinned and sparse arrays have been extensively investigated to reduce the number of active antennas while preserving system performance \cite{Haupt1994thinned, Cen2010sparse}. Additionally, beamspace MIMO offers a cost-effective implementation by employing lens antenna arrays to minimize the number of radio frequency (RF) chains and phase shifters at the transceiver \cite{Brady2013beamspace, Zeng2014Lens, zeng2016millim}. With recent advancements in metamaterials, intelligent reflecting surfaces (IRSs) can be fabricated more cost-efficiently by integrating large numbers of semi-passive and tunable reflective elements \cite{wu2019IRS, Huang2019RIS, mei2022intelligent, zheng2022survey, wu2024ISfor6G}, which allow for dynamic control of electromagnetic wave propagation to enhance signal coverage and spectral efficiency without substantial hardware investments. Moreover, dynamic metasurface antennas (DMAs) and reconfigurable holographic surfaces (RHSs) utilize metamaterial elements as active radiating antennas \cite{Shlezinger2021DMAs, deng2021RHS}, which are densely arranged to achieve high aperture and beamforming gains without the need for conventional phase shifters.

However, all the implementations mentioned above are based on fixed antennas (FAs), which generally have fixed antenna positions and/or orientations once manufactured. In practical wireless systems with varying signal propagation environments, these FAs cannot fully exploit the spatial variation of wireless channels, which thus limits their degrees of freedom (DoFs) in the spatial domain. In addition, their antenna costs and energy consumption also scale with the number of antennas and the degree of reconfigurability for each element. As a result, existing wireless networks still face significant challenges in balancing spectrum efficiency and hardware costs due to the inflexible utilization of antenna resources.

Recently, movable antenna (MA) has been recognized as a promising technology to enhance wireless communication performance by enabling the local movement of antennas at the transmitter (Tx)/receiver (Rx) \cite{zhu2023MAMag}. Different from conventional FAs undergoing random fading channels, MAs can dynamically change their placements to improve channel conditions. Therefore, superior performance can be achieved by MA systems using the same or even smaller number of antennas and their associated RF chains as compared to traditional FA-based systems. This paradigm shift coexists with most of the existing antenna technologies based on different fabrications while enhancing their capability to adapt to signal propagation environments and system requirements. The significant performance advantages of MA-aided communication systems over conventional FA systems in terms of received signal power improvement, interference mitigation, flexible beamforming, and multiplexing enhancement have been well-validated in the literature \cite{zhu2022MAmodel,ma2022MAmimo,zhu2023MAarray,zhu2023MAmultiuser}\footnote{The MA technology can coexist with other advanced antenna technologies, such as active phased arrays and MIMO. Given the same number of antennas, MA systems can achieve superior performance compared to FA systems, which will be demonstrated in detail in Section \ref{Sec_Movement}.}. Moreover, recent studies have also demonstrated that MAs can substantially boost wireless sensing performance by fully exploiting the array geometry gain \cite{ma2024MAsensing}, which thus offers great potential for integrated sensing and communication (ISAC) in the forthcoming sixth-generation (6G) wireless networks \cite{you2024next}.

It is worth noting that the antenna selection (AS) technology achieves equivalent antenna repositioning by selectively activating a subset of antennas in a large-scale array \cite{Molisch2004MIMOsys,Sanayei2004antennas}. However, the MA technology offers several distinct advantages over AS. First, MA enables continuous antenna movement within spatial regions to fully exploit channel variations \cite{zhu2022MAmodel,ma2022MAmimo}. In contrast, AS is limited to discrete antenna placement, which constrains its spatial DoFs. Second, the cost of AS scales with the array size as it requires additional antenna elements. Conversely, MA systems achieve repositioning by moving a fixed number of antennas, avoiding the expense of extra elements. This superiority of MA over AS becomes more pronounced as the size of the antenna movement region increases \cite{zhu2024nearfield,ding2024near}. Third, AS is generally implemented in a one-dimensional (1D) line or two-dimensional (2D) surface, while MA can be efficiently moved in three-dimensional (3D) space \cite{zhu2023MAMag,zhu2023MAmultiuser}. Moreover, by further incorporating 3D rotation, six-dimensional MA (6DMA) systems can achieve the highest DoFs in the spatial domain, significantly outperforming conventional FA systems with/without AS \cite{shao20246DMA,shao2024discrete}.

\begin{figure}[t]
	\begin{center}
		\includegraphics[width=7.0 cm]{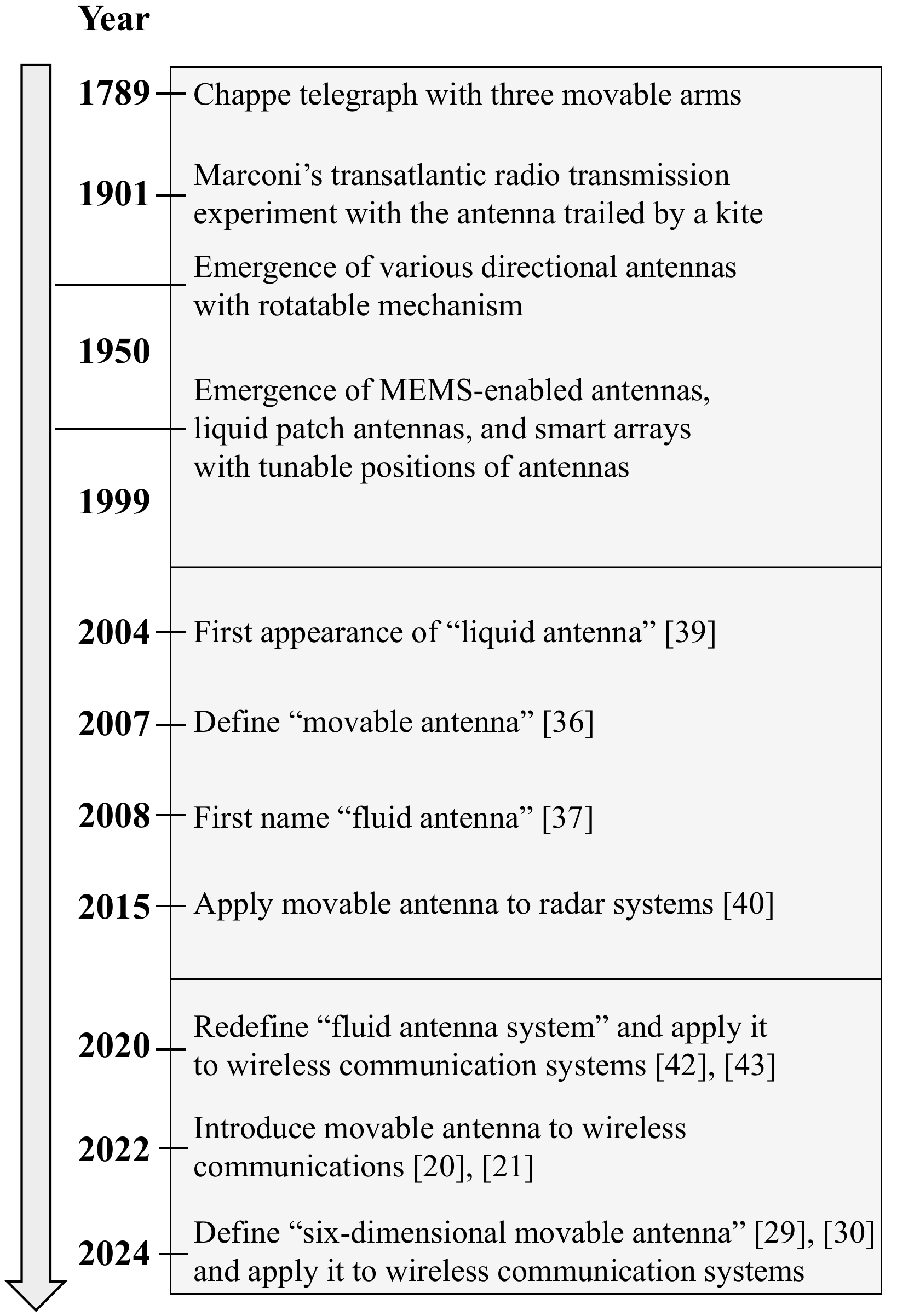}
		\caption{Illustration of the historical development of movable antennas.}
		\vspace{-12pt}
		\label{Fig_history}
	\end{center}
\end{figure}
\subsection{Historical Development}
The movement capabilities of antennas have been exploited to enhance system performance almost throughout the history of communication technologies. The pioneer of MAs can be traced back to the Chappe telegraph invented in the early 1790s, which utilized three movable arms on a tower to facilitate long-distance communication \cite{holzmann1995early}. In this system, information can be conveyed by adjusting the positions of the movable arms, which was the earliest example of spatial index modulation using mechanical reconfigurability. A key milestone of global telecommunications occurred in 1901 when Guglielmo Marconi successfully received the first transatlantic radio signals. In this experiment, the wire antenna was trailed by a kite to adjust its position and height in a trial-and-error manner until the signal was successfully captured. Following this, the development of radio and antenna technologies accelerated. Various directional antennas, such as the loop antenna, Yagi-Uda antenna, and parabolic reflector antenna \cite{balanis2016antenna}, were invented and then applied to communication, radar, or radio astronomy systems. These antennas were usually mounted on rotatable platforms to dynamically adjust their beam directions to enhance system performance. One typical application of this technology in modern cellular networks is the downtilt antennas installed on BSs to improve coverage. In the late twentieth century, the advance in reconfigurable antennas and smart antennas inspired the design of the micro-electromechanical system (MEMS)-enabled antennas with tunable radiation patterns and frequency responses \cite{chiao1999mems}, as well as the optimization of elements' positions in an antenna array for achieving desired beam patterns \cite{Lewis1983sidelobe, Ismail1991nullst}. 

\begin{figure*}[t]
	\begin{center}
		\includegraphics[width=16 cm]{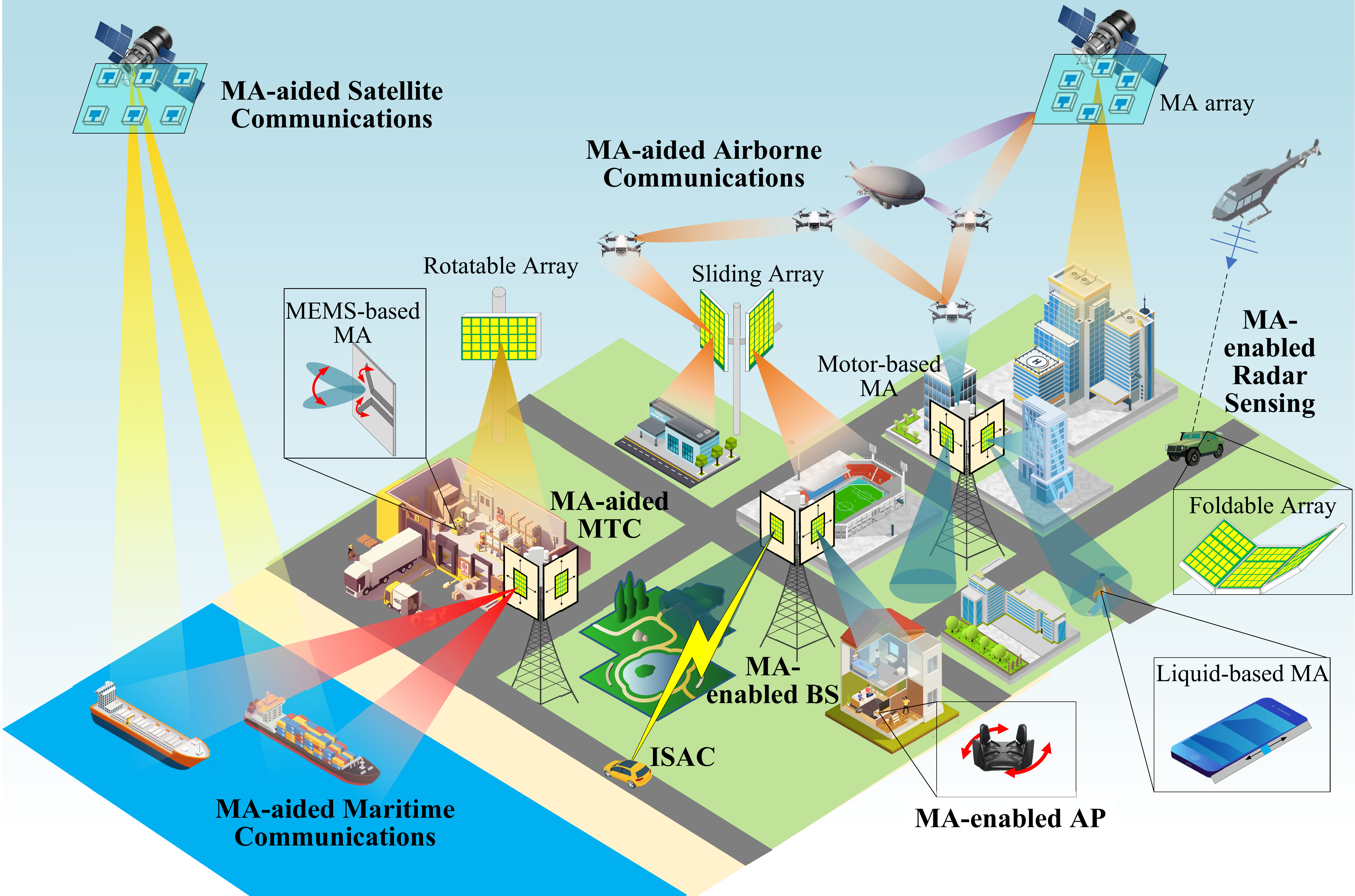}
		\caption{Application scenarios of MA-aided wireless networks.}
		\vspace{-12pt}
		\label{Fig_Applications}
	\end{center}
\end{figure*}

Based on these aforementioned antenna technologies exploiting antenna movement, the term ``movable antenna'' was formally used in a book published in 2007 \cite[Section 17.4]{balanis2007modern}, which reviewed some existing implementation examples of MAs. Besides, the term ``fluid antenna'' first appeared in 2008 \cite{Tam2008fluid}, while it had been previously also called ``liquid antenna'' and originally referred to antennas utilizing fluid or liquid dielectrics as electromagnetic radiators \cite{kosta1989liquid, Kosta2004liquid}. According to the original definition, fluid antennas can be seen as a specific type of MA by enabling the movement of the soft-material antenna in a liquid container. In 2015, an MA-enabled radar prototype demonstrated the advantages of antenna movement for improving radar imaging accuracy \cite{zhuravlev2015experimental}, building on a long history of similar techniques in radar systems known as the synthetic aperture radar (SAR) \cite{Moreira2013SAR}.

Despite its success in radar systems and several applications in wireless communication, MA technology has garnered relatively little attention from the communication community over the past few decades. This is likely because the situation that a few FAs have traditionally been sufficient to fulfill the requirements of many wireless systems. However, as communication and sensing requirements continue to escalate, the limitations of FAs have become increasingly evident, sparking renewed interest in MA technology. Building on its original definition, the concept of MA was introduced and rigorously investigated in the context of wireless communication in 2022 \cite{zhu2022MAmodel,ma2022MAmimo}. In particular, the movement capability of antennas in 3D space has been fully exploited in \cite{zhu2023MAMag,zhu2023MAmultiuser}. Previously, the term ``fluid antenna system (FAS)'' was introduced to the wireless communication field, where the authors in \cite{wong2020limit, wong2020fluid} extended its original definition to include position- or shape-flexible antennas. In particular, existing FAS models mainly focused on 1D or 2D position reconfiguration. Despite their different origins and implementation methods, MAs and FASs are conceptually similar in exploiting flexible antenna positioning and can be sometimes interchangeably used. Especially, over the last two years, there has been a significant surge in research on MA/FAS-assisted wireless systems \cite{zhu2024historical}. Recently, a general 6DMA system was proposed in \cite{shao20246DMA, shao2024discrete, shao2024Mag6DMA} by incorporating the DoFs in 3D position and 3D orientation/rotation of antennas. Practical implementation architectures of 6DMAs were later introduced in \cite{ning2024movable}. From the perspective of hardware implementation, MA and 6DMA emphasize physical movement for practical realization. From the perspective of the system modeling, 6DMA includes the existing FAS models with antenna position adjustments as special cases, and provides additional design flexibility to further improve wireless system performance. The historical development of MA technologies within the antenna and communication communities is summarized in Fig. \ref{Fig_history}.

\begin{table*}[]\small 
	\centering
	\caption{Comparison of the contributions of this paper with other related overview, survey, and tutorial articles.}
	\label{Tab_overview_article}
	\begin{tabular}{|c|>{\centering\arraybackslash}m{1.2cm}|>{\centering\arraybackslash}m{14cm}|} 
		\hline
		\textbf{Ref.}                           & \textbf{Type}        & \textbf{Main Contributions}                                                                                                                                                                                                                                                                                                                                                                                                                                                                                                                                                                                                                        \\ \hline
		\cite{zhu2023MAMag}         & Magazine       & An overview of MA-aided wireless communication, including the application scenarios, hardware architectures, channel characterization, performance advantages, and design issues.                                                                                                                                                                                                                                                                                                                                                                                                                                        \\ \hline
		\cite{shao2024Mag6DMA}          & Magazine      & An overview of 6DMA-enhanced wireless network, including the system and channel model, performance enhancement, implementation, challenges, and potential solutions.                                                                                                                                                                                                                                                                                                                                                                                                                                                                       \\ \hline
		\cite{ning2024movable}       & Magazine      & An overview of the general architectures and practical implementations of MA/6DMA.                                                                                                                                                                                                                                                                                                                                                                                                                                                           \\ \hline
		\cite{zheng2024flexible}      & Magazine     & An overview of flexible-position MIMO for wireless communications, including the hardware implementations, potential applications, channel hardening, and trajectory optimization.                                                                                                                                                                                                                                                                                                                                                                                                                                        \\ \hline
		\cite{WangC_FAS_RIS_AI_survey} & Magazine & An overview of artificial intelligence (AI)-empower FAS-MIMO, which summarized the opportunities and design challenges, with a highlight on the utilization of AI approaches.                                                                                                                                                                                                                                                                                                                                                                                                                                                                                        \\ \hline
		\cite{Shah2024survey}       & Survey       & A survey of the FAMA technology, introducing its concept, realization, advantages, and combination with other emerging technologies.                                                                                                                                                                                                                                                                                                                                                                                                                                                                                  \\ \hline
		\cite{wong2022bruce}       & Tutorial       & A review of channel models, performance characterization, and hardware implementations of FAS and several promising topics for 6G.                                                                                                                                                                                                                                                                                                                                                                                                                                                                                                                        \\ \hline
		\cite{new2024tutorial}      & Tutorial       & A tutorial on FAS for 6G networks that covers channel modeling, channel estimation, performance analysis, and multiple access, mainly based on spatial-correlation channel models under rich-scattering environments.                                                                                                                                                                                                                      \\ \hline
		\makecell[c]{This \\paper}                  & Tutorial                       & Provide a \emph{more comprehensive review} of the historical development, implementation architectures, and applications of MAs in future wireless communication and/or sensing systems; Introduce the \emph{general field-response channel model} applicable to the narrowband/wideband systems, far-field/near-field conditions, and 6DMA with flexible antenna position/orientation reconfigurations; Formulate a \emph{generic optimization framework} for MA-aided wireless systems and address their two \emph{major design issues}, namely, antenna movement optimization and channel acquisition; Present the state-of-the-art \emph{prototypes and experimental results}; Discuss the \emph{extensions} to other wireless systems and \emph{future research directions}. \\ \hline
	\end{tabular}
\end{table*}

\subsection{Applications}
By leveraging the adaptive antenna movement, MAs offer enormous opportunities to unlock new dimensions for improving the adaptability, resilience, and efficiency of future wireless networks. As shown in Fig. \ref{Fig_Applications}, MAs can be applied to various communication and sensing systems for enhancing their performance.  

\subsubsection{Mobile Networks}
Conventional FA-enabled BSs/access points (APs) usually suffer from random channel fading due to the mobility of users and scatterers. Given the fixed positions and orientations of antennas, these systems can only allocate limited resources in the time, frequency, or power domain to meet communication demands. As a result, their ability to maintain high spectrum and energy efficiency is constrained in the case of poor channel conditions. In contrast, MA/6DMA-aided BSs/APs offer a new DoF by dynamically reconfiguring the positions and/or orientations of antennas to enhance channel conditions for mobile users \cite{xiao2023multiuser,wu2023movable, Yang2024movable,shao20246DMA,shao2024discrete,shao2024Mag6DMA,zhu2024nearfield}. This adaptability allows for more efficient utilization of time, frequency, and power resources, and thus improves the network performance effectively. In practice, the movement period of the antennas can be flexibly designed to adapt to either instantaneous or statistical channel variations, allowing for customizable trade-offs between the communication performance and the overhead associated with antenna movement. For example, for low-mobility users with slowly varying channels, the antenna movement can be conducted based on instantaneous channels \cite{zhu2023MAmultiuser,xiao2023multiuser,wu2023movable, Yang2024movable}. While for high-mobility users with fast fading channels, the antenna position can be reconfigured based on statistical channels over a long time period \cite{Hu20242024twotimeMA,zhu2024nearfield,zheng2024twotimeMA}, which circumvents the frequent antenna movement in practice. Moreover, by adjusting both 3D positions and 3D rotations of antennas based on channel spatial distributions \cite{shao20246DMA,shao2024discrete,shao2024Mag6DMA}, 6DMA-enabled BSs/APs can better respond to dynamic user distributions and varying environmental conditions, significantly enhancing wireless coverage and long-term throughput in future mobile networks.

\subsubsection{Machine-Type Communications}
With the ongoing development of the Internet of Things (IoT), future wireless networks are expected to support a vast number of machine-type communication (MTC) devices deployed in environments such as smart cities, automated industries, and smart homes. These devices are typically positioned in confined areas with low or no mobility, and their surrounding environments usually exhibit slow temporal variations. Under such conditions, the deployment of MAs on MTC devices can effectively exploit the spatial diversity of the wireless channel \cite{zhu2023MAMag}, especially in scenarios where conventional time or frequency diversities are insufficient, such as narrowband IoT (NB-IoT) or long-range (LoRa) communication systems. In industrial IoT, where efficient spatial interference suppression is crucial, MAs/6DMAs can be relocated and/or rotated to optimize channel conditions between desired transceivers, while minimizing interference from undesired sources, even with a limited number of antennas \cite{WangHH_interference_MA}. 

\subsubsection{Satellite Communications}
In satellite communication systems, particularly for low-earth orbit (LEO) satellites, FA arrays are typically employed to synthesize directional beams, which can compensate for the significant path loss resulting from the long signal propagation distance to ground users. As LEO satellites travel in their orbits, the beam coverage requirements change over time. Conventional FA arrays are limited to adapting to these variations through analog or digital beamforming alone. In comparison, MA arrays introduce a reconfigurable geometry, which enables more flexible and optimized beam patterns by jointly adjusting antenna positions and beamforming weights \cite{ZhuLP_satellite_MA,lin2024power}. This capability allows MA arrays to enhance beam coverage for terrestrial users as well as mitigate interference efficiently. Moreover, due to the high altitude of the satellites, the angle of departure/arrival (AoD/AoA) for line-of-sight (LoS) links between LEO satellites and terrestrial users exhibit slow variation and periodicity over time, which makes the reconfiguration of MA positions both feasible and efficient.

\subsubsection{Airborne Communications}
With the rapid growth of the low-altitude economy, various manned and unmanned aerial platforms are emerging to support applications such as transportation, logistics, and surveillance. Meanwhile, high-altitude platforms are gaining prominence due to their ability to deliver long endurance and wide-area coverage in these scenarios. Wireless communication plays a crucial role in enabling control and payload data transmission for these aerial platforms. Given the distinct altitudes of such platforms, airborne communication features unique full 3D coverage requirements. The integration of MA/6DMA can further enhance 3D coverage efficiency and ensure seamless connectivity between airborne platforms and space/ground stations \cite{kuang2024movableISAC,ren20246DMAUAV,liu2024uav6DMA}. Additionally, airborne communications often face spatially asymmetric and temporally varying demands due to the inherent mobility of these platforms. MAs introduce additional DoFs for flexible antenna and beam resource allocation, enabling improved communication performance and effective interference mitigation across different links.

\subsubsection{Maritime Communications}
In maritime communication systems, reliable connectivity is essential for vessels traveling across vast oceanic regions. Conventional FA systems on ships or offshore platforms often struggle to maintain consistent communication links due to the dynamic nature of the maritime environment, including frequent movement of vessels, as well as environmental factors such as waves and weather \cite{Wang2018maritime, Alqurashi2023maritime}. In comparison, MA/6DMA systems offer a significant advantage in these scenarios by dynamically adjusting antenna positions and/or orientations to improve channel conditions with satellites, shore stations, or other vessels. By reconfiguring the geometry of the antenna array in response to the varying signal conditions, MA systems can improve link reliability, enhance coverage, and reduce communication outages. Additionally, the capability of MAs to track moving targets and mitigate interference allows for more efficient use of the available spectrum, particularly in crowded maritime environments where multiple communication links are required for navigation, safety, and operational coordination.

\subsubsection{Wireless Sensing}
Wireless sensing is essential for applications like environmental monitoring, smart cities, and autonomous systems, providing real-time data and improved decision-making. As the demand for more accurate and efficient sensing grows, large-scale antenna arrays are usually employed to improve the sensing resolution and coverage, which, however, result in higher hardware complexity and energy consumption. In this context, MA arrays offer a new DoF through antenna position reconfiguration \cite{ma2024MAsensing, Chen2024moving, shao2024exploiting}. By expanding the movement region of the antennas, MA systems can effectively increase the array aperture and thus enhance sensing performance without increasing the number of antennas. Additionally, the flexible geometry of MA arrays reduces the correlation between steering vectors over different directions, which can efficiently mitigate interference and ambiguity in the angular domain. Furthermore, real-time antenna movement adjustments allow MA-enabled sensing systems to adapt to dynamic environments and diverse sensing task requirements, thus providing higher design flexibility as compared to conventional FA-based radar or SAR systems \cite{Moreira2013SAR}.

\subsubsection{ISAC}
As the demand for more efficient and versatile communication networks increases, ISAC has emerged as a critical trend in the development of the 6G mobile networks. Specifically, ISAC systems combine sensing and communication functionalities into a unified framework, thereby enhancing resource utilization and enabling advanced applications such as autonomous driving and smart cities. MAs can play a pivotal role in this context by enhancing performance in both sensing and communication systems. This adaptability not only facilitates the dynamic allocation of resources but also improves the trade-offs between communication quality and sensing accuracy \cite{ma2025MAISAC,li2024MAISACMag, WuHS_MA_RIS_ISAC, lyu2024flexibleISAC, peng2024jointISAC, kuang2024movableISAC, wang2024multiuser}. Moreover, by reconfiguring antenna positions and/or orientations in response to varying operational requirements, the antenna movement capability allows for flexible function switching between communication and sensing tasks. As such, MA/6DMA serves as an important candidate technology for meeting the diverse and evolving demands of next-generation wireless networks.

\begin{figure}[t]
	\begin{center}
		\includegraphics[width=\figwidth cm]{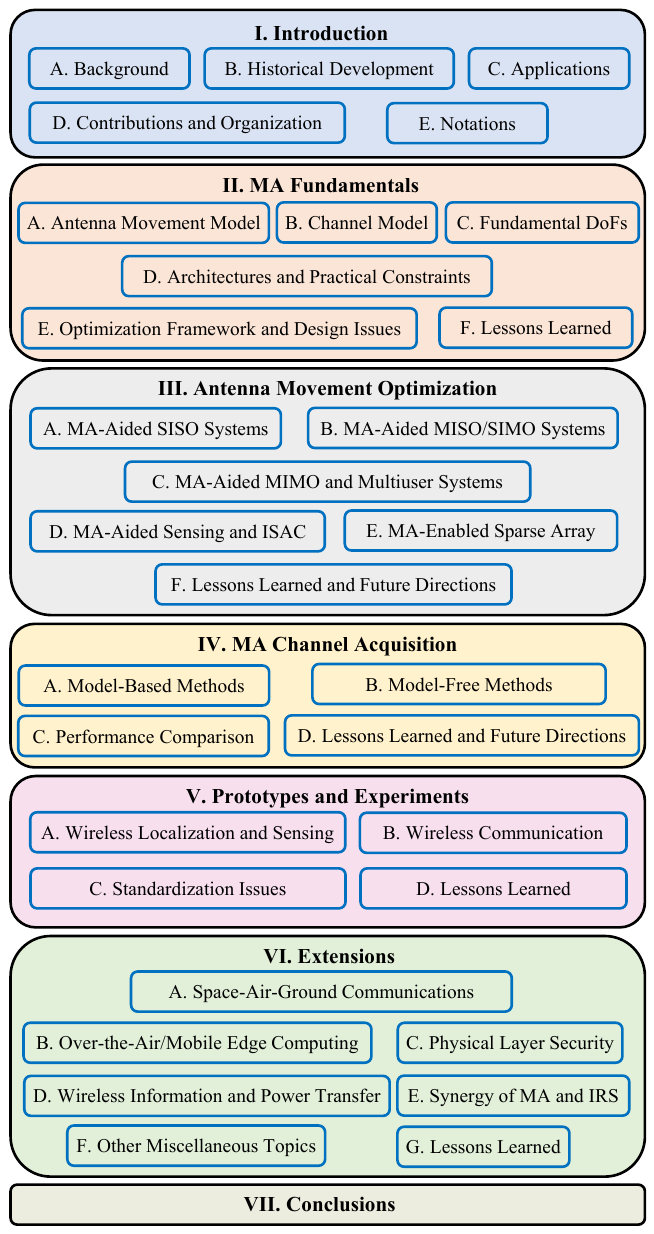}
		\caption{Organization of this paper.}
		\vspace{-12pt}
		\label{Fig_organization}
	\end{center}
\end{figure}

\subsection{Contributions and Organization}
Given the significant technical potential and fertile application prospects of MAs, this paper offers a comprehensive tutorial on the fundamentals and advancements for MA-aided wireless networks. Specifically, it covers the implementation architectures, general channel models, major design issues, proof-of-concept prototypes, and various extensions of MA systems, aiming to provide foundational knowledge for researchers in this field and to inspire future investigations. The comparison of this paper and other related overview articles is summarized in Table \ref{Tab_overview_article}. In particular, two tutorial articles \cite{wong2022bruce} and \cite{new2024tutorial} mainly investigated FAS for 6G communication systems, while the survey paper \cite{Shah2024survey} only focused on a specific branch named fluid antenna multiple access (FAMA). The hardware architectures of FAS introduced in \cite{wong2022bruce,new2024tutorial,Shah2024survey} mainly concentrated on liquid-based and pixel-based antennas, while the mechanically movable architectures were not thoroughly reviewed. From the channel modeling perspective, they mainly studied the spatial-correlation channel models under rich-scattering environments, limited to narrowband systems and 1D/2D flexibility in antenna position. Thus, the corresponding analytical results and channel estimation methods for FAS in \cite{wong2022bruce,new2024tutorial,Shah2024survey} are generally inefficient for scenarios with finite channel paths, wideband systems, or 3D antenna movement/rotation. In addition, the magazine papers \cite{zhu2023MAMag,zheng2024flexible,WangC_FAS_RIS_AI_survey,ning2024movable,shao2024Mag6DMA} all lack the comprehensive introduction and comparison of different channel modeling methods, various hardware architectures, detailed optimization approaches, as well as the state-of-the-art prototypes of MA-aided wireless systems. In comparison, the main contributions of this paper over these existing works are summarized as follows:
\begin{itemize}
	\item  We provide an overview of the historical development of MA technologies across the antenna and communication communities. The promising application scenarios of MAs in future wireless networks are envisioned, such as mobile networks, MTC, satellite communications, maritime communication, wireless sensing, ISAC, and so on. 
	\item The state-of-the-art hardware architectures for implementing MAs are comprehensively reviewed, utilizing the taxonomy of element-level and array-level movements. The advantages and shortcomings of different architectures are compared to inspire customized designs for different wireless systems.
	\item We present the general field-response channel model for characterizing the continuous spatial variation of wireless channels with respect to (w.r.t.) antenna movement, which is applicable to narrowband/wideband systems, far-field/near-field propagation scenarios, and antenna position/orientation reconfigurations.
	\item We characterize the fundamental DoFs for antenna movement in reconfiguring wireless channels and formulate a general optimization framework to jointly optimize the antenna movement and resource allocation for MA-aided wireless systems.
	\item We delve into the major design issues for MA systems from the perspective of antenna movement optimization and channel acquisition. Under the new field-response channel models, efficient algorithms are presented to attain performance gains over FA-based systems under various application scenarios, environmental conditions, and design objectives.
	\item We review the existing prototypes of the MA-aided wireless communication/sensing systems and show the experimental results to validate their performance advantages. Moreover, the extension of MA design to other wireless systems and its synergy with other emerging wireless technologies are discussed.
\end{itemize}

\begin{table*}[]\footnotesize  
	\centering
	\caption{List of acronyms.}
	\label{Tab_acronym}
	\begin{tabular}{|c|c|c|c|}
		\hline
		\textbf{Acronyms} & \textbf{Meaning}                                                                                               & \textbf{Acronyms} & \textbf{Meaning}                                                                                \\ \hline
		1D                & One-dimensional                                                                                                & LEO               & Low-earth orbit                                                                                 \\ \hline
		2D                & Two-dimensional                                                                                                & LoRa              & Long-range                                                                                      \\ \hline
		3D                & Three-dimensional                                                                                              & LoS               & Line-of-sight                                                                                   \\ \hline
		3GPP              & 3rd Generation Partnership Project                                                                             & LS                & Least square                                                                                    \\ \hline
		6D                & Six-dimensional                                                                                                & MA                & Movable antenna                                                                                 \\ \hline
		6DMA              & Six-dimensional movable antenna                                                                                & MEC               & Mobile edge computing                                                                           \\ \hline
		6G                & Sixth generation                                                                                               & MEMS              & Micro-electromechanical system                                                                  \\ \hline
		ACCS              & Antenna-centric coordinate system                                                                              & MIMO              & Multiple-input multiple-output                                                                  \\ \hline
		AF                & Amplify-and-forward                                                                                            & MISO              & Multiple-input single-output                                                                    \\ \hline
		AI                & Artificial intelligence                                                                                        & MMSE              & Minimum mean squared error                                                                      \\ \hline
		AirComp           & Over-the-air computation                                                                                       & mmWave            & Millimeter-wave                                                                                 \\ \hline
		ALS               & Alternating least square                                                                                       & MoA               & Modular array                                                                                   \\ \hline
		AN                & Artificial noise                                                                                               & MRA               & Minimum redundant array                                                                         \\ \hline
		AO                & Alternating optimization                                                                                       & MRT               & Maximal ratio transmitting                                                                      \\ \hline
		AoA               & Angle of arrival                                                                                               & MSE               & Mean square error                                                                               \\ \hline
		AoD               & Angle of departure                                                                                             & MTC               & Machine-type communication                                                                      \\ \hline
		AOM               & Antenna orientation matrix                                                                                     & MUSIC             & Multiple signal classification                                                                  \\ \hline
		AP                & Access point                                                                                                   & NA                & Nested array                                                                                    \\ \hline
		APV               & Antenna position vector                                                                                        & NB-IoT            & Narrowband IoT                                                                                  \\ \hline
		AS                & Antenna selection                                                                                              & NFRV              & Near-FRV                                                                                        \\ \hline
		AWGN              & Additive white Gaussian noise                                                                                  & NLoS              & Non-LoS                                                                                         \\ \hline
		AWV               & Antenna weight vector                                                                                          & NMSE              & Normalized mean square error                                                                    \\ \hline
		BnB               & Branch and bound                                                                                               & NOMA              & Non-orthogonal multiple access                                                                  \\ \hline
		BS                & Base station                                                                                                   & NUSA              & Non-uniform sparse array                                                                        \\ \hline
		CDL               & Clustered delay line                                                                                           & OFDM              & Orthogonal frequency division multiplexing                                                      \\ \hline
		CFR               & Channel frequency response                                                                                     & OMP               & Orthogonal matching pursuit                                                                     \\ \hline
		CIR               & Channel impulse response                                                                                       & PLS               & Physical layer security                                                                         \\ \hline
		CKM               & Channel knowledge map                                                                                          & PPRM              & Path polarization response matrix                                                               \\ \hline
		CMSE              & Computation mean square error                                                                                  & PRM               & Path response matrix                                                                            \\ \hline
		CP                & Canonical polyadic                                                                                             & PSI               & Path state information                                                                          \\ \hline
		CPA               & Co-prime array                                                                                                 & PSO               & Particle swarm optimization                                                                     \\ \hline
		CR                & Cognitive radio                                                                                                & PU                & Primary user                                                                                    \\ \hline
		CRB               & Cramer-Rao bound                                                                                               & RF                & Radio frequency                                                                                 \\ \hline
		CSCG              & Circularly symmetric complex Gaussian                                                                          & RFID              & RF identification                                                                               \\ \hline
		CSI               & Channel state information                                                                                      & RHS               & Reconfigurable holographic surface                                                              \\ \hline
		DF                & Decode-and-forward                                                                                             & RSMA              & Rate-splitting multiple access                                                                  \\ \hline
		DFT               & Discrete Fourier transform                                                                                     & Rx                & Receiver                                                                                        \\ \hline
		DMA               & Dynamic metasurface antenna                                                                                    & SAR               & Synthetic aperture radar                                                                        \\ \hline
		DoFs              & Degrees of freedom                                                                                             & SCA               & Successive convex approximation                                                                 \\ \hline
		EDoF              & Effective DoF                                                                                                  & SIMO              & Single-input multiple-output                                                                    \\ \hline
		ESPRIT            & \begin{tabular}[c]{@{}c@{}}Estimation of signal parameters via \\ rotational invariance technique\end{tabular} & SINR              & Signal-to-interference-plus-noise ratio                                                         \\ \hline
		Eve               & Eavesdropper                                                                                                   & SISO              & Single-input single-output                                                                      \\ \hline
		FA                & Fixed antenna                                                                                                  & SNR               & Signal-to-noise ratio                                                                           \\ \hline
		FAMA              & Fluid antenna multiple access                                                                                  & SU                & Secondary user                                                                                  \\ \hline
		FAS               & Fluid antenna system                                                                                           & SVO               & Steering vector orthogonality                                                                   \\ \hline
		FPA               & Fixed-position antenna                                                                                         & SWIPT             & \begin{tabular}[c]{@{}c@{}}Simultaneous wireless information \\ and power transfer\end{tabular} \\ \hline
		FRI               & Field response information                                                                                     & THz               & Terahertz                                                                                       \\ \hline
		FRM               & Field response matrix                                                                                          & Tx                & Transmitter                                                                                     \\ \hline
		FRV               & Field response vector                                                                                          & UAV               & Unmanned aerial vehicle                                                                         \\ \hline
		GMA               & Group MA                                                                                                       & ULA               & Uniform linear array                                                                            \\ \hline
		GNSS              & Global navigation satellite system                                                                             & UPA               & Uniform planar array                                                                            \\ \hline
		INS               & Inertial navigation system                                                                                     & USA               & Uniform sparse array                                                                            \\ \hline
		IoT               & Internet of Things                                                                                             & WPT               & Wireless power transfer                                                                         \\ \hline
		IRS               & Intelligent reflecting surface                                                                                 & w.r.t.            & With respect to                                                                                 \\ \hline
		ISAC              & Integrated sensing and communication                                                                           & XL-MIMO           & Extremely large-scale MIMO                                                                      \\ \hline
		LCS               & Local coordinate system                                                                                        & ZF                & Zero-forcing                                                                                    \\ \hline
	\end{tabular}
\end{table*}

The organization of this paper is shown in Fig. \ref{Fig_organization}. Specifically, we show in Section \ref{Sec_Intro} the motivation of MA-enhanced wireless networks, followed by their historical development and future applications. Section \ref{Sec_Fundamental} presents the fundamentals of MA systems, including the antenna movement models, channel models, hardware architectures, and a unified optimization framework. Section \ref{Sec_Movement} investigates the antenna movement optimization under various MA-aided wireless systems, while Section \ref{Sec_Channel} studies the MA channel acquisition. Section \ref{Sec_Prototype} demonstrates the prototypes of the MA systems and the corresponding experimental results. In Section \ref{Sec_Extension}, we discuss the extensions of MA to other wireless systems and technologies. Finally, this paper is concluded in Section \ref{Sec_Conclusion}. The acronyms used in this paper are summarized in Table \ref{Tab_acronym}.

\subsection{Notations}
$a$, $\mathbf{a}$, $\mathbf{A}$, and $\mathcal{A}$ denote a scalar, a vector, a matrix, and a set, respectively. $(\cdot)^{\rm{T}}$ and $(\cdot)^{\rm{H}}$ denote transpose and conjugate transpose, respectively. $\mathbf{Q} \succeq \mathbf{0}$ indicates that $\mathbf{Q}$ is a positive semi-definite matrix. $\mathrm{Tr}(\cdot)$ and $\det(\cdot)$ denote the trace and determinant of a square matrix, respectively. $[\mathbf{a}]_i$ and $[\mathbf{A}]_{i,j}$ denote the $i$-th entry of vector $\mathbf{a}$ and the entry in the $i$-th row and $j$-th column of matrix $\mathbf{A}$, respectively. $\mathcal{A} \setminus \mathcal{B}$ and $\mathcal{A} \cap \mathcal{B}$ represent the subtraction set and intersection set of $\mathcal{A}$ and $\mathcal{B}$, respectively. $\mathcal{CN}(\mathbf{0},\mathbf{\Lambda})$ denotes the circularly symmetric complex Gaussian (CSCG) distribution with mean zero and covariance matrix $\mathbf{\Lambda}$. $\mathbb{E}\{\cdot\}$ denotes the expectation of a random variable. $\mathbb{R}^{M \times N}$ and $\mathbb{C}^{M \times N}$ represent the sets of real and complex matrices/vectors of dimension $M \times N$, respectively. $\mathbb{Z}$ denotes the set of integers. $|\cdot|$ and $\angle(\cdot)$ denote the amplitude and the phase of a complex number or complex vector, respectively. $\|\cdot\|_{1}$ and $\|\cdot\|_{2}$ denote the 1-norm and 2-norm of a vector, respectively. $\|\cdot\|_{\mathrm{F}}$ represents the Frobenius norm of a matrix. $\mathbf{1}_{N}$ and $\mathbf{0}_{N}$ denote an $N$-dimensional column vector with all elements equal to 1 and 0, respectively. $\mathbf{I}_{N}$ represents the $N$-dimensional identical matrix. Operators $\times$, $\odot$, and $\otimes$ denote the cross product, Khatri-Rao product, and Kronecker product, respectively. $\mathrm{e}$ and $\mathrm{j}=\sqrt{-1}$ denote the Euler's number and the imaginary unit, respectively.

\section{MA Fundamentals} \label{Sec_Fundamental}
This section introduces the fundamentals of MA-aided wireless systems. First, a general antenna movement model is provided to demonstrate the flexible position and orientation of an antenna in 3D space \cite{zhu2023MAMag,shao2024Mag6DMA}. Then, the field-response channel model tailored for MA systems is presented, which characterizes the spatial variation of wireless channels w.r.t. antenna position and/or orientation. Based on this channel model, the fundamental DoFs provided by antenna movement in improving wireless channels are discussed. Subsequently, existing hardware architectures for implementing MAs and their associated constraints are reviewed. Finally, a unified optimization framework is developed for antenna movement optimization to improve the performance of wireless systems.

\subsection{Antenna Movement Model}
\begin{figure}[t]
	\begin{center}
		\includegraphics[width=\figwidth cm]{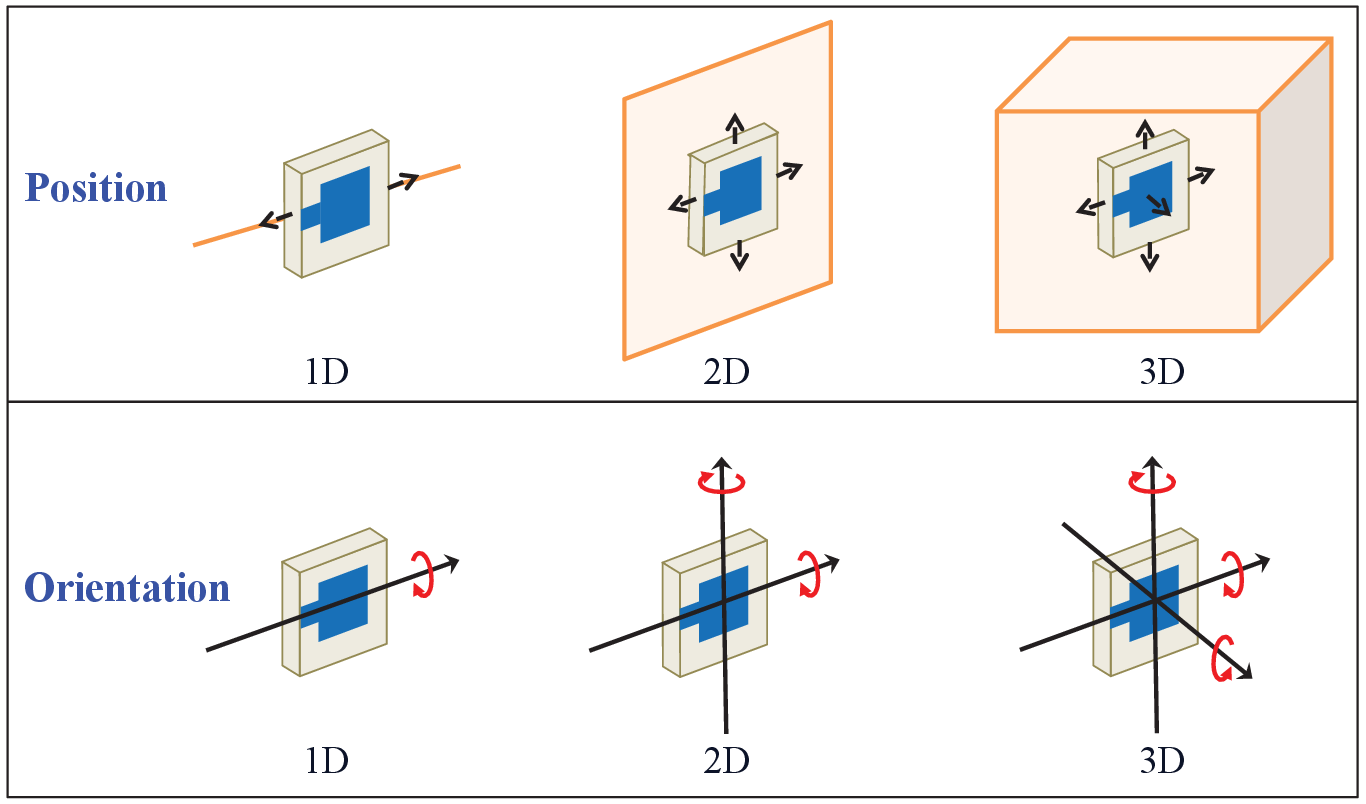}
		\caption{Antenna movement model for a 6DMA system with flexible position and orientation \cite{shao2024Mag6DMA}.}
		\vspace{-12pt}
		\label{Fig_Antenna_Movement}
	\end{center}
\end{figure}
In practice, considering the specific implementation architectures, system requirements, and operating environments, the system can be flexibly designed by leveraging different DoFs in antenna movement from one-dimensional (1D) to 6D. In this section, we illustrate the antenna movement model based on the general 6DMA framework \cite{shao2024Mag6DMA}, while the system with lower DoFs in antenna movement can be realized by fixing some dimensions or imposing some constraints on antenna position and/or orientation. The antenna movement model for a 6DMA is shown in Fig. \ref{Fig_Antenna_Movement}, which includes changing the antenna position and orientation in 1D, two-dimensional (2D), or 3D spatial space. 

\begin{figure}[t]
	\begin{center}
		\includegraphics[width=\figwidth cm]{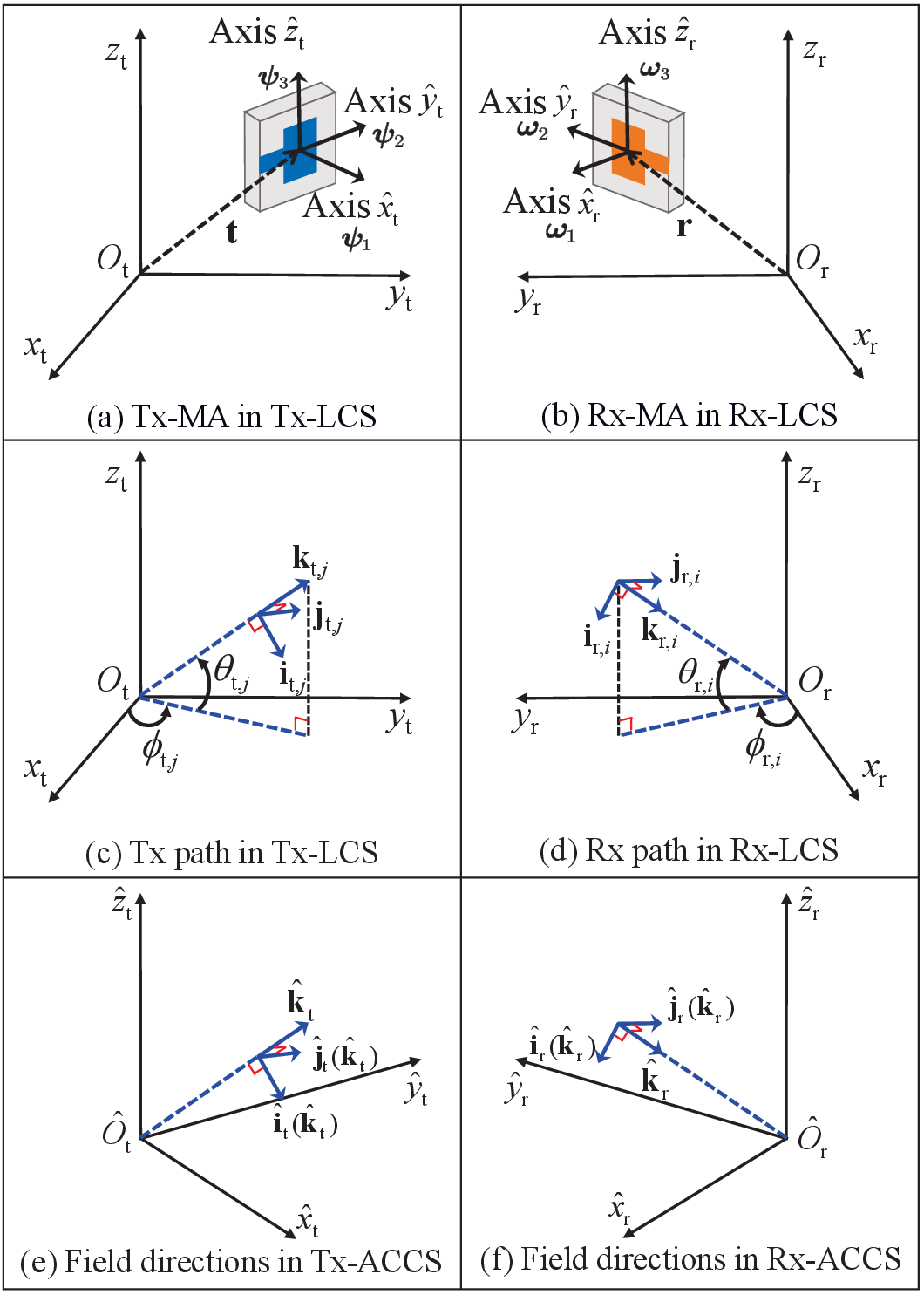}
		\caption{Illustration of the APVs, AOMs, AoDs, AoAs, and wave vectors defined in the LCSs at the Tx and Rx.}
		\vspace{-12pt}
		\label{Fig_Coordinate}
	\end{center}
\end{figure}

\begin{figure*}[t]
	\begin{center}
		\includegraphics[width=12 cm]{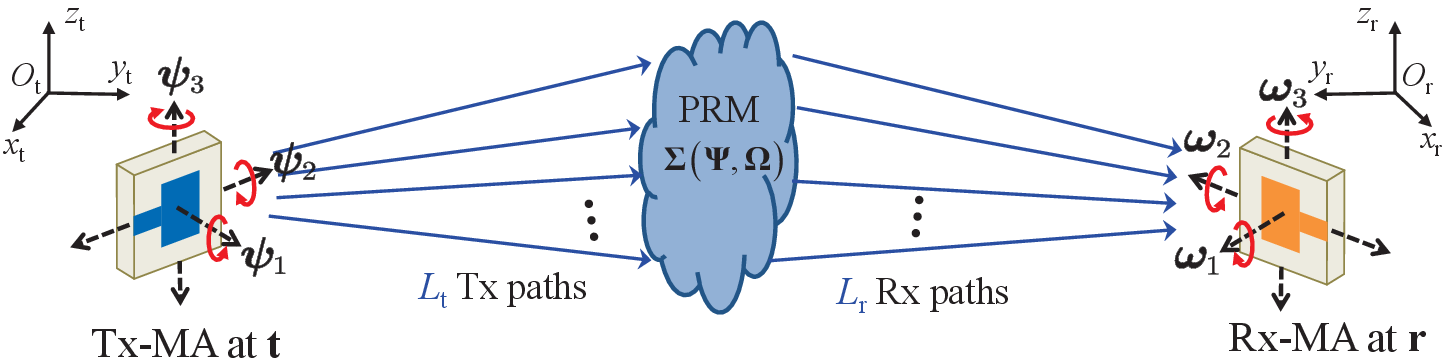}
		\caption{Illustration of the field-response channel model between the Tx-MA and Rx-MA under the far-field condition.}
		\vspace{-12pt}
		\label{Fig_Channel}
	\end{center}
\end{figure*}

As shown in Fig. \ref{Fig_Coordinate}, we establish a 3D local coordinate system (LCS) to depict the position and orientation of a 6DMA at the Tx/Rx \cite{shao20246DMA,shao2024discrete}. Due to the independent antenna movement at the Tx and Rx, we adopt different LCSs with different notations at the Tx and Rx for distinction. Specifically, the position of a Tx-MA in the Tx-LCS shown in Fig. \ref{Fig_Coordinate}(a) is denoted as $\mathbf{t} = [x_{\mathrm{t}}, y_{\mathrm{t}}, z_{\mathrm{t}}]^{\mathrm{T}} \in \mathbb{R}^{3 \times 1}$, which refers to the phase center of the Tx-MA and is termed as the antenna position vector (APV). Besides, to characterize the orientation of the Tx-MA in the Tx-LCS, we establish a 3D antenna-centric coordinate system (ACCS) at the Tx with the origin at the phase center of the antenna. The three axes of the Tx-ACCS are respectively denoted as $\hat{x}_{\mathrm{t}}$, $\hat{y}_{\mathrm{t}}$, and $\hat{z}_{\mathrm{t}}$, which can be defined as any three orthonormal directions associated with the Tx-MA. For different shapes of antennas, we may adopt different definitions of the directions of axes for convenience. For example, for a line-shaped antenna, we can define axis $\hat{x}_{\mathrm{t}}$ along the direction of the antenna line, while axes $\hat{y}_{\mathrm{t}}$ and $\hat{z}_{\mathrm{t}}$ are the corresponding two orthogonal directions perpendicular to axis $\hat{x}_{\mathrm{t}}$. For a rectangular-shaped antenna shown in Fig. \ref{Fig_Coordinate}(a), we can define axes $\hat{x}_{\mathrm{t}}$ and $\hat{y}_{\mathrm{t}}$ as the two directions parallel to the two sides of the rectangular, while axis $\hat{z}_{\mathrm{t}}$ is perpendicular to the antenna surface. Denote $\boldsymbol{\psi}_{1}$, $\boldsymbol{\psi}_{2}$, and $\boldsymbol{\psi}_{3}$ as the coordinates of the directions of axes $\hat{x}_{\mathrm{t}}$, $\hat{y}_{\mathrm{t}}$, and $\hat{z}_{\mathrm{t}}$ in the Tx-LCS, respectively. Then, the transform matrix between the Tx-LCS and the Tx-ACCS is given by $\mathbf{\Psi}=[\boldsymbol{\psi}_{1},\boldsymbol{\psi}_{2},\boldsymbol{\psi}_{3}] \in \mathbb{R}^{3 \times 3}$, which uniquely determines the orientation of the Tx-MA w.r.t. the Tx-LCS and is termed as the antenna orientation matrix (AOM). Similarly, as shown in Fig. \ref{Fig_Coordinate}(b), the APV and AOM of the Rx-MA in the Rx-LCS are denoted as $\mathbf{r} = [x_{\mathrm{r}}, y_{\mathrm{r}}, z_{\mathrm{r}}]^{\mathrm{T}} \in \mathbb{R}^{3 \times 1}$ and $\mathbf{\Omega}=[\boldsymbol{\omega}_{1},\boldsymbol{\omega}_{2},\boldsymbol{\omega}_{3}] \in \mathbb{R}^{3 \times 3}$, respectively.

%where $\boldsymbol{\psi}_{1}$ denotes the unit orientation vector of the antenna with the maximum radiation gain over this direction, $\boldsymbol{\psi}_{2}$ is the unit polarization vector orthogonal to $\boldsymbol{\psi}_{1}$, and $\boldsymbol{\psi}_{3} = \boldsymbol{\psi}_{1} \times \boldsymbol{\psi}_{2}$ is orthogonal to both $\boldsymbol{\psi}_{1}$ and $\boldsymbol{\psi}_{2}$. Thus, $\mathbf{\Psi}$ is an orthogonal matrix that uniquely determines the orientation of the Tx-MA in its LCS.

\subsection{Channel Model}
Given the antenna movement model, we present the field-response channel model for MA systems in this subsection\footnote{The field-response channel model characterizes the response between the transmit and receive antennas, which is determined by the propagation environments and remains independent of specific system functionalities. Thus, it applies to both communication and sensing systems.}. For ease of exposition, we begin with the basic field-response channel model w.r.t. the positions of MAs for narrowband single-input single-output (SISO) systems under the far-field condition. Then, its extensions to MIMO systems, wideband systems, and near-field channel models are provided. Moreover, the field-response channel model is further extended to the general case as a function of both the APVs and AOMs of the Tx-MA and Rx-MA. Finally, other existing channel models for MA systems are comprehensively reviewed.
\subsubsection{Basic Field-Response Channel Model}
We temporarily assume fixed orientations of both the Tx-MA and Rx-MA and focus on the channel variation w.r.t. the positions of antennas. Denoting the length of the antenna moving region in each dimension at the Tx/Rx as $A$ and the carrier wavelength as $\lambda$, the corresponding Rayleigh distance is given by $2A^{2}/\lambda$. For example, for $A=5\lambda$ and $\lambda=0.05$ meter (m) at 6 GHz carrier frequency, the Rayleigh distance equals $2.5$ m. If the distance between the Tx and Rx as well as that between the Tx/Rx and scatterers exceeds the Rayleigh distance, the far-field condition holds and the planar wave model can be adopted. As shown in Fig. \ref{Fig_Channel}, the channel response between the Tx-MA and Rx-MA can be characterized by the superposition of $L_{\mathrm{t}}$ Tx paths and $L_{\mathrm{r}}$ Rx paths. In particular, denote the elevation and azimuth AoDs for the $j$-th ($1 \leq j \leq L_{\mathrm{t}}$) Tx path as $\theta_{\mathrm{t},j}$ and $\phi_{\mathrm{t},j}$, respectively, as demonstrated in Fig. \ref{Fig_Coordinate}(c). Then, we can obtain the wave vector for the $j$-th Tx path as $\mathbf{k}_{\mathrm{t},j} = [\cos{\theta_{\mathrm{t},j}}\cos{\phi_{\mathrm{t},j}},\cos{\theta_{\mathrm{t},j}}\sin{\phi_{\mathrm{t},j}}, \sin{\theta_{\mathrm{t},j}}]^{\mathrm{T}}$. The difference in the signal propagation distance for the $j$-th Tx path between position $\mathbf{t}$ and the reference point (i.e., the origin $O_{\mathrm{t}}$) is thus given by $\mathbf{k}_{\mathrm{t},j}^{\mathrm{T}}\mathbf{t}$. As such, we can define the field response vector (FRV) to account for the phase variation of all Tx channel paths w.r.t. the position of the Tx-MA as \cite{zhu2022MAmodel}
\begin{equation}\label{eq_FRV_Tx}
	\mathbf{g}(\mathbf{t})=\left[\mathrm{e}^{\mathrm{j}\frac{2\pi}{\lambda}\mathbf{k}_{\mathrm{t},1}^{\mathrm{T}}\mathbf{t}}, \mathrm{e}^{\mathrm{j}\frac{2\pi}{\lambda}\mathbf{k}_{\mathrm{t},2}^{\mathrm{T}}\mathbf{t}}, \dots, \mathrm{e}^{\mathrm{j}\frac{2\pi}{\lambda}\mathbf{k}_{\mathrm{t},L_{\mathrm{t}}}^{\mathrm{T}}\mathbf{t}}\right]^{\mathrm{T}} \in \mathbb{C}^{L_{\mathrm{t}} \times 1},
\end{equation}
where $\lambda$ denotes the signal wavelength. 

Similarly, as shown in Fig. \ref{Fig_Coordinate}(d), we denote the elevation and azimuth AoAs for the $i$-th ($1 \leq i \leq L_{\mathrm{r}}$) Rx path as $\theta_{\mathrm{r}, i}$ and $\phi_{\mathrm{r}, i}$, respectively. The wave vector for the $i$-th Rx path can be obtained as $\mathbf{k}_{\mathrm{r},i} = [\cos{\theta_{\mathrm{r},i}}\cos{\phi_{\mathrm{r},i}},\cos{\theta_{\mathrm{r},i}}\sin{\phi_{\mathrm{r},i}}, \sin{\theta_{\mathrm{r},i}}]^{\mathrm{T}}$. The FRV to account for the phase variation of all Rx channel paths w.r.t. the position of the Rx-MA can thus be defined as \cite{zhu2022MAmodel}
\begin{equation}\label{eq_FRV_Rx}
	\mathbf{f}(\mathbf{r})=\left[\mathrm{e}^{\mathrm{j}\frac{2\pi}{\lambda}\mathbf{k}_{\mathrm{r},1}^{\mathrm{T}}\mathbf{r}}, \mathrm{e}^{\mathrm{j}\frac{2\pi}{\lambda}\mathbf{k}_{\mathrm{r},2}^{\mathrm{T}}\mathbf{r}}, \dots, \mathrm{e}^{\mathrm{j}\frac{2\pi}{\lambda}\mathbf{k}_{\mathrm{r},L_{\mathrm{r}}}^{\mathrm{T}}\mathbf{r}}\right]^{\mathrm{T}} \in \mathbb{C}^{L_{\mathrm{r}} \times 1}.
\end{equation}

Furthermore, we define the path response matrix (PRM) $\mathbf{\Sigma} \in \mathbb{C}^{L_{\mathrm{r}} \times L_{\mathrm{t}}}$, with the entry in the $i$-th row and the $j$-th column representing the response coefficient between the $j$-th Tx path and the $i$-th Rx path. It is worth noting that the PRM is determined by various factors, including the orientation, radiation, and polarization of the Tx-MA and Rx-MA as well as the carrier frequency and the propagation environment between the Tx and Rx. An accurate modeling of the PRM under different scenarios is complicated and remains an open problem. Nevertheless, for any given $\mathbf{\Sigma}$, we can express the baseband equivalent channel between the Tx-MA and the Rx-MA as a function of their positions \cite{zhu2022MAmodel}, i.e.,
\begin{equation}\label{eq_channel_basic}
	h(\mathbf{t}, \mathbf{r}) = \mathbf{f}(\mathbf{r})^{\mathrm{H}} \mathbf{\Sigma}\mathbf{g}(\mathbf{t}).
\end{equation}
It is worth noting that the field-response channel model shown in \eqref{eq_channel_basic} is consistent with the channel models for conventional FA systems. Under specific conditions on the FRVs and PRMs, the field-response channel model can be degraded into the pure LoS channel (with a single channel path) and the geometric channel (with a diagonal PRM) as well as Rayleigh fading and Rician fading channels (with infinite channel paths) \cite{zhu2022MAmodel}.

For the case with a continuum set of (infinite) channel paths distributed in the angular domain, the channel response in \eqref{eq_channel_basic} can be represented by a general integral form as
\begin{equation}\label{eq_channel_inf}\small
	\iiiint_{\mathcal{A}} f^{*}(\mathbf{r}, \theta_{\mathrm{t}}, \phi_{\mathrm{t}})\sigma(\theta_{\mathrm{t}}, \phi_{\mathrm{t}}, \theta_{\mathrm{r}}, \phi_{\mathrm{r}})g(\mathbf{t}, \theta_{\mathrm{t}}, \phi_{\mathrm{t}}) \dif \theta_{\mathrm{t}} \dif \phi_{\mathrm{t}} \dif \theta_{\mathrm{r}} \dif \phi_{\mathrm{r}},
\end{equation}
where $\mathcal{A}=[-\pi/2, \pi/2] \times [-\pi, \pi] \times [-\pi/2, \pi/2] \times [-\pi, \pi]$ is the integral interval of the AoDs and AoAs. $g(\mathbf{t}, \theta_{\mathrm{t}}, \phi_{\mathrm{t}}) = \mathrm{e}^{\mathrm{j}\frac{2\pi}{\lambda}\mathbf{k}(\theta_{\mathrm{t}}, \phi_{\mathrm{t}})^{\mathrm{T}} \mathbf{t}}$ and 
$f(\mathbf{r}, \theta_{\mathrm{r}}, \phi_{\mathrm{r}}) = \mathrm{e}^{\mathrm{j}\frac{2\pi}{\lambda}\mathbf{k}(\theta_{\mathrm{r}}, \phi_{\mathrm{r}})^{\mathrm{T}} \mathbf{r}}$ denote the field-response functions at the Tx and Rx, respectively, with $\mathbf{k}(\theta_{\mathrm{t}}, \phi_{\mathrm{t}}) = [\cos{\theta_{\mathrm{t}}}\cos{\phi_{\mathrm{t}}},\cos{\theta_{\mathrm{t}}}\sin{\phi_{\mathrm{t}}}, \sin{\theta_{\mathrm{t}}}]^{\mathrm{T}}$ and $\mathbf{k}(\theta_{\mathrm{r}}, \phi_{\mathrm{r}}) = [\cos{\theta_{\mathrm{r}}}\cos{\phi_{\mathrm{r}}},\cos{\theta_{\mathrm{r}}}\sin{\phi_{\mathrm{r}}}, \sin{\theta_{\mathrm{r}}}]^{\mathrm{T}}$ being the wave vectors. $\sigma(\theta_{\mathrm{t}}, \phi_{\mathrm{t}}, \theta_{\mathrm{r}}, \phi_{\mathrm{r}})$ denotes the path response function, which characterizes the response between transmit and receive paths with arbitrary AoDs and AoAs. In practice, it is usually unnecessary to determine the exact value of the path response function across the continuum set of AoDs/AoAs, as it can be effectively approximated using the PRM over a discrete set of AoDs/AoAs \cite{TseFundaWC}. Specifically, the finite size of the antenna moving region in the spatial domain, i.e., $A$, results in a limited resolution in the angular domain. Physical paths with virtual AoAs/AoDs (i.e., components of the wave vectors) differing by less than $\frac{\lambda}{2A}$ are not resolvable by the MA \cite[Sec. 7.3.4]{TseFundaWC}. Consequently, the infinite set of physical paths can be grouped into angularly resolvable bins, each with an angular width of $\frac{\lambda}{2A}$, and aggregated to form a finite number of resolvable paths \cite[Sec. 7.3.4]{TseFundaWC}. Each dimension of the integral in \eqref{eq_channel_inf} can then be efficiently approximated by the superposition of these resolvable paths across $2A$ angular bins, resulting in a representation analogous to \eqref{eq_channel_basic}.

\subsubsection{Extension to MIMO Channel Model}
For MIMO systems with $N_{\mathrm{t}}$ Tx-MAs and $N_{\mathrm{r}}$ Rx-MAs, we can define the field response matrices (FRMs) for all antennas at the Tx and Rx as \cite{ma2022MAmimo}
\begin{subequations}
	\begin{align}\label{eq_FRM_Tx_Rx}
		&\mathbf{G}(\tilde{\mathbf{t}})=\left[\mathbf{g}(\mathbf{t}_{1}), \mathbf{g}(\mathbf{t}_{2}), \dots, \mathbf{g}(\mathbf{t}_{N_{\mathrm{t}}})\right]  \in \mathbb{C}^{L_{\mathrm{t}} \times N_{\mathrm{t}}},\\
		&\mathbf{F}(\tilde{\mathbf{r}})=\left[\mathbf{f}(\mathbf{r}_{1}), \mathbf{f}(\mathbf{r}_{2}), \dots, \mathbf{f}(\mathbf{r}_{N_{\mathrm{r}}})\right]  \in \mathbb{C}^{L_{\mathrm{r}} \times N_{\mathrm{r}}},
	\end{align}
\end{subequations}
where $\tilde{\mathbf{t}} \triangleq [\mathbf{t}_{1}^{\mathrm{T}}, \mathbf{t}_{2}^{\mathrm{T}}, \dots, \mathbf{t}_{N_{\mathrm{t}}}^{\mathrm{T}}]^{\mathrm{T}}$ and $\tilde{\mathbf{r}} \triangleq [\mathbf{r}_{1}^{\mathrm{T}}, \mathbf{r}_{2}^{\mathrm{T}}, \dots, \mathbf{r}_{N_{\mathrm{r}}}^{\mathrm{T}}]^{\mathrm{T}}$ are the stacked APVs for all Tx-MAs and Rx-MAs, respectively. Then, the channel matrix from the Tx to Rx is given by \cite{ma2022MAmimo}
\begin{equation}\label{eq_channel_MIMO}
	\mathbf{H}(\tilde{\mathbf{t}}, \tilde{\mathbf{r}}) = \mathbf{F}(\tilde{\mathbf{r}})^{\mathrm{H}} \mathbf{\Sigma}\mathbf{G}(\tilde{\mathbf{t}}) \in \mathbb{C}^{N_{\mathrm{r}} \times N_{\mathrm{t}}}.
\end{equation}

Note that the MIMO channel shown in \eqref{eq_channel_MIMO} assumes that mutual coupling effects among the antennas at the Tx and Rx are negligible. This assumption is valid when the antennas are sufficiently spaced apart and well isolated. However, if the antennas are closely spaced and not ideally isolated, mutual coupling effects must be taken into account for MIMO systems. In general, such effects on wireless channels can be characterized by the mutual coupling matrix using the network theory \cite{Wallace2004mutual, Chen2018mutual, Amani2022sparse}. Specifically, the mutual coupling matrices for the Tx-MAs and Rx-MAs are given by $\mathbf{C}_{\mathrm{t}}(\tilde{\mathbf{t}}) = (\mathbf{Z}_{\mathrm{t}}(\tilde{\mathbf{t}}) + \mathbf{Z}_{\mathrm{s}})^{-1}  \in \mathbb{C}^{N_{\mathrm{t}} \times N_{\mathrm{t}}}$ and $\mathbf{C}_{\mathrm{r}}(\tilde{\mathbf{r}}) = \mathbf{Z}_{\mathrm{l}}(\mathbf{Z}_{\mathrm{l}} + \mathbf{Z}_{\mathrm{r}}(\tilde{\mathbf{r}}))^{-1} \in \mathbb{C}^{N_{\mathrm{r}} \times N_{\mathrm{r}}}$, respectively, where $\mathbf{Z}_{\mathrm{t}}(\tilde{\mathbf{t}}) \in \mathbb{C}^{N_{\mathrm{t}} \times N_{\mathrm{t}}}$, $\mathbf{Z}_{\mathrm{r}}(\tilde{\mathbf{r}}) \in \mathbb{C}^{N_{\mathrm{r}} \times N_{\mathrm{r}}}$, $\mathbf{Z}_{\mathrm{s}} \in \mathbb{C}^{N_{\mathrm{t}} \times N_{\mathrm{t}}}$, and $\mathbf{Z}_{\mathrm{l}} \in \mathbb{C}^{N_{\mathrm{r}} \times N_{\mathrm{r}}}$ represent the self- and mutual-impedance matrix at the Tx, the self- and mutual-impedance matrix at the Rx, the source impedance matrix, and the load impedance matrix, respectively. As such, the effective MIMO channel matrix between the Tx and Rx is given by 
\begin{equation}\label{eq_channel_MIMO_eff}
	\mathbf{H}_\mathrm{eff}(\tilde{\mathbf{t}}, \tilde{\mathbf{r}}) = \mathbf{C}_{\mathrm{r}}(\tilde{\mathbf{r}}) \mathbf{H}(\tilde{\mathbf{t}}, \tilde{\mathbf{r}}) \mathbf{C}_{\mathrm{t}}(\tilde{\mathbf{t}})  \in \mathbb{C}^{N_{\mathrm{r}} \times N_{\mathrm{t}}}.
\end{equation}
Therein, the source and load impedance matrices, $\mathbf{Z}_{\mathrm{s}}$ and $\mathbf{Z}_{\mathrm{l}}$, are constant diagonal matrices determined by the source and load circuits, respectively. The impedance matrices at the Tx and Rx, $\mathbf{Z}_{\mathrm{t}}(\tilde{\mathbf{t}})$ and $\mathbf{Z}_{\mathrm{r}}(\tilde{\mathbf{r}})$, are impacted by the positions of the Tx-MAs and Rx-MAs, respectively. For some simple cases, such as parallel half-wavelength dipoles, the self- and mutual-impedance can be derived in closed form as a function of the inter-antenna distance \cite{Chen2018mutual}. However, for the general cases with arbitrary antenna configurations, it is challenging to obtain the expressions of the impedance matrices in closed form, whereas the full-wave simulation can be adopted to calculate the impedance and mutual coupling \cite{Amani2022sparse}.

\subsubsection{Extension to Wideband Channel Model}
For MA-aided wideband communications, we consider a typical orthogonal frequency division multiplexing (OFDM) system with bandwidth $B$ and the number of subcarriers $M$. Then, we can define multiple (clustered) delay taps to characterize the baseband equivalent channel, where each delay tap spans a time interval of $1/B$ and the maximum number of delay taps between the Tx and Rx is denoted as $T$. For the $\tau$-th delay tap, $1 \leq \tau \leq T$, we denote the total number of Tx paths as $L_{\mathrm{t},\tau}$. The elevation and azimuth AoDs for the $j$-th Tx path over the $\tau$-th delay tap are denoted as $\theta_{\mathrm{t},\tau, j}$ and $\phi_{\mathrm{t},\tau, j}$, $1 \leq j \leq L_{\mathrm{t},\tau}$, respectively. Then, the FRV for all the Tx paths over the $\tau$-th (clustered) delay tap, $1 \leq \tau \leq T$, is given by
\begin{equation}\label{eq_FRV_Tx_tap}
	\mathbf{g}_{\tau}(\mathbf{t})=\left[\mathrm{e}^{\mathrm{j}\frac{2\pi}{\lambda}\mathbf{k}_{\mathrm{t},\tau, 1}^{\mathrm{T}}\mathbf{t}}, \mathrm{e}^{\mathrm{j}\frac{2\pi}{\lambda}\mathbf{k}_{\mathrm{t},\tau, 2}^{\mathrm{T}}\mathbf{t}}, \dots, \mathrm{e}^{\mathrm{j}\frac{2\pi}{\lambda}\mathbf{k}_{\mathrm{t},\tau, L_{\mathrm{t},\tau}}^{\mathrm{T}}\mathbf{t}}\right]^{\mathrm{T}} \in \mathbb{C}^{L_{\mathrm{t},\tau} \times 1},
\end{equation}
with {\small $\mathbf{k}_{\mathrm{t},\tau, j} = [\cos \theta_{\mathrm{t},\tau, j} \cos \phi_{\mathrm{t},\tau, j}, \cos \theta_{\mathrm{t},\tau, j} \sin \phi_{\mathrm{t},\tau, j}, \sin \theta_{\mathrm{t},\tau, j}]^{\mathrm{T}}$}, $1 \leq j \leq L_{\mathrm{t},\tau}$, denoting the wave vector for each Tx path. Similarly, denote the total number of Rx paths over the $\tau$-th delay tap as $L_{\mathrm{r},\tau}$, $1 \leq \tau \leq T$. The elevation and azimuth AoAs for the $i$-th Rx path over the $\tau$-th delay tap are denoted as $\theta_{\mathrm{r},\tau, i}$ and $\phi_{\mathrm{r},\tau, i}$, $1 \leq i \leq L_{\mathrm{r},\tau}$, respectively. Then, the FRV for all the Rx paths over the $\tau$-th (clustered) delay tap, $1 \leq \tau \leq T$, is given by
\begin{equation}\label{eq_FRV_Rx_tap}
	\mathbf{f}_{\tau}(\mathbf{r})=\left[\mathrm{e}^{\mathrm{j}\frac{2\pi}{\lambda}\mathbf{k}_{\mathrm{r},\tau, 1}^{\mathrm{T}}\mathbf{r}}, \mathrm{e}^{\mathrm{j}\frac{2\pi}{\lambda}\mathbf{k}_{\mathrm{r},\tau, 2}^{\mathrm{T}}\mathbf{r}}, \dots, \mathrm{e}^{\mathrm{j}\frac{2\pi}{\lambda}\mathbf{k}_{\mathrm{r},\tau, L_{\mathrm{r},\tau}}^{\mathrm{T}}\mathbf{r}}\right]^{\mathrm{T}} \in \mathbb{C}^{L_{\mathrm{r},\tau} \times 1},
\end{equation}
with {\small $\mathbf{k}_{\mathrm{r},\tau, i} = [\cos \theta_{\mathrm{r},\tau, i} \cos \phi_{\mathrm{r},\tau, i}, \cos \theta_{\mathrm{r},\tau, i} \sin \phi_{\mathrm{r},\tau, i}, \sin \theta_{\mathrm{r},\tau, i}]^{\mathrm{T}}$}, $1 \leq i \leq L_{\mathrm{r},\tau}$, denoting the wave vector for each Rx path. 

Denoting by $\mathbf{\Sigma_{\tau}}$ the PRM between all Tx paths and all Rx paths over the $\tau$-th (clustered) delay tap, $1 \leq \tau \leq T$, the baseband equivalent channel impulse response (CIR) can be obtained as \cite{zhu2024wideband}
\begin{equation}\label{eq_channel_wideband_CIR}
	h_{\tau}(\mathbf{t}, \mathbf{r}) = \mathbf{f}_{\tau}(\mathbf{r})^{\mathrm{H}} \mathbf{\Sigma}_{\tau} \mathbf{g}_{\tau}(\mathbf{t}).
\end{equation}
Moreover, denote $\mathbf{h}(\mathbf{t}, \mathbf{r}) = [h_{1}(\mathbf{t}, \mathbf{r}), \dots, h_{T}(\mathbf{t}, \mathbf{r}), \mathbf{0}_{M-T}^{\mathrm{T}}]^{\mathrm{T}}$ as the $M$-dimensional zero-padded baseband equivalent CIR vector and $\mathbf{D}_{M}$ as the $M$-dimensional discrete Fourier transform (DFT) matrix. The channel frequency response (CFR) over all the subcarriers between the Tx-MA at position $\mathbf{t}$ and the Rx-MA at position $\mathbf{r}$ is thus given by \cite{zhu2024wideband}
\begin{equation}\label{eq_channel_wideband_CFR}
	\mathbf{c}(\mathbf{t}, \mathbf{r}) = \mathbf{D}_{M} \mathbf{h}(\mathbf{t}, \mathbf{r}) \in \mathbb{C}^{M \times 1}.
\end{equation}

\subsubsection{Extension to Near-Field Channel Model}
As the size of the antenna moving region or the carrier frequency increases, the Tx and Rx may be located within the Rayleigh distance from each other. In this case, the spherical wave model should be adopted to characterize the channel between the Tx-MA and Rx-MA under the near-field condition \cite{zhu2024nearfield,ding2024near,chen2024joint}. Specifically, let $\mathbf{r}_{0} \in \mathbb{R}^{3 \times 1}$ and $\mathbf{T} \in \mathbb{R}^{3 \times 3}$ denote the coordinates of the reference point for the Rx region, i.e., $O_{\mathrm{r}}$, in the Tx-LCS and the coordinate transform matrix between the Tx-LCS and Rx-LCS, respectively. Then, the distance between the Tx-MA and Rx-MA is given by $\|\mathbf{r}_{0}+\mathbf{T}^{\mathrm{T}}\mathbf{r}-\mathbf{t}\|_{2}$. The channel coefficient of the LoS path between the Tx-MA and Rx-MA can thus be obtained as
\begin{equation}\label{eq_channel_near_LoS}
	h_{\mathrm{LoS}}(\mathbf{t}, \mathbf{r}) = \sigma_{0} \mathrm{e}^{\mathrm{j}\frac{2\pi}{\lambda}\|\mathbf{r}_{0}+\mathbf{T}^{\mathrm{T}}\mathbf{r}-\mathbf{t}\|_{2}},
\end{equation}
where $\sigma_{0}$ is the amplitude of the LoS channel gain, which equals zero if the LoS path does not exist. The non-LoS (NLoS) paths between the Tx-MA and Rx-MA are generated by scatterers in the environment. Let $\mathbf{s}_{\mathrm{t},j}$, $1 \leq j \leq L_{\mathrm{t}}$, denote the 3D coordinates of the $j$-th point scatterer at the Tx side in its LCS. The $L_{\mathrm{t}}$-dimensional near-FRV (NFRV) at the Tx is given by
\begin{equation}\label{eq_NFRV_Tx}
	\mathbf{g}_{\mathrm{near}}(\mathbf{t})=\left[\mathrm{e}^{\mathrm{j}\frac{2\pi}{\lambda}\|\mathbf{t}-\mathbf{s}_{\mathrm{t},1}\|_{2}}, \mathrm{e}^{\mathrm{j}\frac{2\pi}{\lambda}\|\mathbf{t}-\mathbf{s}_{\mathrm{t},2}\|_{2}}, \dots, \mathrm{e}^{\mathrm{j}\frac{2\pi}{\lambda}\|\mathbf{t}-\mathbf{s}_{\mathrm{t},L_{\mathrm{t}}}\|_{2}}\right]^{\mathrm{T}}.
\end{equation}
Similarly, denoting the 3D coordinates of the $i$-th point scatterer at the Rx side in its LCS as $\mathbf{s}_{\mathrm{r},i}$, $1 \leq i \leq L_{\mathrm{r}}$, the $L_{\mathrm{r}}$-dimensional NFRV at the Rx is given by
\begin{equation}\label{eq_NFRV_Rx}
	\mathbf{f}_{\mathrm{near}}(\mathbf{r})=\left[\mathrm{e}^{\mathrm{j}\frac{2\pi}{\lambda}\|\mathbf{r}-\mathbf{s}_{\mathrm{r},1}\|_{2}}, \mathrm{e}^{\mathrm{j}\frac{2\pi}{\lambda}\|\mathbf{r}-\mathbf{s}_{\mathrm{r},2}\|_{2}}, \dots, \mathrm{e}^{\mathrm{j}\frac{2\pi}{\lambda}\|\mathbf{r}-\mathbf{s}_{\mathrm{r},L_{\mathrm{r}}}\|_{2}}\right]^{\mathrm{T}}.
\end{equation}
The channel coefficient of the NLoS paths between the Tx-MA and Rx-MA can be obtained as
\begin{equation}\label{eq_channel_near_NLoS}
	h_{\mathrm{NLoS}}(\mathbf{t}, \mathbf{r}) = \mathbf{f}_{\mathrm{near}}(\mathbf{r})^{\mathrm{T}} \mathbf{\Sigma}_{\mathrm{NLoS}} \mathbf{g}_{\mathrm{near}}(\mathbf{t}),
\end{equation}
where $\mathbf{\Sigma}_{\mathrm{NLoS}}$ is the PRM of the NLoS paths between $O_{\mathrm{t}}$ and $O_{\mathrm{r}}$. As a result, the channel coefficient between the Tx-MA and Rx-MA is given by
\begin{equation}\label{eq_channel_near}
	h_{\mathrm{near}}(\mathbf{t}, \mathbf{r}) = h_{\mathrm{LoS}}(\mathbf{t}, \mathbf{r}) + h_{\mathrm{NLoS}}(\mathbf{t}, \mathbf{r}),
\end{equation}
which is an extension of the basic field-response channel model in \eqref{eq_channel_basic} under the near-field condition.

Note that the near-field-response channel model in \eqref{eq_channel_near} assumes uniform spherical waves under stationary propagation environments. If the antenna moving regions are sufficiently large such that their size is comparable to the Tx-Rx distance, non-uniform spherical waves should be adopted to characterize the channel variation w.r.t. the positions of antennas, where the change of the channel amplitude gain for LoS/NLoS paths and the non-stationary field response information (including the changed number and locations of scatterers as well as the different PRMs) need to be considered.

\subsubsection{Extension to 6DMA Channel Model}
Next, we extend the field-response channel model in \eqref{eq_channel_basic} to 6DMA systems with flexible antenna position and orientation \cite{shao20246DMA,shao2024discrete,shao2024Mag6DMA,pi20246DMAcoordi,zhang20246DMAhybrid}. Note that the different orientations of the Tx-MA and Rx-MA only influence the response coefficient for each channel path between the Tx and Rx considering the polarization. Thus, we can express the PRM in \eqref{eq_channel_basic} as a function of the Tx-AOM and Rx-AOM, i.e., $\mathbf{\Sigma}(\mathbf{\Psi}, \mathbf{\Omega})\in \mathbb{C}^{L_{\mathrm{r}} \times L_{\mathrm{t}}}$. To derive the PRM in closed form, we need to define the radiation field pattern for the Tx-MA, which is a function of the wave vector in the Tx-ACCS. To this end, we denote $\hat{\mathbf{k}}_{\mathrm{t}} \in \mathbb{R}^{3 \times 1}$ as the coordinates of a wave vector in the Tx-ACCS, which satisfies $\hat{\mathbf{k}}_{\mathrm{t}}=\mathbf{\Psi}^{\mathrm{T}}\mathbf{k}_{\mathrm{t}}$, with $\mathbf{k}_{\mathrm{t}}$ being the coordinates of the wave vector in the Tx-LCS. $\hat{\mathbf{k}}_{\mathrm{t}}$, $\hat{\mathbf{i}}_{\mathrm{t}}(\hat{\mathbf{k}}_{\mathrm{t}}) \in \mathbb{R}^{3 \times 1}$ and $\hat{\mathbf{j}}_{\mathrm{t}}(\hat{\mathbf{k}}_{\mathrm{t}}) \in \mathbb{R}^{3 \times 1}$ are defined as a set of orthonormal basis vectors in the Tx-ACCS shown in Fig. \ref{Fig_Coordinate}(e). For any given $\hat{\mathbf{k}}_{\mathrm{t}}$, there exist infinite pairs of $\hat{\mathbf{i}}_{\mathrm{t}}(\hat{\mathbf{k}}_{\mathrm{t}})$ and $\hat{\mathbf{j}}_{\mathrm{t}}(\hat{\mathbf{k}}_{\mathrm{t}})$ that are orthogonal to $\hat{\mathbf{k}}_{\mathrm{t}}$. To ensure uniqueness, we adopt the convention commonly used in the antenna community, where $\hat{\mathbf{i}}_{\mathrm{t}}(\hat{\mathbf{k}}_{\mathrm{t}})$ is parallel to the plane spanned by axis $\hat{z}_{\mathrm{t}}$ and wave vector $\hat{\mathbf{k}}_{\mathrm{t}}$, while $\hat{\mathbf{j}}_{\mathrm{t}}(\hat{\mathbf{k}}_{\mathrm{t}})$ is parallel to the $\hat{x}_{\mathrm{t}}$-$\hat{O}_{\mathrm{t}}$-$\hat{y}_{\mathrm{t}}$ plane. Based on this definition, the radiation field patterns for the Tx-MA are denoted as $F_{\mathrm{t},1}(\hat{\mathbf{k}}_{\mathrm{t}})$ and $F_{\mathrm{t},2}(\hat{\mathbf{k}}_{\mathrm{t}})$, which correspond to the field coefficients over basis vectors $\hat{\mathbf{i}}_{\mathrm{t}}(\hat{\mathbf{k}}_{\mathrm{t}})$ and $\hat{\mathbf{j}}_{\mathrm{t}}(\hat{\mathbf{k}}_{\mathrm{t}})$, respectively. Then, we can obtain the radiation gain of the Tx-MA for the $j$-th Tx path, $1 \leq j \leq L_{\mathrm{t}}$, as
\begin{equation} \label{eq_radiation_Tx}
	G_{\mathrm{t},j}(\mathbf{\Psi}) = \sqrt{|F_{\mathrm{t},1}(\mathbf{\Psi}^{\mathrm{T}} \mathbf{k}_{\mathrm{t},j})|^{2} + |F_{\mathrm{t},2}(\mathbf{\Psi}^{\mathrm{T}} \mathbf{k}_{\mathrm{t},j})|^{2}}.
\end{equation}
Similarly, the radiation field patterns for the Rx-MA are respectively denoted as $F_{\mathrm{r},1}(\hat{\mathbf{k}}_{\mathrm{r}})$ and $F_{\mathrm{r},2}(\hat{\mathbf{k}}_{\mathrm{r}})$ over orthonormal basis vectors $\hat{\mathbf{i}}_{\mathrm{r}}(\hat{\mathbf{k}}_{\mathrm{r}})$ and $\mathbf{j}_{\mathrm{r}}(\hat{\mathbf{k}}_{\mathrm{r}})$ in the 3D Rx-ACCS shown in Fig. \ref{Fig_Coordinate}(f). Then, we obtain the radiation gain of the Rx-MA for the $i$-th Rx path, $1 \leq i \leq L_{\mathrm{r}}$, as
\begin{equation} \label{eq_radiation_Rx}
	G_{\mathrm{r},i}(\mathbf{\Omega}) = \sqrt{|F_{\mathrm{r},1}(\mathbf{\Omega}^{\mathrm{T}} \mathbf{k}_{\mathrm{r},i})|^{2} + |F_{\mathrm{r},2}(\mathbf{\Omega}^{\mathrm{T}} \mathbf{k}_{\mathrm{r},i})|^{2}}.
\end{equation}

Next, we model the polarization loss between the $j$-th Tx path and the $i$-th Rx path, $1 \leq j \leq L_{\mathrm{t}}$, $1 \leq i \leq L_{\mathrm{r}}$. Given the wave vector of the $j$-th Tx path, $\mathbf{k}_{\mathrm{t},j}$, in the 3D Tx-LCS, we define two reference direction vectors as $\mathbf{i}_{\mathrm{t},j}$ and $\mathbf{j}_{\mathrm{t},j}$, which are orthogonal to $\mathbf{k}_{\mathrm{t},j}$ shown in Fig. \ref{Fig_Coordinate}(c). Similarly, the two reference direction vectors for the $i$-th Rx path (w.r.t. wave vector $\mathbf{k}_{\mathrm{r},i}$) in the 3D Rx-LCS are defined as $\mathbf{i}_{\mathrm{r},i}$ and $\mathbf{j}_{\mathrm{r},i}$ shown in Fig. \ref{Fig_Coordinate}(d). Then, we define $\mathbf{\Lambda}_{i,j} \in \mathbb{C}^{2 \times 2}$ as the path polarization response matrix (PPRM) between the $j$-th Tx path and the $i$-th Rx path, which corresponds to the response coefficients between field directions $(\mathbf{i}_{\mathrm{t},j}, \mathbf{j}_{\mathrm{t},j})$ and $(\mathbf{i}_{\mathrm{r},i}, \mathbf{j}_{\mathrm{r},i})$ and is only determined by the propagation environment. Thus, the polarization gain between the $j$-th Tx path and the $i$-th Rx path is given by \eqref{eq_polarization} shown at the top of the next page. For the special case of linear polarization, we have $F_{\mathrm{t},2}(\mathbf{\Psi}^{\mathrm{T}} \mathbf{k}_{\mathrm{t},j})=0$ and $F_{\mathrm{r},2}(\mathbf{\Omega}^{\mathrm{T}} \mathbf{k}_{\mathrm{r},i})=0$, and thus the polarization gain in \eqref{eq_polarization} can be simplified as

\begin{figure*}[!t]
	\centering 
	\begin{equation} \label{eq_polarization} \footnotesize
		G_{\mathrm{p},i,j}(\mathbf{\Psi}, \mathbf{\Omega}) = 
		\underbrace{\left[ \begin{aligned}
				\frac{F_{\mathrm{r},1}(\mathbf{\Omega}^{\mathrm{T}} \mathbf{k}_{\mathrm{r},i})}{G_{\mathrm{r},i}(\mathbf{\Omega})}\\ \frac{F_{\mathrm{r},2}(\mathbf{\Omega}^{\mathrm{T}} \mathbf{k}_{\mathrm{r},i})}{G_{\mathrm{r},i}(\mathbf{\Omega})} \end{aligned} \right]^{\mathrm{T}}}_{\text{Polarization of Rx-MA}} 
		\underbrace{\left[\begin{aligned}
				\hat{\mathbf{i}}_{\mathrm{r}}(\mathbf{\Omega}^{\mathrm{T}} \mathbf{k}_{\mathrm{r},i})^{\mathrm{T}} \mathbf{\Omega}^{\mathrm{T}}\mathbf{i}_{\mathrm{r},i}, ~  
				\hat{\mathbf{i}}_{\mathrm{r}}(\mathbf{\Omega}^{\mathrm{T}} \mathbf{k}_{\mathrm{r},i})^{\mathrm{T}} \mathbf{\Omega}^{\mathrm{T}} \mathbf{j}_{\mathrm{r},i} \\
				\hat{\mathbf{j}}_{\mathrm{r}}(\mathbf{\Omega}^{\mathrm{T}} \mathbf{k}_{\mathrm{r},i})^{\mathrm{T}} \mathbf{\Omega}^{\mathrm{T}} \mathbf{i}_{\mathrm{r},i}, ~ 
				\hat{\mathbf{j}}_{\mathrm{r}}(\mathbf{\Omega}^{\mathrm{T}} \mathbf{k}_{\mathrm{r},i})^{\mathrm{T}} \mathbf{\Omega}^{\mathrm{T}} \mathbf{j}_{\mathrm{r},i}
			\end{aligned}\right]}_{\text{Coordinate transform of field directions for Rx path}}
		\mathbf{\Lambda}_{i,j}
		\underbrace{\left[\begin{aligned} 
				\mathbf{i}_{\mathrm{t},j}^{\mathrm{T}} \mathbf{\Psi} 
				\hat{\mathbf{i}}_{\mathrm{t}}(\mathbf{\Psi}^{\mathrm{T}} \mathbf{k}_{\mathrm{t},j}) , ~ 
				\mathbf{i}_{\mathrm{t},j}^{\mathrm{T}} \mathbf{\Psi}
				\hat{\mathbf{j}}_{\mathrm{t}}(\mathbf{\Psi}^{\mathrm{T}} \mathbf{k}_{\mathrm{t},j}) \\
				\mathbf{j}_{\mathrm{t},j}^{\mathrm{T}} \mathbf{\Psi}
				\hat{\mathbf{i}}_{\mathrm{t}}(\mathbf{\Psi}^{\mathrm{T}} \mathbf{k}_{\mathrm{t},j}) , ~ 
				\mathbf{j}_{\mathrm{t},j}^{\mathrm{T}} \mathbf{\Psi}
				\hat{\mathbf{j}}_{\mathrm{t}}(\mathbf{\Psi}^{\mathrm{T}} \mathbf{k}_{\mathrm{t},j})
			\end{aligned}\right]}_{\text{Coordinate transform of field directions for Tx path}}
		\underbrace{\left[\begin{aligned}
				\frac{F_{\mathrm{t},1}(\mathbf{\Psi}^{\mathrm{T}} \mathbf{k}_{\mathrm{t},j})}{G_{\mathrm{t},j}(\mathbf{\Psi})}, \\ \frac{F_{\mathrm{t},2}(\mathbf{\Psi}^{\mathrm{T}} \mathbf{k}_{\mathrm{t},j})}{G_{\mathrm{t},j}(\mathbf{\Psi})} \end{aligned}\right]}_{\text{Polarization of Tx-MA}}
	\end{equation}
	\vspace*{2pt}
	\hrulefill
\end{figure*}
{\small
	\begin{equation} \label{eq_polarization_linear} 
		G_{\mathrm{p},i,j}^{\mathrm{linear}}(\mathbf{\Psi}, \mathbf{\Omega}) = 
		\left[ \begin{aligned}
			\hat{\mathbf{i}}_{\mathrm{r}}(\mathbf{\Omega}^{\mathrm{T}} \mathbf{k}_{\mathrm{r},i})^{\mathrm{T}} \mathbf{\Omega}^{\mathrm{T}}\mathbf{i}_{\mathrm{r},i}\\ 
			\hat{\mathbf{i}}_{\mathrm{r}}(\mathbf{\Omega}^{\mathrm{T}} \mathbf{k}_{\mathrm{r},i})^{\mathrm{T}} \mathbf{\Omega}^{\mathrm{T}}\mathbf{j}_{\mathrm{r},i} \end{aligned} \right]^{\mathrm{T}}
		\mathbf{\Lambda}_{i,j}
		\left[\begin{aligned} 
			\mathbf{i}_{\mathrm{t},j}^{\mathrm{T}} \mathbf{\Psi}
			\hat{\mathbf{i}}_{\mathrm{t}}(\mathbf{\Psi}^{\mathrm{T}} \mathbf{k}_{\mathrm{t},j}) \\
			\mathbf{j}_{\mathrm{t},j}^{\mathrm{T}} \mathbf{\Psi}
			\hat{\mathbf{i}}_{\mathrm{t}}(\mathbf{\Psi}^{\mathrm{T}} \mathbf{k}_{\mathrm{t},j})
		\end{aligned}\right].
	\end{equation}
}

Based on the above derivations, we obtain the path response coefficient between the $j$-th Tx path, $1 \leq j \leq L_{\mathrm{t}}$, and the $i$-th Rx path, $1 \leq i \leq L_{\mathrm{r}}$, as 
\begin{equation} \label{eq_path_response}
	\left[\mathbf{\Sigma}(\mathbf{\Psi}, \mathbf{\Omega})\right]_{i,j} = G_{\mathrm{r},i}(\mathbf{\Omega}) G_{\mathrm{p},i,j}(\mathbf{\Psi}, \mathbf{\Omega}) G_{\mathrm{t},j}(\mathbf{\Psi}).
\end{equation}
Finally, the channel between the Tx-MA and Rx-MA can be expressed as a function of their positions and orientations, i.e.,
\begin{equation}\label{eq_channel_rotatable}
	h(\mathbf{t}, \mathbf{r}, \mathbf{\Psi}, \mathbf{\Omega}) = \mathbf{f}(\mathbf{r})^{\mathrm{H}} \mathbf{\Sigma}(\mathbf{\Psi}, \mathbf{\Omega}) \mathbf{g}(\mathbf{t}),
\end{equation}
which provides a general channel model for 6DMA systems considering the impact of both antenna positions and orientations \cite{shao20246DMA,shao2024discrete,shao2024Mag6DMA,pi20246DMAcoordi,zhang20246DMAhybrid}, and is applicable to antennas with any arbitrary radiation pattern (omnidirectional or directional) and polarization (linear, circular, or elliptical). In fact, the field-response channel model presented above is consistent with the clustered delay line (CDL) channel model from 0.5 to 100 GHz standardized by the 3rd Generation Partnership Project (3GPP) \cite{3gpp2019studyo}. In particular, since the CDL channel model pertains to the one-to-one correspondence between the transmit and receive paths, it can be regarded as a special case of the field-response channel model with diagonal PRMs.

\begin{table*}[t]
	\renewcommand{\arraystretch}{0.8}
	\caption{A summary of existing channel models for MA systems.} \label{Tab_channel_model}
	\centering 
	\small
	\begin{tabular}{|>{\centering\arraybackslash}m{1.3cm}|>{\centering\arraybackslash}m{1.6cm}|>{\centering\arraybackslash}m{1.7cm}|>{\centering\arraybackslash}m{0.4cm}|>{\centering\arraybackslash}m{1.7cm}|>{\centering\arraybackslash}m{1.5cm}|>{\centering\arraybackslash}m{3cm}|>{\centering\arraybackslash}m{1.4cm}|>{\centering\arraybackslash}m{1.5cm}|}
		\hline
		\textbf{Catagory} & \textbf{Advantages}    & \textbf{Limitations}        & \textbf{Ref.} & \textbf{System Setup}        & \textbf{Movement Mode}      & \textbf{Channel Model}  & \textbf{Bandwidth}  & \textbf{Field Condition} \\ \hline
		\multirow{6}{*}{\makecell[c] {\\ \\  \\ \\ \\ \\ Field-\\response\\ channel \\model} }     & \multirow{6}{*}{\begin{tabular}[c]{@{}c@{}}\makecell[c]{\\  \\ \\ \makecell[c]{1) General \\model is \\adaptive \\to different \\propagation \\environments;}\\ 2) Facilitate \\antenna\\ movement\\ optimization}\\ \end{tabular}} & \multirow{6}{*}{\begin{tabular}[c]{@{}c@{}}\makecell[c]{\\ \\ \\ 1) Rely on \\accurate \\channel \\structures;}\\ \makecell[c]{2) Highly\\ non-linear \\forms are \\intractable for \\performance \\analysis}\end{tabular}}                                                             
		& \cite{zhu2022MAmodel}       & SISO with Tx-MA and Rx-MA    & 2D continuous position              & Field-response of finite/infinite channel paths                  & Narrow & Far-field             \\ \cline{4-9} 
		&  & & \cite{ma2022MAmimo}        & MIMO with Tx-MA and Rx-MA    & 2D continuous position              & Field-response of finite channel paths                           & Narrow & Far-field             \\ \cline{4-9} 
		&  & & \cite{zhu2024wideband}      & SISO with Tx-MA and Rx-MA    & 3D continuous position               & Field-response of finite channel paths                           & Wide   & Far-field             \\ \cline{4-9} 
		&  &  & \cite{zhu2024nearfield}      & MU-MISO with Tx-MA only   & 2D continuous position                & Field-response of finite channel paths                           & Narrow & Near-field            \\ \cline{4-9} 
		&  &  & \cite{ding2024near}      & MU-MISO with Tx-MA and Rx-MA & 2D continuous position              & Field-response of finite channel paths                           & Narrow & Near-field            \\ \cline{4-9} 
		&  &  & \cite{shao20246DMA}      & MU-MISO with Rx-6DMA only      & 6D continuous position \& rotation & Field-response of finite channel paths                           & Narrow & Far-field             \\ \hline
		\multirow{6}{*}{\makecell[c]{\\ \\  \\ \\ \\ \\ Spatial-\\correlation\\ channel \\model}} & \multirow{6}{*}{\begin{tabular}[c]{@{}c@{}}\makecell[c]{\\ \\  \\ \\ \\ \makecell[c]{1) Robust to \\modeling \\errors;}\\ 2) Facilitate\\ performance\\ analysis}\end{tabular}}                               & \multirow{6}{*}{\begin{tabular}[c]{@{}c@{}}\makecell[c]{1) Rely on \\channel \\fading \\assumption;}\\ \makecell[c]{ 2) Neglect \\the impact \\of antenna \\radiation;}\\ \makecell[c]{3) Abstract \\random \\variables\\ have an \\inadequate \\interpretation \\of physical \\propagation \\environments}\end{tabular}} 
		& \cite{Wong2021fluid}      & SISO with Rx-MA only         & 1D discrete position                & Simplified Jake's model using auxiliary Gaussian variables       & Narrow & Far-field             \\ \cline{4-9} 
		&  &  & \cite{Khammassi2023approx} & SISO with Rx-MA only         & 1D discrete position              & Approximate correlation model to fix inaccuracy in \cite{Wong2021fluid} & Narrow & Far-field        \\ \cline{4-9} 
		&  &  & \cite{Psomas2023continuous}    & SISO with Rx-MA only         & 1D continuous position              & Jake's model applied to continuous antenna position              & Narrow & Far-field             \\ \cline{4-9} 
		&  &  & \cite{New2024fluid}       & MIMO with Tx-MA and Rx-MA    & 2D discrete position               & Correlation model assuming 3D rich scattering          & Narrow & Far-field             \\ \cline{4-9} 
		&  &  & \cite{Espinosa2024block}  & SISO with Rx-MA only         & 1D discrete position                & Approximate correlation by block-diagonal matrices       & Narrow & Far-field             \\ \cline{4-9} 
		&  &  & \cite{Rostami2023copula}   & SISO with Rx-MA only         & 1D discrete position                & Copula-based model to generate arbitrary fading channels         & Narrow & Far-field             \\ \hline
	\end{tabular}
\end{table*}

\subsubsection{Other Existing Channel Models} The field-response channel model explicitly depicts the continuous variation of channel response w.r.t. the positions and orientations of the Tx-MA and Rx-MA under any deterministic propagation environment \cite{zhu2022MAmodel,ma2022MAmimo,zhu2024wideband,zhu2024nearfield,ding2024near,chen2024joint,shao20246DMA,shao2024discrete}. This enables us to directly optimize the antenna position and/or orientation to improve the channel condition as well as communication performance for various wireless systems. Moreover, the field-response channel model is also applicable to stochastic channels given the statistical distributions of the AoDs, AoAs, and PRMs \cite{zhu2022MAmodel,chen2023joint}. However, due to the highly non-linear relationship between the APV/AOM and the channel gain, the performance analysis of MA systems using the field-response channel model is generally difficult. 

In comparison, the spatial-correlation channel model can simplify the complicated channel parameters by extracting the channel statistics and thus facilitate performance analysis \cite{Ozdogan2019MIMOfading, Wang2024cellfree}\footnote{It was shown in \cite{zhu2022MAmodel} that the field-response channel model can be degraded into the spatial-correlation channel model, such as Rayleigh fading and Rician fading, under the condition of infinite channel paths with random coefficients. Therefore, spatial-correlation channel models can be regarded as special cases and simplified approximations of the field-response channel model under rich-scattering scenarios.}. In this context, Jake's model was applied to FAS-aided wireless communication systems in \cite{Wong2021fluid} for characterizing the spatial correlation of the wireless channels between multiple discrete ports, i.e., candidate antenna positions in a 1D line segment, at the Rx. In particular, under the Rayleigh fading assumption with rich scattering, a series of independent Gaussian random variables were introduced to parameterize the channel gains in a simplified form. However, this oversimplified channel model was later proven to be inaccurate because it fails to guarantee the exact correlation among all antenna ports \cite{Khammassi2023approx}.

To address this issue, more parameters were incorporated in \cite{Khammassi2023approx} to accurately capture the correlation between antenna ports, based on which some approximations were introduced to achieve a trade-off between the model accuracy and analytical traceability. Then, the spatial-correlation channel models were extended to the cases with 1D continuous antenna movement at the Rx and with 2D antenna discrete movement at both the Tx and Rx in \cite{New2024fluid} and \cite{Espinosa2024block}, respectively. To further simplify performance analysis, an approximate channel model was proposed in \cite{Espinosa2024block} under the block-fading assumption, where the original spatial correlation matrix for the channels of multiple Rx ports was replaced by a block-diagonal matrix for approximation. Given that the aforementioned channel models \cite{Wong2021fluid,Khammassi2023approx,Psomas2023continuous,New2024fluid,Espinosa2024block} are all predicated on the assumption of Rayleigh fading with a specific correlation function, their generalizability is inherently constrained. In this regard, the copula theory was leveraged in \cite{Rostami2023copula} to develop a general framework to generate channels with arbitrary fading distributions.

In addition to the tractability in performance analysis, the spatial-correlation channel models are robust to modeling errors in terms of non-ideal signal emission, propagation, and reception. However, they highly rely on the assumptions of the propagation environments and channel fading distributions, e.g., uniform scattering for Jake's model and Rayleigh fading in most of the existing works \cite{Wong2021fluid,Khammassi2023approx,Psomas2023continuous,New2024fluid,Espinosa2024block}. In practice, the signal propagation environment can be dynamic, making it challenging to obtain accurate distributions or statistics of wireless channels. Moreover, most of the existing channel models based on spatial correlation can only depict the channels' statistics over different MA positions, while it is difficult to characterize the impact of antenna radiation, e.g., channel gains under different antenna orientations. Representative works on channel modeling for MA systems have been summarized and compared in Table \ref{Tab_channel_model}. Considering the respective advantages of the field-response and spatial-correlation channel models, interested readers are suggested to select appropriate models based on different application scenarios and design objectives. In addition, to overcome their respective limitations, more research efforts are expected in channel modeling to bridge the gap between these two channel models, so as to enable more efficient performance analysis and optimization for MA-aided wireless systems. 

\subsection{Fundamental DoFs}
In this subsection, we elaborate the fundamental DoFs in antenna movement for reshaping wireless channels. Specifically, we analyze the individual and combined gains achieved by reconfiguring antenna position and orientation, highlighting how these adjustments can enhance system performance.

\subsubsection{Antenna Position}
As illustrated in \eqref{eq_channel_basic}, the channel between the Tx-MA and Rx-MA is determined by the superposition of the channel paths. For a fixed antenna orientation, altering the antenna position primarily affects the phases of the complex coefficients for these channel paths. When the Tx-MA/Rx-MA is relocated such that the coefficients of the channel paths constructively combine, a high channel gain can be achieved. Conversely, destructive superposition among the channel paths' coefficients results in a low channel gain between the Tx-MA and Rx-MA. In scenarios with numerous channel paths of comparable power gains and wide angular spread, the channel power gain varies significantly with changes in antenna position \cite{zhu2022MAmodel,zhu2024wideband}, offering more DoFs for reshaping wireless channels. However, if a dominant path with substantially higher power exists, the variation in channel power gain due to positional changes is reduced. Additionally, it is important to note that the PRM is also influenced by the antenna's radiation pattern. Directional antennas amplify the gain of paths within their main lobe while diminishing those of other paths, potentially creating an effective dominant path along the main lobe's direction. Thus, omnidirectional MAs, which lack such directional emphasis, can provide more DoFs for channel reshaping through position reconfiguration compared to directional MAs.

It is also worth noting that the performance gains in MA systems do not always depend on multipath channels. Under pure LoS conditions, the PRM, Tx-FRV, and Rx-FRV are all degraded into scalars. In this case, the amplitude of the channel gain remains unaffected by the positions of the Tx-MA and Rx-MA. Nevertheless, reconfiguring the antenna position can still alter the phase of the channel gain, which is advantageous for MA arrays composed of multiple MAs. Specifically, repositioning multiple MAs changes the array geometry, thereby reconfiguring the steering vector (also known as the array response vector) across different directions. This endows MA arrays with considerable DoFs to switch between different array layouts, facilitating more flexible beamforming for communication and/or sensing systems. For instance, by jointly designing the APV and antenna weight vector (AWV) of an MA array, more versatile beam patterns can be achieved, such as in interference nulling and multi-beamforming scenarios \cite{zhu2023MAarray,ma2024multi,wang2024flexible}. Additionally, optimizing the geometry of an MA array can significantly enhance sensing accuracy \cite{ma2024MAsensing}. Further details on the advantages of MA array-aided communication and sensing systems under LoS conditions will be discussed in Section \ref{Sec_Movement}.

\subsubsection{Antenna Orientation}
As shown in \eqref{eq_path_response} and \eqref{eq_channel_rotatable}, the orientation of each individual antenna impacts the PRM between the Tx and Rx by adjusting the antenna's radiation and polarization gains \cite{shao20246DMA,shao2024discrete,shao2024Mag6DMA,pi20246DMAcoordi,zhang20246DMAhybrid}. When only an LoS path exists, aligning the main lobe of the radiation pattern of a Tx-MA/Rx-MA with the LoS direction and matching their polarization can maximize the channel gain \cite{pi20246DMAcoordi,zhang20246DMAhybrid,zhang2024polarization,zheng2025rotatable}. Thus, rotating directional antennas can yield more DoFs for reshaping wireless channels than rotating omnidirectional antennas under the LoS condition. However, as the number of channel paths increases, the orientation of the Tx-MA/Rx-MA must balance the radiation and polarization gains across multiple paths, which reduces the available DoFs for channel reconfiguration through antenna rotation.

It is important to note that for MAs considered in this paper, the radiation pattern and polarization characteristics are fixed in the ACCS and only change in the LCS due to antenna rotation. Similar functionality can be achieved with conventional reconfigurable antennas \cite{Christodoulou2012reconfig,tandel2023reconfigurable}, which can reconfigure radiation patterns, polarization, and/or frequency responses. In general, electronically reconfigurable antennas offer faster response times and more flexible radiation patterns. In contrast, physical antenna rotation is slower but comes at a lower hardware cost. Additionally, since antenna orientation can be adjusted flexibly within the full 3D space, rotating an MA may enhance coverage performance compared to conventional FAs with reconfigurable or non-reconfigurable radiation patterns \cite{shao2024Mag6DMA,zheng2025rotatable}.

\subsubsection{Antenna Position and Orientation}
By combining the DoFs in both antenna position and orientation, the FRVs and PRM in \eqref{eq_channel_rotatable} can be jointly reconfigured, which allows the wireless channel between the Tx-MA and Rx-MA to achieve the highest flexibility in 6DMA systems \cite{shao20246DMA,shao2024discrete,shao2024Mag6DMA,pi20246DMAcoordi,zhang20246DMAhybrid}. Considering the dynamic nature of wireless propagation environments, the number of channel paths and their field responses may change under different application scenarios and vary over time. Thus, the combined DoFs in tuning antenna position and orientation can adapt to different channel conditions. Such an integrated approach maximizes the potential for channel shaping and performance optimization in 6DMA-aided wireless communication systems.

\begin{figure}[t]
	\begin{center}
		\includegraphics[width=\figwidth cm]{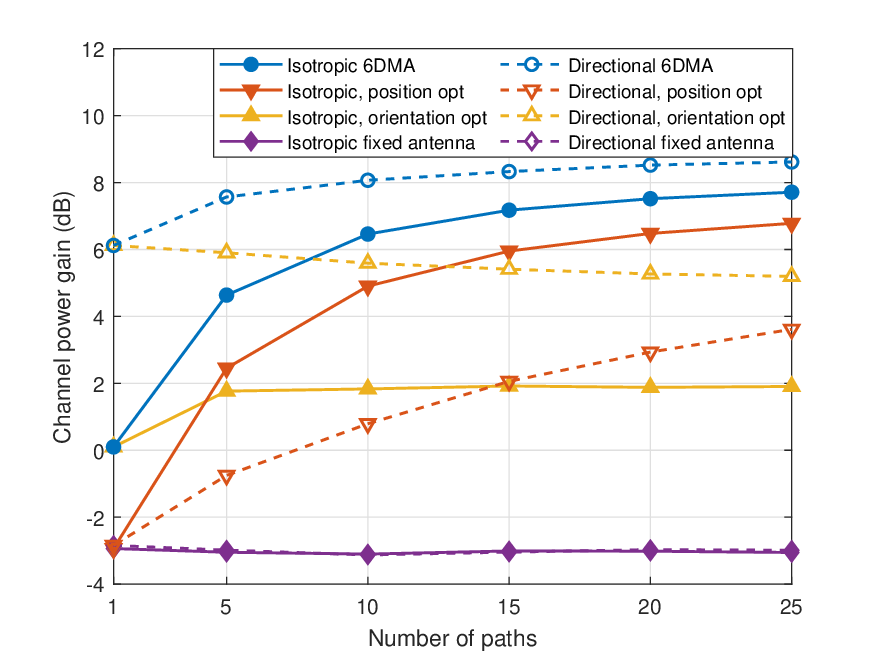}
		\caption{Channel power gains achieved by MAs/6DMAs and FAs.}
		\vspace{-12pt}
		\label{Fig_fundamental_DoFs}
	\end{center}
\end{figure}
To evaluate the individual and combined gains of reconfiguring antenna position and orientation, we conduct a case study where a single FA at the Tx transmits signals to an Rx equipped with a single MA/6DMA. The Rx-MA employs linear polarization and two types of radiation patterns: isotropic and directional. For isotropic radiation, the antenna radiation gain is equal to unity across all directions in the 3D space. For directional radiation, the antenna maintains a constant gain of 6 dBi within the main lobe and a zero gain outside the main lobe. The width of the main lobe is calibrated so that the total energy captured by the directional antenna matches that of the isotropic antenna over the 3D space. The AoAs for each Rx path follow the uniform distribution over the 3D space with the joint probability density function given by $p(\theta, \phi)=\frac{\cos \theta}{4\pi},~\theta \in [-\pi/2, \pi/2],~\phi \in [-\pi, \pi]$, and the path response coefficient is modeled as a CSCG random variable, with variance normalized by the number of Rx paths, i.e., $1/L_{\mathrm{r}}$ \cite{zhu2022MAmodel}. The Rx-MA is allowed to move within a 3D cubic region of size $10 \lambda$, and its orientation can be adjusted freely in 3D space. For each channel realization, the optimal antenna position and/or orientation is determined through an exhaustive search to maximize the channel power gain between the Tx-MA and Rx-MA. 

Fig. \ref{Fig_fundamental_DoFs} illustrates the achievable channel power gains for MA systems with position optimization, 6DMA with orientation optimization only, and 6DMA with joint position and orientation optimization, with each point averaged over $10^4$ random channel realizations. As can be observed, the MA/6DMA systems can always achieve a higher channel power gain compared to FAs, especially for a large number of Rx paths. In the case of a single path, antenna position optimization cannot achieve any performance gain, where the $-3$ dB channel power gain (or $3$ dB channel power loss) is caused by the polarization loss. In comparison, the isotropic antenna and directional antenna can achieve their respective maximum channel power gains, i.e., $0$ and $6$ dB, by aligning their polarization directions with the single channel path via orientation optimization. For isotropic antennas, the performance gain provided by antenna position optimization is higher than that of antenna orientation if the number of paths exceeds 10. This is because an isotropic antenna cannot increase the gain for all paths simultaneously via rotation, while the antenna position optimization can align the phases of the complex coefficients for multiple paths. For directional antennas, antenna orientation optimization offers more performance improvement over antenna position optimization because the former can amplify the path gain by tuning the direction of the main lobe. However, as the number of channel paths increases, antenna orientation optimization has to balance the gains of multiple paths, which decreases the performance improvement over FAs. By integrating the DoFs in both antenna position and orientation optimization, the 6DMA can achieve the best performance among all considered schemes, with over 10 dB improvement in the channel power gain over FAs. Note that such performance gain of MAs/6DMAs over FAs requires additional hardware modules to enable antenna movement and computational overheads to conduct antenna movement optimization. Nonetheless, there are alternative ways to balance the hardware cost and computational complexity, which will be discussed in detail in Sections \ref{MA_arch} and \ref{Sec_Movement}, respectively.

\begin{table*}[t]
	\centering
	\caption{Representative architectures for mechanically movable and electronically reconfigurable antennas.}
	\label{Tab_architecture}\small
	\begin{tabular}{|c|c|c|c|c|}
		\hline
		\textbf{Movement Scale} & \textbf{Control Strategy} & \textbf{Implementation} & \textbf{Movement Mode} & \textbf{Ref.} \\ \hline
		\multirow{8}{*}{\centering Element-level} & \multirow{4}{*}{\centering\begin{tabular}[m]{@{}m{5em}@{}}Mechanical\end{tabular}} & Stepper motor & \begin{tabular}[c]{@{}l@{}}Circular track\end{tabular} & \cite{Basbug2017design}  \\ \cline{3-5} 
		&  & MEMS & \begin{tabular}[m]{@{}l@{}}Flip mode\end{tabular} & \cite{marnat2013new} \\ \cline{3-5} 
		&  & Motor & \begin{tabular}[c]{@{}l@{}}3D position \& orientation \end{tabular} &  \cite{zhu2023MAMag,shao2024Mag6DMA} \\ \cline{2-5} 
		& \multirow{3}{*}{\centering\begin{tabular}[c]{@{}l@{}} Liquid-based\end{tabular}} & Syringe & 1D position &  \cite{Morishita2013Liquid} \\ \cline{3-5} 
		&  & Nano-pump & 1D position & \cite{Shen2021Reconfigurable} \\ \cline{3-5} 
		&  & Electrowetting & 1D position &  \cite{Wang2022Continuous}\\ \cline{2-5} 
		& \multirow{2}{*}{\centering\begin{tabular}[m]{@{}m{5em}@{}}Electronic\end{tabular}} & Dual-mode microstrip patch & 1D position &  \cite{ning2024movable} \\ \cline{3-5} 
		&  & Pin diode biasing network & 2D rotation &  \cite{costantine2015reconfigurable}\\ \hline
		\multirow{4}{*}{\centering Array-level} & \multirow{4}{*}{\centering\begin{tabular}[m]{@{}m{5em}@{}}Mechanical\end{tabular}} & Wheel gear and motor & Sliding mode & \cite{ning2024movable}\\ \cline{3-5} 
		&  & Motor & Rotatable mode &  \cite{ning2024movable}\\ \cline{3-5} 
		&  & Rod and motor & \begin{tabular}[c]{@{}l@{}}Sliding \& rotatable \& deployable modes \end{tabular} &  \cite{shao2024Mag6DMA} \\ \cline{3-5} 
		&  & Origami & Deployable mode & \cite{venkatesh2022origami} \\ \cline{3-5} 
		&  & Inflatable structure & Deployable  mode &  \cite{babuscia2013inflatable}\\ \hline
	\end{tabular}%
\end{table*}

\begin{figure*}[t]
	\begin{center}
		\includegraphics[width=0.98\textwidth]{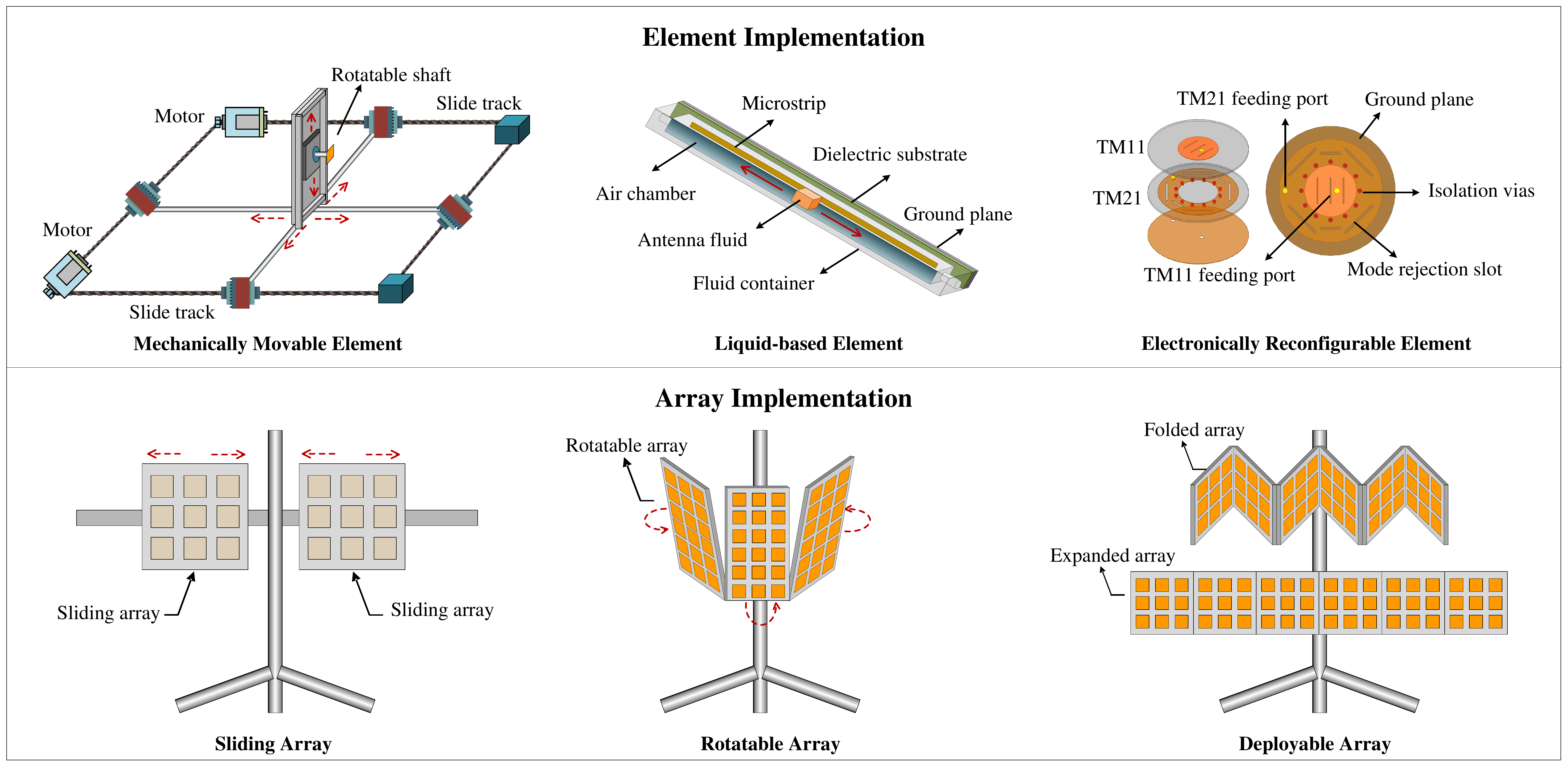}
		\caption{Illustration of typical implementation methods for realizing antenna movement and reconfiguration.}
		\label{Fig_implemantation}
	\end{center}
	\vspace{-12pt}
\end{figure*}

\subsection{Architectures and Practical Constraints}\label{MA_arch}
% 1. Movable element, Mechanically, stepper motor, 1D movement + 1D rotation, \cite{Basbug2017design} 
% 2. Movable element, Mechanically, MEMS, 1D movement + 1D rotation, \cite{marnat2013new}
% 3. Movable element, Mechanically, slide track, servo motor, 3D movement + 3D rotation \cite{zhu2023MAMag}
% 4. Movable element, Liquid-driven, PCB substrate, 1D movement,  \cite{Wang2022Continuous}
% 5. Movable element, Liquid-driven, Pipe container, 1D movement, \cite{Wong2021fluid}
% 5. Movable element, Liquid-driven, Air chamber, 1D movement, \cite{9715064}
% 6. Movable element, Electrically, Dual-mode microstrip patch, 1D movement, \cite{mitha2021principles}
% 7. Movable element, Electrically, Pin diode biasing network, 2D rotation, \cite{costantine2015reconfigurable}
% 8. Movable array, Mechanically, wheel gear, motor, sliding array, \cite{ning2024movable}
% 9. Movable array, Mechanically,  motor, rotatable array, \cite{ning2024movable}
% 10. Movable array, Mechanically,  motor, foldable array, \cite{ning2024movable}
% 10. Movable array, Mechanically,  origami, transformer array, \cite{venkatesh2022origami}

The hardware architectures and their associated constraints are essential factors in implementing practical MA-aided wireless systems. This subsection provides a comprehensive overview of the state-of-the-art architectures for realizing element-level and array-level antenna movement as well as those based on electronic reconfiguration for achieving equivalent movement. %Each architecture is examined in terms of its operational principles, design features, and practical constraints. 

\subsubsection{Mechanically Movable Element}
Mechanically movable elements utilize external mechanical structures with the aid of actuators, such as electric motors or precision gears, to achieve physical movement \cite{zhu2023MAMag}. These actuators convert control signaling and energy into mechanical motion, allowing for precise positioning and/or rotation of the antenna elements. By mounting antennas directly on motor-driven shafts \cite{ning2024movable} or using MEMS \cite{marnat2013new}, mechanical movement can place the antennas at the possible best positions and/or orientations for signal transmission or reception. In general, the response time of a motor-based MA ranges from milliseconds to seconds, while the compact MEMS-based MA may have a faster response time in the order of microseconds to milliseconds. The main constraints for mechanically movable elements are related to the mechanism complexity, the maintenance requirements, and the physical limitations of the actuators. The coupling between antenna elements is also affected by their physical movement, requiring careful consideration of the antenna placement to minimize this adverse effect. Additionally, the wiring and feed network for these antennas may limit the range of motion due to the need to maintain electrical connections. Energy efficiency and latency in movement are also critical issues that need to be addressed to ensure that the antennas can quickly and accurately respond to changes in the wireless environment.

\subsubsection{Liquid-based Element}
Liquid-based elements leverage the flow property of liquid/fluid materials within a container, which can be driven by a syringe \cite{Morishita2013Liquid}, a nano-pump \cite{Shen2021Reconfigurable}, or electrowetting \cite{Wang2022Continuous}. For example, by manually applying pressure to the syringe or digitally controlling the micro-pump/nano-pump, liquid metal can move in the air chamber \cite{Morishita2013Liquid,Shen2021Reconfigurable}. In addition, utilizing electrowetting techniques, a voltage applied to electrodes causes charge redistribution on the surface of the fluid metal. This alters the surface tension and generates Marangoni forces, which move the liquid metal and thereby change the shape and position of the antenna within the container \cite{Wang2022Continuous,wu2024fluidMag}. The typical response time of a liquid-based MA is in the order of milliseconds to seconds. The selection of a suitable fluid material is a significant challenge for fluid antennas, as it must meet various criteria including cost, safety, physical and chemical stability, melting point, evaporation, and viscosity. The electromagnetic properties of the fluid, such as permittivity, permeability, conductivity, and loss, are crucial for antenna performance and must be optimized. Additionally, achieving rotational functionality in fluid antennas to enable omnidirectional coverage remains a challenging problem that requires further research and innovation.

% \footnote{Reconfigure antennas have the broadest scope encompassing MA, FAS, and other forms of tunable antennas driven by either mechanical or electronic components \cite{haupt2013reconfigurable}. These electronically reconfigurable antennas with flexible positions can be classified as FAS if we follow its definition in \cite{Wong2022bruce,wong2020fluid}.} 
\subsubsection{Electronically Reconfigurable Element}
Electronically reconfigurable antennas, such as dual-mode patch antennas, can adjust the phase center to achieve equivalent movement of active antenna positions. For example, by exciting different modes in a stacked circular patch antenna, e.g., TM11 and TM21, the phase center can be displaced from the physical center, which effectively changes the antenna position and radiation pattern without involving mechanical movement \cite{ning2024movable}. Besides, the PIN diode biasing network is another efficient way for achieving rapid reconfiguration of antenna radiation patterns \cite{costantine2015reconfigurable}. In addition, the pixel antenna consists of massive pixels that can be reconfigured using electronic switches \cite{jiang2021pixel}. In this way, the position and/or radiation characteristics of the antenna are changed, which can be regarded as an equivalent way of implementing active antenna movement. Benefiting from the electronic architecture, the electronically reconfigurable antenna has a response time ranging from nanoseconds to milliseconds. Electronically reconfigurable antennas circumvent issues such as wear and the space required for drive mechanisms, making them a promising approach for implementing compact wireless systems with flexible antenna positions and/or radiation patterns. Nevertheless, the practical constraints for electronically reconfigurable antennas lie in the high hardware cost and the complex circuit for realizing reconfigurability \cite{ning2024movable}. Moreover, the displacement of the phase center and the coverage range of the radiation pattern for electronically reconfigurable antennas are usually limited as compared to mechanical MAs.

\subsubsection{Sliding Array}
A sliding array consists of one array or multiple sub-arrays that can slide along predefined paths or within specific regions \cite{yichi2024movable}. This allows for a certain degree of spatial reconfiguration to enhance wireless transmission and reception. The array can dynamically adjust the distances between sub-arrays, which enables more effective beam coverage and aperture enhancement for better signal transmission. Compared to element-level MAs, the sliding array architecture can effectively reduce the hardware cost because multiple antenna elements within each sub-array can be collectively moved using the same drive component, at the sacrifice of DoFs in their movement. Similarly, the practical constraints of sliding arrays revolve around the mechanical systems that enable the movement of sub-arrays. These systems must be robust, reliable, and capable of precise control. The design should also account for the potential for increased structural complexity and the need for efficient control algorithms to manage the positions of the sub-arrays without excessive energy consumption or latency.

% \footnote{Both sliding and rotatable arrays can be regarded as simplified implementations of 6DMA \cite{shao2024Mag6DMA}, achieving different trade-offs between the hardware complexity and movement flexibility.}
\subsubsection{Rotatable Array}
A rotatable array is a single array with flexible rotation/orientation adjustment, which is a special case of 6DMA suitable for environments with nonuniform and time varying user distributions by fixing the antenna position and allowing flexible rotation \cite{shao20246DMA,shao2024discrete,shao2024Mag6DMA}. For example, the yaw and pitch rotations of an array at the BS can tune the main lobe of antennas’ radiation pattern towards target user clusters, which thus helps improve their effective channel gains and rate performance \cite{shao20246DMA,shao2024discrete,shao2024Mag6DMA,ning2024movable}. In addition, based on the spatially varying distribution of users, such as those in skyscrapers or on the street, a rectangular rotatable array (i.e., 6DMA with fixed array position) can be strategically placed vertically or horizontally via roll rotation to improve coverage performance. Even with fixed antenna position for reducing the movement cost, the optimization of antenna rotation for rotatable arrays is still a non-trivial problem in the setup of multiuser communications \cite{shi2024capacity}, which calls for further investigation in future work.

\subsubsection{Deployable Array}
Deployable structures, capable of expanding and contracting, have enabled the development of deployable (or foldable) arrays. These arrays utilize internal mechanical structures to adjust their geometry, including the position and orientation of each antenna array. Common approaches for realizing transformable arrays include origami techniques and inflatable structures \cite{venkatesh2022origami, babuscia2013inflatable}. Deployable arrays have been widely used in spacecraft applications \cite{Chahat2017deployable}, where they can be folded for compact storage during launch and expanded during operation. This adaptability can also benefit terrestrial systems. For instance, a deployable array equipped at the BS can be folded to reduce wind resistance in extreme weather, while its extendable structure can be adapted to provide communication services. Compared to external mechanisms, deployable arrays offer a simpler and more cost-effective solution for implementing MAs. However, their positioning and rotational ranges are constrained by the mechanical properties of the structure, limiting flexibility compared to MAs with external machinery. Additionally, the durability and reliability of these mechanisms must be carefully considered in practical wireless systems.

The representative architectures of mechanically movable and electronically reconfigurable antennas are summarized in Table \ref{Tab_architecture}. Based on the scale of movement, we can categorize the aforementioned architectures into two types, i.e., element-level and array-level movement. The element-level movement can be actuated in three ways: mechanically movable element, liquid-based element, and electronically reconfigurable element. In contrast, array-level movements are primarily realized by mechanical structures, which can be further divided into sliding array, rotatable array, and deployable array based on the movement mode. The examples of implementing these movable and reconfigurable antennas are illustrated in Fig. \ref{Fig_implemantation}. Note that these architectures of MAs can be applied across different frequency bands, including sub-6 GHz, millimeter-wave (mmWave), and terahertz (THz) bands. Since the size of an antenna element is usually in direct proportion to the carrier wavelength, the size of the movable element may range from micrometers to decimeters. Besides, depending on the number of antenna elements, the size of a movable array may vary from millimeters to meters or even dozens of meters. Considering the unique advantages and limitations of various hardware architectures, it is crucial to select the most suitable structures (or their combinations) based on specific system requirements, operating environments, and practical constraints, such that a superior balance between hardware cost, energy efficiency, system reliability, and overall performance is achieved. For example, mobile devices with a limited space are more likely to adopt MEMS-enabled MAs or electronically reconfigurable antennas due to their compact size and rapid reconfiguration capabilities, which allow for real-time improvements in instantaneous channel conditions. Conversely, motor-based MAs are more suitable for BSs or large-scale machines. A common example is downtilt antennas widely deployed in existing cellular networks. While their movement speed is limited, motor-based MAs can perform gradual adjustments to accommodate slowly-varying instantaneous channels or long-term statistical channel characteristics. To further reduce hardware costs associated with antenna movement, array-level movement architectures can be implemented, where a small number of motors drive multiple antennas collectively. Additionally, various hybrid architectures, such as hybrid sliding and rotatable structures or combinations of motor-based arrays and electronically reconfigurable elements, can be developed to leverage the advantages of different implementation methods while mitigating their respective limitations. With improvements in mechanical and electronic processes, it is anticipated that more advanced architectures will emerge in the rapidly evolving field of MAs.

\subsection{Optimization Framework and Design Issues}
To fully exploit the DoFs in reconfiguring antenna positions and orientations, we can formulate a general optimization framework for MA/6DMA-aided wireless systems, which is given by \cite{zhu2023MAMag,zhu2022MAmodel,ma2022MAmimo,ma2024MAsensing,shao20246DMA,shao2024discrete,shao2024Mag6DMA,shao2024exploiting}
\begin{subequations}\label{eq_framework}
	\begin{align}
		\mathop{\max}\limits_{\{\mathbf{t}, \mathbf{r}, \mathbf{\Psi}, \mathbf{\Omega}\}, \{\mathbf{c}\}}~
		&U(\{\mathbf{t}, \mathbf{r}, \mathbf{\Psi}, \mathbf{\Omega}\}, \{\mathbf{c}\}) \\
		\mathrm{s.t.}~~~~~~  &f_{i}(\{\mathbf{t}, \mathbf{r}, \mathbf{\Psi}, \mathbf{\Omega}\}) \geq 0, ~ 1 \leq i \leq I_{1},\\
		&g_{i}(\{\mathbf{c}\}) \geq 0, ~ 1 \leq i \leq I_{2},\\
		&q_{i}(\{\mathbf{t}, \mathbf{r}, \mathbf{\Psi}, \mathbf{\Omega}\}, \{\mathbf{c}\}) \geq 0, ~ 1 \leq i \leq I_{3},
	\end{align}
\end{subequations}
where $\{\mathbf{t}, \mathbf{r}, \mathbf{\Psi}, \mathbf{\Omega}\}$ and $\{\mathbf{c}\}$ denote the sets of variables for antenna movement  (including the positions and orientations of both Tx-MAs and Rx-MAs) and communication/sensing resource (e.g., power, bandwidth, beamformer/waveform, user association), respectively. Specifically, $U(\cdot)$ is the utility function (e.g., achievable rate, secrecy rate, energy efficiency, coverage probability, detection probability, sensing accuracy); $f_{i}(\cdot)$ represents the constraint on antenna movement (e.g., finite movable region, discrete antenna movement, minimum antenna spacing, maximum antenna moving speed); $g_{i}(\cdot)$ is the constraint on communication/sensing resource (e.g., maximum transmit power, maximum system bandwidth, beamformer/waveform constraints, maximum number of associated users); and $q_{i}(\cdot)$ is the coupled constraints on antenna movement and communication/sensing resource (e.g., minimum received signal power, maximum interference leakage, minimum signal-to-interference-plus-noise ratio (SINR), maximum Cramer-Rao bound (CRB)).

The utility function in \eqref{eq_framework} defines the key performance metrics of MA/6DMA-aided wireless systems. Its formulation depends on specific system setups and practical requirements, such as achievable rate, energy efficiency, and coverage probability in wireless communication systems, or detection probability, Cramer-Rao bound, and estimation accuracy in wireless sensing systems. Leveraging the additional DoFs provided by antenna movement optimization, MA-aided wireless systems can significantly enhance these performance metrics compared to their FA counterparts. However, incorporating an antenna movement module introduces additional hardware complexity and energy consumption. Practical constraints, including the movement range and speed, movement frequency and duration, and associated energy consumption, critically affect the achievable performance gains. Consequently, efficient design of hardware architectures and movement mechanisms (i.e., addressing how to move, when to move, and where to move) under specific system setups and practical requirements is essential to unleash the potential of MAs in future wireless networks.

In practice, to maximize the utility function, it is required to develop efficient algorithms for antenna movement optimization. However, as shown in \eqref{eq_channel_basic}, \eqref{eq_channel_MIMO}, \eqref{eq_channel_MIMO_eff}, \eqref{eq_channel_wideband_CFR}, \eqref{eq_channel_near}, and \eqref{eq_channel_rotatable}, due to the highly non-linear relationship between the antenna position/orientation and the channel gain, the antenna movement optimization is challenging in general. Moreover, for MA-aided wireless communication systems, accurate channel state information (CSI) between the Tx and Rx regions is crucial for antenna movement optimization. Different from conventional systems which only require the CSI between FAs, the mapping between the antenna position/orientation to the channel gain should be acquired for MA/6DMA systems, which involves a large (or even infinite) number of states of the antenna position and orientation. Thus, channel acquisition incurs another challenge for MA systems. Next, we will delve into these two design issues for MA systems, i.e., antenna movement optimization and channel acquisition. Due to the page limit and for ease of exposition, we mainly focus on MAs with flexible positions, which is in the majority of the existing literature, while the system design involving antenna rotations can be referred to some recent works \cite{shao20246DMA,shao2024discrete,shao2024Mag6DMA,pi20246DMAcoordi,zhang20246DMAhybrid,zhang2024polarization, zheng2025rotatable, tominaga2024spectral,ren20246DMAUAV,shao2024exploiting,shao2024distributed, sun2025rotatable,yi2024performance,shi20246DMAcellfree, jin2024movableOAM,zhang2025roma,guo2025flexibleMA}.

\subsection{Lessons Learned}
In this section, we introduced the general field-response channel models tailored for MA systems to characterize the spatial variations of wireless channels w.r.t. antenna movement. These models serve as a foundation for optimizing antenna positions and/or orientations in various MA/6DMA-aided wireless communication and sensing systems. Additionally, different implementation architectures present distinct trade-offs among movement flexibility, speed, range, hardware cost, and energy consumption. They impose specific constraints on MA systems, necessitating customized designs to fully harness their performance benefits.

\section{Antenna Movement Optimization} \label{Sec_Movement}
For MA systems, finding the optimal antenna positions is critical to achieving the best communication/sensing performance. Due to the highly non-linear relationship between the wireless channels and antenna positions, antenna movement optimization is challenging. In this section, we present efficient solutions for antenna positioning in various MA-aided wireless communication and/or sensing systems and characterize their performance advantages over conventional fixed-position antenna (FPA) systems. % Use FPA in the following

\begin{figure*}[!t]
	\begin{center}
		\includegraphics[width=18 cm]{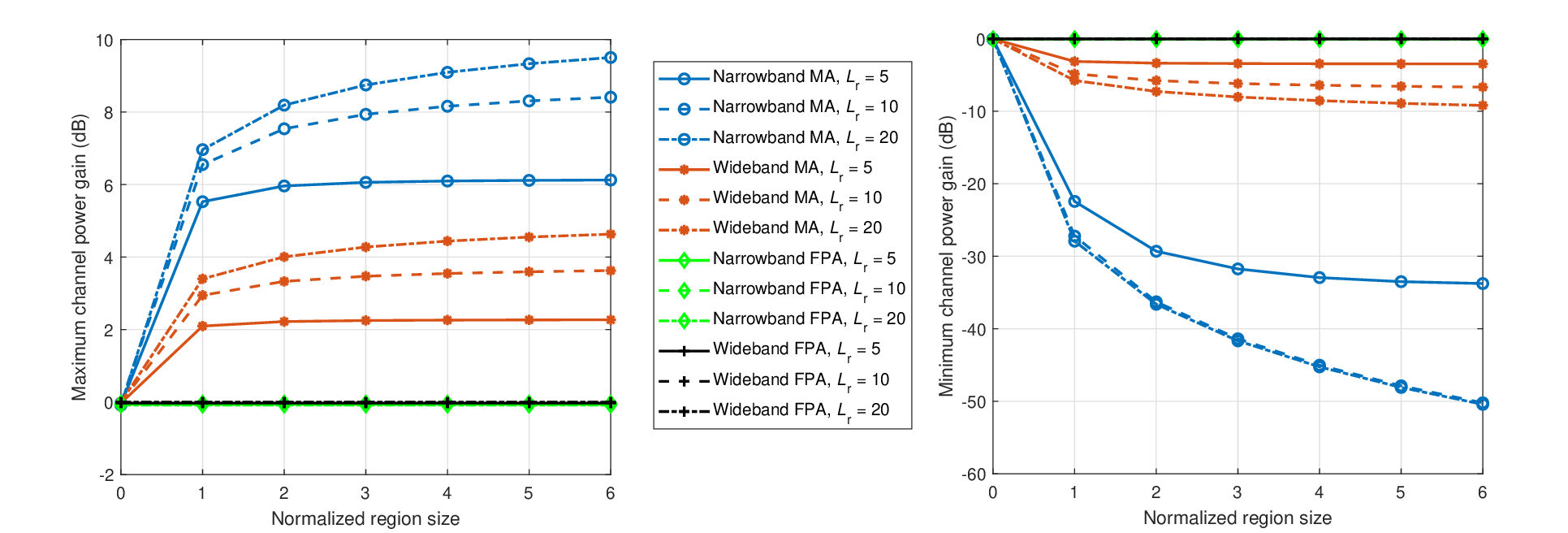}
		\caption{Maximum and minimum channel power gains achieved by an Rx-MA under narrowband and wideband systems.}
		\vspace{-12pt}
		\label{Fig_SISO_gain}
	\end{center}
\end{figure*}
\subsection{MA-Aided SISO Systems}
In MA-aided SISO systems, adjusting the MA position effectively tunes the phases of the coefficients for multiple channel paths. In the spatial domain, these path coefficients can either constructively superimpose to maximize the channel power gain or destructively superimpose to minimize it. Consequently, optimizing the antenna position primarily enhances communication performance in two aspects, i.e., increasing the received signal power and decreasing the interference power. We start with the basic narrowband MA system to demonstrate its performance advantages and then explore the potential gains achievable in wideband MA systems.
\subsubsection{Narrowband Systems}
For ease of exposition, we considered the simplified scenario with an Rx-MA only, while the corresponding results are also applicable to the Tx-MA. Given any fixed position of the Tx antenna, we can simplify the field-response channel model in \eqref{eq_channel_basic} as $h(\mathbf{r})=\mathbf{f}(\mathbf{r})^{\mathrm{H}} \mathbf{b}$, where $\mathbf{b}=\mathbf{\Sigma} \mathbf{g}(\mathbf{t}) \triangleq [b_1, b_2, \dots, b_{L_{\mathrm{r}}}]^{\mathrm{T}}$ is a constant vector. It was shown in \cite{zhu2022MAmodel} that the channel power gain $|h(\mathbf{r})|^{2}$ exhibits periodicity in a 2D antenna moving region if the number of Rx paths is no larger than three. This conclusion can be easily extended to the case of 3D antenna moving regions, in which $|h(\mathbf{r})|^{2}$ exhibits periodicity if the number of Rx paths is no larger than four. Then, we can derive an upper bound on the channel power gain as 
\begin{equation} \label{eq_channel_power_gain_ub}
	\left|h(\mathbf{r})\right|^{2} \leq \left \| \mathbf{b} \right \|_{1}^{2},
\end{equation}
which can be achieved if the phases of the complex coefficients for all the $L_{\mathrm{r}}$ Rx paths are aligned. According to the periodic property, this upper bound is tight for 3D moving regions when the number of Rx paths is no larger than four and the position of the Rx-MA for achieving this bound can be derived in closed form \cite{zhu2022MAmodel}. Similarly, we can obtain a lower bound on the channel power gain as 
\begin{equation} \label{eq_channel_power_gain_lb}
	\left|h(\mathbf{r})\right|^{2} \geq \left(\max \left\{0, 2|b_{l^{\star}}|- \left \| \mathbf{b} \right \|_{1}\right\}\right)^{2},
\end{equation}
where $b_{l^{\star}}$ is the element of $\mathbf{b}$ with the largest amplitude. Note $2|b_{l^{\star}}|- \left \| \mathbf{b} \right \|_{1} = |b_{l^{\star}}| - \sum_{l=1, l \neq l^{\star}}^{L_{\mathrm{r}}} |b_{l}|$ indicates that the coefficient of the $l^{\star}$-th path is destructively superimposed with all other Rx paths' coefficients. This lower bound is also tight for 3D moving regions when the number of Rx paths is no larger than four. 

As the number of Rx paths increases, there is no explicit periodicity of the channel power gain in the spatial domain. Nevertheless, for sufficiently large moving regions, the upper bound in \eqref{eq_channel_power_gain_ub} and the lower bound in \eqref{eq_channel_power_gain_lb} can still be approached with an arbitrarily small gap \cite{zhu2024wideband}. In comparison, the FPA at the reference point of the Rx can only achieve a fixed channel power gain, i.e., $|\sum_{l=1}^{L_{\mathrm{r}}} b_{l}|^2$. Thus, the antenna position optimization for MA-SISO systems can significantly increase the received signal power by maximizing the channel power gain between the Tx and the Rx and decrease the interference by minimizing the channel power gain between the interference source and the Rx, especially for a large number of channel paths. Since only one MA is adopted for SISO systems, the optimal antenna position can be simply obtained by exhaustive search \cite{zhu2022MAmodel} or gradient ascent/descent search \cite{zhu2024wideband}, given the perfect CSI between the Tx and Rx regions. To further facilitate the MA position optimization in practical systems, a recent work \cite{zeng2024csi} has proposed a zeroth-order gradient approximation algorithm to predict the position with the maximum channel power gain, by sequentially measuring the received signal powers at several training positions without explicit channel estimation.

\subsubsection{Wideband Systems}
The performance gain in terms of received signal power improvement and interference mitigation also applies to MA-aided wideband SISO systems. Specifically, given any fixed position of the Tx antenna, we can simplify CIR in \eqref{eq_channel_wideband_CIR} as $h_{\tau}(\mathbf{r})=\mathbf{f}_{\tau}(\mathbf{r})^{\mathrm{H}} \mathbf{b}_{\tau}$, $1 \leq \tau \leq T$, where $\mathbf{b}_{\tau}=\mathbf{\Sigma}_{\tau} \mathbf{g}_{\tau}(\mathbf{t}) \triangleq [b_{\tau,1}, b_{\tau,2}, \dots, b_{\tau, L_{\mathrm{r},\tau}}]^\mathrm{T}$ is a constant vector. Thus, we can also obtain the upper bound and lower bound on the channel power gain over each delay tap as \cite{zhu2024wideband}
\begin{subequations} \label{eq_CIR_bound}
	\begin{align}
		&\left|h_{\tau}(\mathbf{r})\right|^{2} \leq \left \| \mathbf{b}_{\tau} \right \|_{1}^{2},\\
		&\left|h_{\tau}(\mathbf{r})\right|^{2} \geq \left(\max \left\{0, 2|b_{\tau,l_{\tau}^{\star}}|- \left \| \mathbf{b}_{\tau} \right \|_{1}\right\}\right)^{2},
	\end{align}
\end{subequations}
where $l_{\tau}^{\star}$ is the element of $\mathbf{b}_{\tau}$ with the largest amplitude. Note that in addition to tuning the amplitude of the CIR, the phase of the CIR over each delay tap can also be adjusted through antenna position optimization. However, since the position of the Rx-MA inherently impacts the CIR over multiple delay taps, a much larger moving region is required to approach the upper-bound/lower-bound for all delay taps \cite{zhu2024wideband}. Thus, for an MA-aided SISO system with a confined moving region, the antenna position optimization must balance the channel gains over multiple delay taps to achieve the desired communication performance.

We show in Fig. \ref{Fig_SISO_gain} the maximum and minimum channel power gains for enhancing the desired signal and suppressing the undesired interference, respectively, which are achievable by a single Rx-MA. The simulation setup is the same as that of Fig. \ref{Fig_fundamental_DoFs} by neglecting the polarization loss of the antenna. The AoAs for each Rx path follow the uniform distribution over the 3D half-space with the joint probability density function given by $p(\theta, \phi)=\frac{\cos \theta}{2\pi},~\theta \in [-\pi/2, \pi/2],~\phi \in [-\pi/2, \pi/2]$. For wideband systems, the total bandwidth is $20$ MHz and the number of subcarriers is $64$. The delays of the Rx paths are assumed to be random variables following the uniform distribution from $0$ to $0.3$ microseconds ($\mu$s). The wideband channel power gain is averaged over all subcarriers, while the narrowband channel power gain is that over the first subcarrier. As can be observed from Fig. \ref{Fig_SISO_gain} (left), the maximum channel power gain achieved by the MA increases with the size of the antenna moving region and the number of Rx paths. Due to the frequency selectivity, the performance gain of the wideband MA system is smaller than its narrowband counterpart. In addition, as shown in Fig. \ref{Fig_SISO_gain} (right), the narrowband MA system can achieve tens of dB decrease in the minimum channel power gain compared to FPA systems for effective interference suppression. However, the wideband MA system is less effective than its narrowband counterpart for interference suppression because it is difficult to achieve low channel gains over all subcarriers. In summary, the MA system can flexibly adjust the antenna position for increasing/decreasing the channel gain of the desired/interference link, especially for scenarios with a large number of channel paths having a small delay spread.

\subsection{MA-Aided Multiple-Input Single-Output (MISO)/Single-Input Multiple-Output (SIMO) Systems}
In MISO/SIMO communication systems, MAs can achieve more efficient multi-path channel reconfiguration by jointly optimizing their positions, allowing the total channel power gain to be enhanced (or reduced) for signal transmission (or interference mitigation), similar to the effects observed in SISO systems \cite{ma2024multi,Lai2024FAScombining}. Moreover, when utilizing an MA array, position adjustments of multiple antenna elements can effectively reshape the array geometry, fundamentally altering the steering vectors (i.e., array response vectors) of the antenna array. This reconfiguration enables the spatial correlations of steering vectors across different directions to be adjusted, facilitating more flexible beamforming and advanced array signal processing, as presented in the following. For ease of exposition, we focus on MISO systems with transmit beamforming only in the subsequent discussion, while it can be similarly applied to SIMO systems with receive beamforming.

\subsubsection{LoS Channels}\label{los_Ch} 
Consider a linear MA array with $N_{\mathrm{t}}$ antenna elements, with the position of the $n$-th antenna element denoted as $x_n$, $1 \leq n \leq N_{\mathrm{t}}$. The total length of the antenna moving region is denoted as $A$, with $0 \le x_n \le A, 1 \leq n \leq N_{\mathrm{t}}$. Let $\mathbf{x}=[x_1,x_2,\dots,x_{N_{\mathrm{t}}}]^{\mathrm{T}} \in\mathbb{R}^{N_{\mathrm{t}}\times 1}$ denote the APV of the MA array. For a given steering angle $\theta$, the steering vector (also known as array response vector) of the MA array can be expressed as a function of the APV $\mathbf{x}$ and $\theta$, i.e.,
\begin{align}
	\mathbf{a}{(\mathbf{x},\theta)}=[\mathrm{e}^{\mathrm{j}\frac{2\pi}{\lambda}x_1\cos\theta},\dots,\mathrm{e}^{\mathrm{j}\frac{2\pi}{\lambda}x_{N_{\mathrm{t}}}\cos\theta}]^{\mathrm{T}}\in\mathbb{C}^{N_{\mathrm{t}}\times 1}.
\end{align}
Denote by $\mathbf{w} \in\mathbb{C}^{N_{\mathrm{t}} \times 1}$ the normalized AWV of the MA array, with $\Vert \mathbf{w}\Vert_2^2=1$. Then, for any given $\mathbf{x}$ and $\mathbf{w}$, the beam gain over the steering angle $\theta$ can be expressed as 
\begin{align}
	G(\mathbf{x},\mathbf{w},\theta)=\vert\mathbf{a}(\mathbf{x},\theta)^{\mathrm{H}}\mathbf{w}\vert^2.
\end{align}

\begin{figure*}[!t]
	\centering
	\includegraphics[width=12 cm]{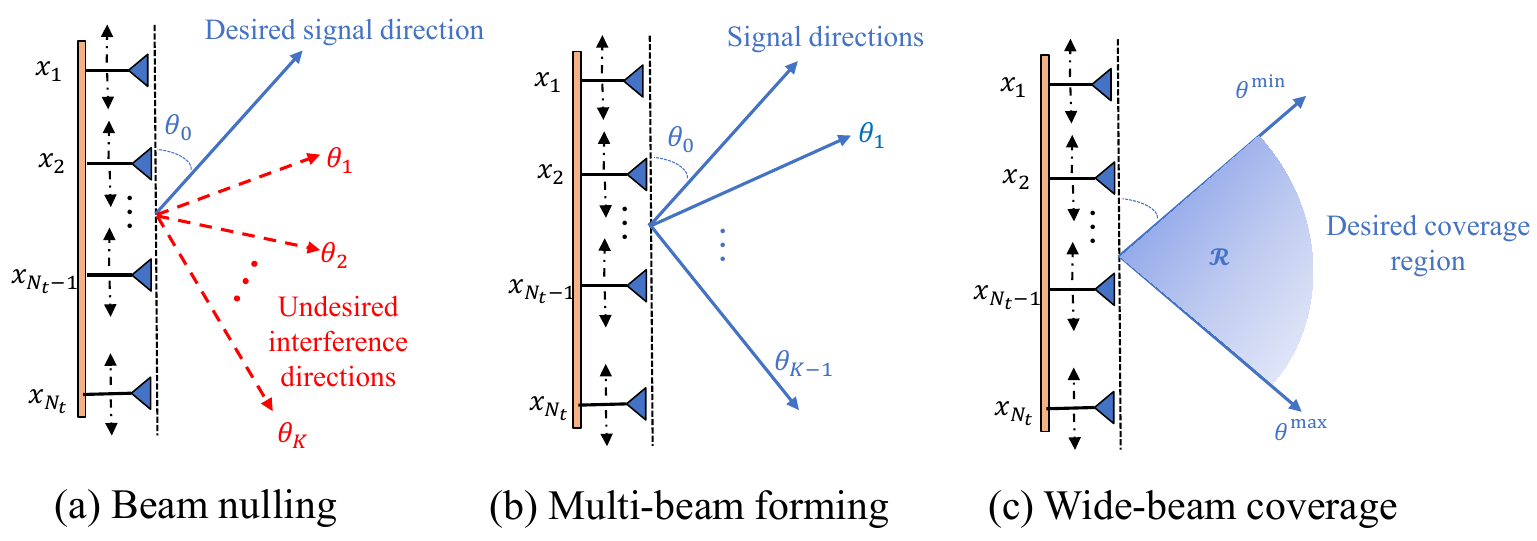}
	\DeclareGraphicsExtensions.
	\vspace{-9pt}
	\caption{MA array for flexible beamforming.}\label{MA_beam}
	\vspace{-12pt}
\end{figure*}

\begin{figure*}[!t]
	\centering
	\subfigure[MA for beam nulling]{\includegraphics[width=0.30\textwidth]{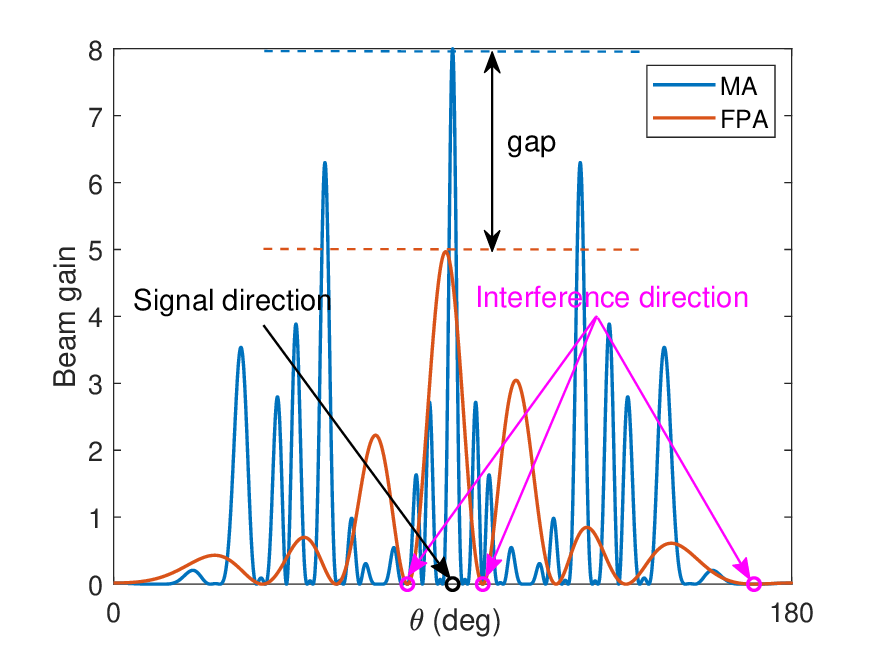}}
	\subfigure[MA for multi-beam forming]{\includegraphics[width=0.30\textwidth]{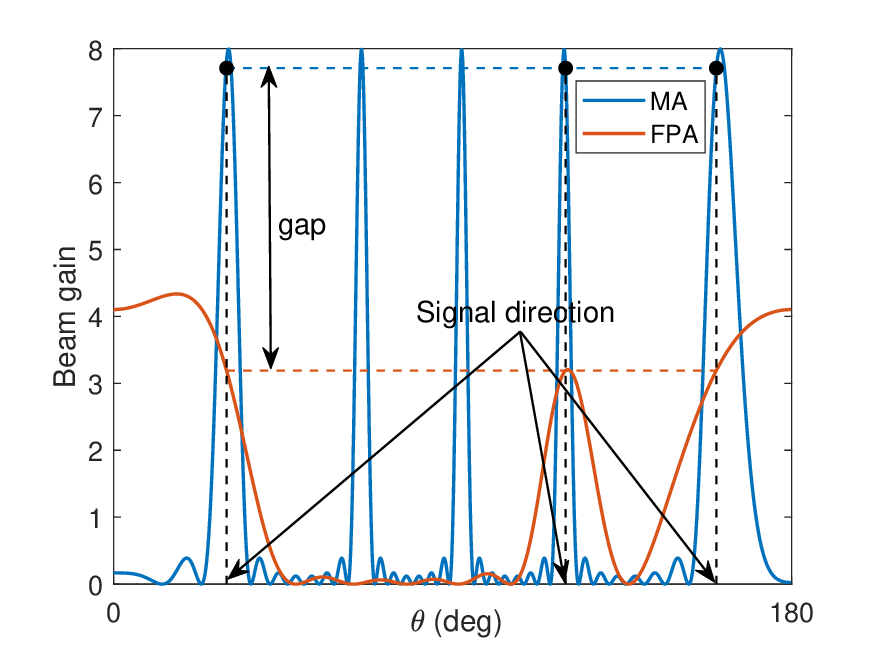}}
	\subfigure[MA for wide-beam coverage]{\includegraphics[width=0.30\textwidth]{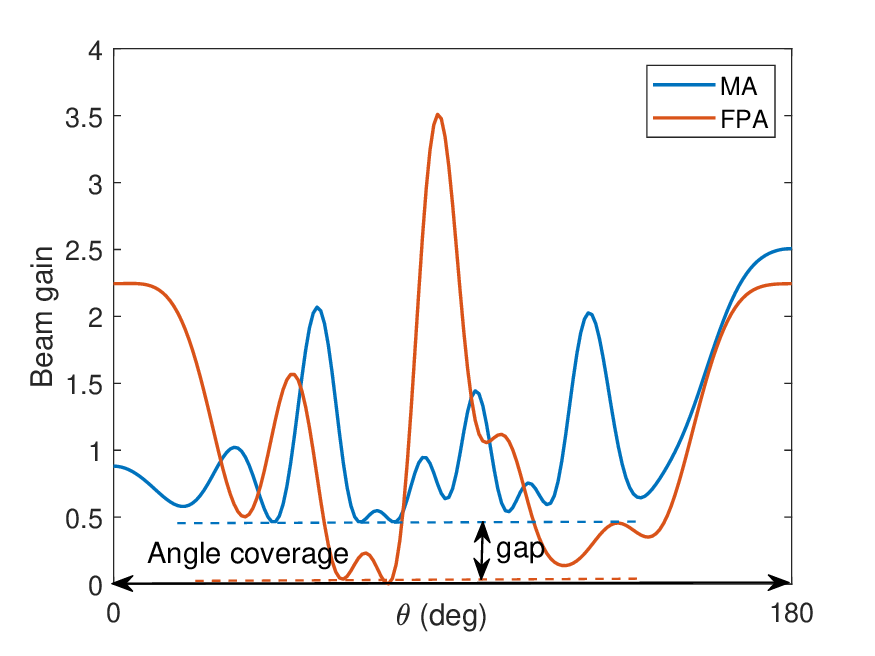}}
	\caption{Comparison of the beam patterns between MA and FPA arrays.}\label{beampattern}
	\vspace{-12pt}
\end{figure*}

By jointly optimizing the AWV (i.e., $\mathbf{w}$) and the APV (i.e., $\mathbf{x}$), the MA array can achieve more efficient beam nulling \cite{zhu2023MAarray, Hu2024MAarrayleak}, multi-beam forming \cite{ma2024multi}, and wide-beam coverage \cite{wang2024flexible} compared to the conventional FPA array that optimizes the AWV only, as shown in Fig.\,\ref{MA_beam}. Specifically, for {\it beam nulling}, we aim to jointly optimize $\mathbf{w}$ and $\mathbf{x}$ to maximize the beamforming gain of the MA array over a desired direction while achieving null steering over multiple undesired directions (e.g., for interference avoidance), as shown in Fig.\,\ref{MA_beam}(a). Mathematically, let $\theta_0$ denote the angle of the desired direction and $\{\theta_k\}_{1\leq k \leq K}$ denote the angles of all undesired interference directions, where $K$ denotes the total number of null directions. As such, the joint optimization of AWV and APV for beam nulling can be formulated as 
\begin{subequations}\label{beamnull}
	\begin{align}
		{\text{(P-Null)}}\;\max_{\mathbf{x},\mathbf{w}}\; &G(\mathbf{x},\mathbf{w},\theta_0)
		\\
		\text{s.t.} \quad & G(\mathbf{x},\mathbf{w},\theta_k)=0, \quad 1\leq k\leq K,
		\label{null_cons}\\ 
		&0 \le x_n \le A, \quad 1 \leq n \leq N_{\mathrm{t}},\label{region}
		\\
		&\vert x_m-x_n\vert\geq d_{\min},\quad m\neq n,
		\label{d_min}\\
		& \Vert \mathbf{w}\Vert_2^2=1,\label{bf_cons}
	\end{align}
\end{subequations}
where $d_{\min}$ in constraint (\ref{d_min}) is the minimum distance between any two MAs to avoid the mutual coupling effect.

For any given APV $\mathbf{x}$, the optimal AWV solution for (P-Null) can be obtained as zero-forcing (ZF) beamforming\cite{zhu2023MAarray}. However, this renders the APV optimization problem difficult to solve due to the intricate nature of ZF beamforming expressed in terms of the APV. To circumvent this difficulty, it is noted that in the absence of the nulling constraints in (\ref{null_cons}), the optimal value of (P-Null) can be obtained as $N_{\mathrm{t}}$ by setting 
\begin{equation}\label{idealBF}
	\mathbf{w}^{\star}(\mathbf{x})=\frac{\mathbf{a}(\mathbf{x},\theta_0)}{\Vert\mathbf{a}(\mathbf{x},\theta_0)\Vert_2}.
\end{equation}

The AWV in (\ref{idealBF}) may not satisfy the nulling constraints in (\ref{null_cons}). Nonetheless, by judiciously optimizing the APV, it was revealed in \cite{zhu2023MAarray} that under certain numbers of MAs and null-steering directions, there always exists an APV solution such that (\ref{null_cons}) can be satisfied by applying the AWV in (\ref{idealBF}), i.e., ${\mathbf{a}}(\mathbf{x},\theta_k)^{\mathrm{H}}{\mathbf{a}}(\mathbf{x},\theta_0)=0, 1 \le k \le K$, referred to as {\it steering vector orthogonality (SVO) condition} \cite{zhu2023MAarray}. For example, if the size of the MA moving region, i.e., $A$, is sufficiently large and $K=1$, an APV satisfying the SVO condition and constraint (\ref{d_min}) always exists for any $N_{\mathrm{t}} \ge 2$. Furthermore, it was shown in \cite{zhu2023MAarray} that the SVO condition can also be achieved for $K > 1$ by utilizing a sequential APV construction method, provided that $K$ does not exceed a threshold determined by the prime factorization of $N_{\mathrm{t}}$. However, in other more general cases, the SVO condition may not be satisfied, and a suboptimal solution to (P-Null) was also proposed in \cite{zhu2023MAarray}.

For {\it multi-beam forming}, we aim to maximize the minimum beamforming gain over multiple desired directions, as shown in Fig.\,\ref{MA_beam}(b). For notational simplicity, we assume $K$ desired directions and denote their steering angles as $\theta_k, 0 \le k \le K-1$, respectively. The associated optimization problem can be formulated as 
\begin{subequations}\label{multibeam}
	\begin{align}
		{\text{(P-MBF)}}\;\max_{\mathbf{x},\mathbf{w},\delta}\; &\delta
		\\
		\text{s.t.} \quad &{\text{(\ref{region}), (\ref{d_min}), (\ref{bf_cons})}},
		\\
		&G(\mathbf{x},\mathbf{w},\theta_k)\geq \delta,\quad  0 \le k \le K-1.\label{delta}
	\end{align}
\end{subequations}
Note that additional constraints can be incorporated into (P-MBF) to mitigate interference towards undesired directions, as studied in \cite{ma2024multi}. However, (P-MBF) is a non-convex optimization problem that is challenging to be solved optimally. Nonetheless, it can be easily seen that to maximize the beam gain at any specific direction (e.g., $\theta_0$), the optimal AWV is given by (\ref{idealBF}). Evidently, if there exists an APV solution such that the APV in (\ref{idealBF}) can also maximize the beam gain at the other $K-1$ directions, i.e., $|{\mathbf{a}}(\mathbf{x},\theta_k)^{\mathrm{H}}{\mathbf{a}}(\mathbf{x},\theta_0)|=N_{\mathrm{t}}, 1 \le k \le K-1$, then it must be an optimal solution to (P-MBF). It is worth noting that this condition leads to grating lobes at the other $K-1$ directions; hence, we refer to it as {\it grating-lobe condition}. 

Let $\Delta_k=\cos\theta_k-\cos\theta_0, 1 \le k \le K-1$. By performing some mathematical manipulations, it can be shown that there always exists an APV such that the grating-lobe condition can be met if all $\Delta_k$'s are rational numbers. Moreover, an optimal APV admits an equal spacing between any two adjacent MAs, which can be constructed based on the irreducible fractions of $\Delta_k$'s. However, due to the random distribution of users or targets in the spatial domain, it is difficult to fully achieve the grating-lobe condition in practice. To solve (P-MBF) in the general case, the authors in \cite{ma2024multi} proposed an alternating optimization (AO) algorithm to iteratively optimize one of the APV $\mathbf{x}$ and the AWV $\mathbf{w}$ with the other being fixed, where the successive convex approximation (SCA) technique is leveraged to find a suboptimal solution for each subproblem.

For {\it wide-beam coverage}, we aim to achieve a uniform beam gain over one (or multiple) continuous region in the spatial domain, ensuring signal coverage at any position within it, as shown in Fig.\,\ref{MA_beam}(c). Mathematically, let ${\cal R}=\{\theta | \theta_{\min} \le \theta \le \theta_{\max}\}$ denote the region of interest, where $\theta_{\min}$ and $\theta_{\max}$ denote its boundaries with $\theta_{\min} < \theta_{\max}$. Note that by changing the width of the region (i.e., $\theta_{\max}-\theta_{\min}$), flexible beam coverage can be achieved accordingly. Assuming analog beamforming with a constant magnitude, the authors in \cite{wang2024flexible} investigated the beam coverage problem with MAs, aiming to maximize the minimum beam gain among all possible angles within $\cal R$, i.e.,
\begin{equation}\label{g_min}
	G_{\min}(\mathbf{x},\mathbf{w})=\min_{\theta\in\mathcal{R}}\;G(\mathbf{x},\mathbf{w},\theta),
\end{equation}
by jointly optimizing the APV and AWV subject to (\ref{region}), (\ref{d_min}), and (\ref{bf_cons}). However, due to the continuous nature of the coverage region and the unit-modulus constraints on analog beamforming, this problem is difficult to be optimally solved. To overcome this difficulty, the authors in \cite{wang2024flexible} first discretized the continuous target region $\cal R$ into multiple subregions, assuming that the beam gain within each subregion is approximately constant. Then, an AO algorithm was proposed to obtain a high-quality suboptimal solution to solve this discrete problem.

In Figs.\,\ref{beampattern}(a)-\ref{beampattern}(c), we compare the beam patterns between MA and FPA arrays for beam nulling, multi-beam forming, and wide-beam coverage, respectively, with $N_{\mathrm{t}}=8$, $d_{\min}=\lambda/2$, and $A=20\lambda$. In Fig.\,\ref{beampattern}(a), we set $K=3$, $\theta_0=90^{\circ}$, $\theta_1=78^{\circ}$, $\theta_2=98^{\circ}$ and $\theta_3=170^{\circ}$. It is observed that a full array gain of the MA array and null steering over all three undesired directions are achieved concurrently. In contrast, the FPA array applies ZF-based AWV and suffers a $40\%$ loss in beamforming gain over $\theta_0$. In Fig.\,\ref{beampattern}(b), we set $K=3$, $\theta_0=30^{\circ}$, $\theta_1=120^{\circ}$, and $\theta_2=160^{\circ}$. It is observed that an MA array can achieve a much higher max-min beamforming gain than an FPA array, which even approaches the full beamforming gain (i.e., $N_{\mathrm{t}}=8$) over each desired direction. In Fig.\,\ref{beampattern}(c), we set $\theta_{\max}=180^{\circ}$ and $\theta_{\min}=0^{\circ}$. It is observed that the MA array results in reduced fluctuation in beamforming gain within the coverage region and achieves a max-min beamforming gain over 20 dB greater than that of the FPA array. The above observations demonstrate that MA arrays can enable more flexible beamforming compared to conventional FPA arrays by exploiting the additional DoFs in antenna position optimization. The beam patterns in Fig. \ref{beampattern} are three typical examples to showcase the advantages of linear MA arrays in flexible beamforming. These techniques are also applicable to MA arrays with 2D antenna movement and/or other design objectives. For example, the studies in \cite{ZhuLP_satellite_MA} have validated the superiority of MA arrays with 2D position adjustments for decreasing the interference leakage, i.e., mitigating the sidelobe level, in LEO satellite communication systems.

\subsubsection{NLoS Channels}
\begin{figure}[!t]
	\centering
	\includegraphics[width=2.8in]{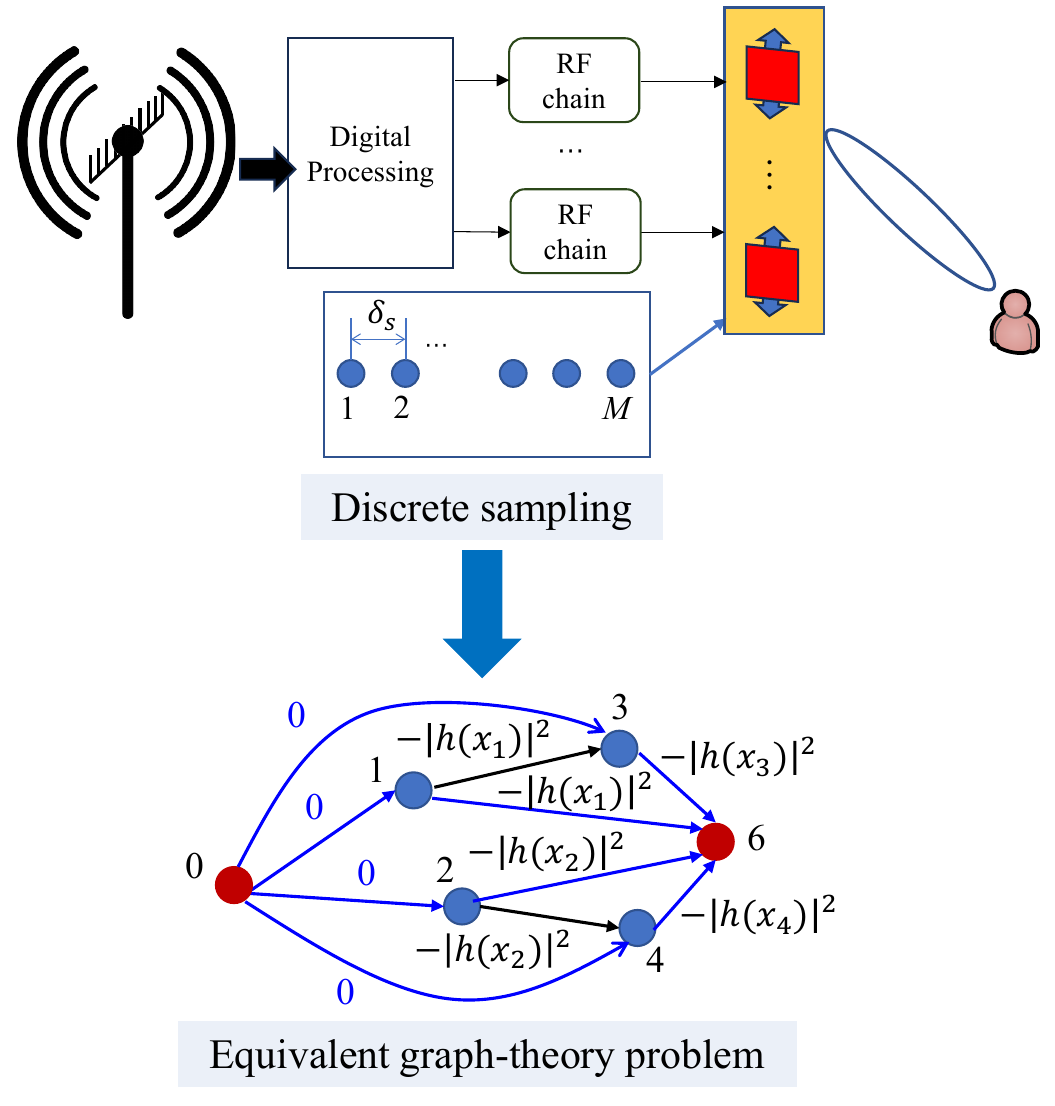}
	\DeclareGraphicsExtensions.
	\vspace{-6pt}
	\caption{MA position optimization in MISO systems using discrete sampling and graph theory.}\label{graph_theory}
	\vspace{-12pt}
\end{figure}

Compared to LoS channels, the joint optimization of AWV and APV becomes more challenging under NLoS channels as more channel paths are involved. For example, consider the same $N_{\mathrm{t}}$-MA Tx with a linear array as described in Section \ref{los_Ch}, transmitting to a single-FPA Rx, as shown in Fig.\,\ref{graph_theory}. Considering the basic field-response channel model in (\ref{eq_channel_basic}) and (\ref{eq_channel_MIMO}), the received signal power at the Rx for any given APV $\mathbf{x}$ and AWV $\mathbf{w}$ is given by
\begin{equation}\label{NLOSPw}
	P_{\text{R}}(\mathbf{x},\mathbf{w}) = \lvert \mathbf{h}(\mathbf{x})^{\mathrm{H}} \mathbf{w} \rvert^2,	
\end{equation}
where $\mathbf{h}(\mathbf{x})^{\mathrm{H}}=\mathbf{f}(\mathbf{r})^{\mathrm{H}} \mathbf{\Sigma}\mathbf{G}(\mathbf{x}) \in {\mathbb C}^{1 \times N_{\mathrm{t}}}$. Since the position of the single Rx-FPA is a constant, the explicit dependence of (\ref{NLOSPw}) on $\mathbf{r}$ is omitted.

To maximize (\ref{NLOSPw}) for any given APV $\mathbf{x}$, the Tx should apply the maximal ratio transmitting (MRT) beamformer, i.e., $\mathbf{w} = \mathbf{h}(\mathbf{x})/\lVert \mathbf{h}(\mathbf{x}) \rVert_{2}$ (assuming a unit transmit power). The resulting maximum received power is given by
\begin{align}\label{recPw}
	P_{\text{R}}(\mathbf{x}) &= \lVert \mathbf{h}(\mathbf{x}) \rVert_{2}^2 = \mathbf{f}_{0}^{\mathrm{H}} \mathbf{\Sigma}\mathbf{G}(\mathbf{x})\mathbf{G}^{\mathrm{H}}(\mathbf{x}){\mathbf{\Sigma}}^{\mathrm{H}}\mathbf{f}_{0} \nonumber\\
	&=\sum\limits_{n=1}^{N_{\mathrm{t}}}{\lvert h(x_n) \rvert^2},
\end{align}
where $h(x_n)=\mathbf{g}(x_n)^{\mathrm{H}}{\mathbf{\Sigma}}^{\mathrm{H}}\mathbf{f}(\mathbf{r})$ denotes the $n$-th entry of $\mathbf{h}(\mathbf{x})$, which only depends on $x_n$.

Next, we aim to optimize the APV $\mathbf{x}$ to maximize (\ref{recPw}), i.e.,
\begin{equation}\label{NLoSOpt_cont}
	{\text{(P-MISO-Cont)}}\;\mathop {\max}\limits_{\mathbf{x}}\; \sum\limits_{n=1}^{N_{\mathrm{t}}}{\lvert h(x_n) \rvert^2},\quad
	\text{s.t.}\;\; {\text{(\ref{region}), (\ref{d_min})}}.
\end{equation}
It is noted that (P-MISO-Cont) is difficult to optimally solve due to its highly nonlinear objective function w.r.t. the APV $\mathbf{x}$. Although a variety of gradient-based algorithms have been proposed to tackle the nonlinear objective function, they are suboptimal in general and may result in performance loss. An alternative approach is by discretizing the linear antenna moving region into a multitude of sampling points, thereby transforming \eqref{NLoSOpt_cont} into a discrete selection problem, as shown in Fig.\,\ref{graph_theory}. Specifically, we uniformly sample the linear region $[0, A]$ into $M\;(M \gg N_{\mathrm{t}})$ discrete positions, with an equal spacing between any two adjacent sampling points given by $\delta_s=A/M$. Hence, the position of the $m$-th sampling point is given by $s_m=\frac{mA}{M}, 1 \leq m \leq M$. As such, the position of each MA can be selected as one of the sampling points in $\{s_m\}$. Let $a_n$ denote the index of the selected sampling point for the $n$-th MA, $1 \leq n \leq N_{\mathrm{t}}$. Thus, the position of the $n$-th MA can be expressed as $x_n=s_{a_n}=\frac{a_nA}{M}, 1 \leq n \leq N_{\mathrm{t}}$. Problem (P-MISO-Cont) can be recast as
\begin{subequations}
	\begin{align}\label{NLoSOpt_disc}
		{\text{(P-MISO-Disc)}}\;\mathop {\max}\limits_{\{a_n\}}\; &\sum\limits_{n=1}^{N_{\mathrm{t}}}{\left| h\Big(\frac{a_nA}{M}\Big) \right|^2} \\
		\text{s.t.}\;\;& 1 \leq a_n \leq M,~  1 \leq n \leq N_{\mathrm{t}},\\
		&\lvert a_m - a_n \rvert \ge a_{\min}, ~ m \ne n,\label{NLoSOpt_disc_index}
	\end{align}
\end{subequations}
where $a_{\min} = d_{\min}/ \delta_s \gg 1$. Note that (P-MISO-Disc) differs from conventional AS problems due to the existence of the constraint on the minimum inter-antenna spacing given by \eqref{NLoSOpt_disc_index} and the much larger number of sampling points compared to the number of candidate antennas in AS. Although (P-MISO-Disc) avoids the highly nonlinear objective function as encountered by (P-MISO-Cont), it is still difficult to optimally solve due to its combinatorial nature. The optimal APV solution to (P-MISO-Disc) can be obtained via an exhaustive search, which, however, incurs an exorbitant complexity in the order of $M-(a_{\min}-1)(N_{\mathrm{t}}-1) \choose N_{\mathrm{t}}$, and may not be applicable to large-scale antenna arrays with large $M$ and/or $N_{\mathrm{t}}$ values in practice. In \cite{mei2024movable}, the authors proposed an efficient graph-based algorithm to solve (P-MISO-Disc) optimally in {\it polynomial} time. Its basic idea is to recast (P-MISO-Disc) as an equivalent fixed-hop shortest path problem by ingeniously modeling the sampling points as vertices in a graph and defining its edge weights based on $\lvert h(x_n) \rvert$'s. Moreover, an edge is added between two vertices if their indices satisfy (\ref{NLoSOpt_disc_index}), as shown in Fig.\,\ref{graph_theory}. It is worth noting that the above graph-optimization method can be applied to both field-response and spatial-correlation channel models as long as the CSI over the sampling positions is given. The MA position optimization with discrete sampling has also been applied to other scenarios, such as multi-antenna broadcasting \cite{wu2023movable}, spectrum sharing \cite{wei2024joint}, physical-layer security \cite{mei2024movable_secure}, IRS-aided wireless communications \cite{WeiX_MA_RIS}, etc.

\begin{figure}[!t]
	\centering
	\includegraphics[width=8.8cm]{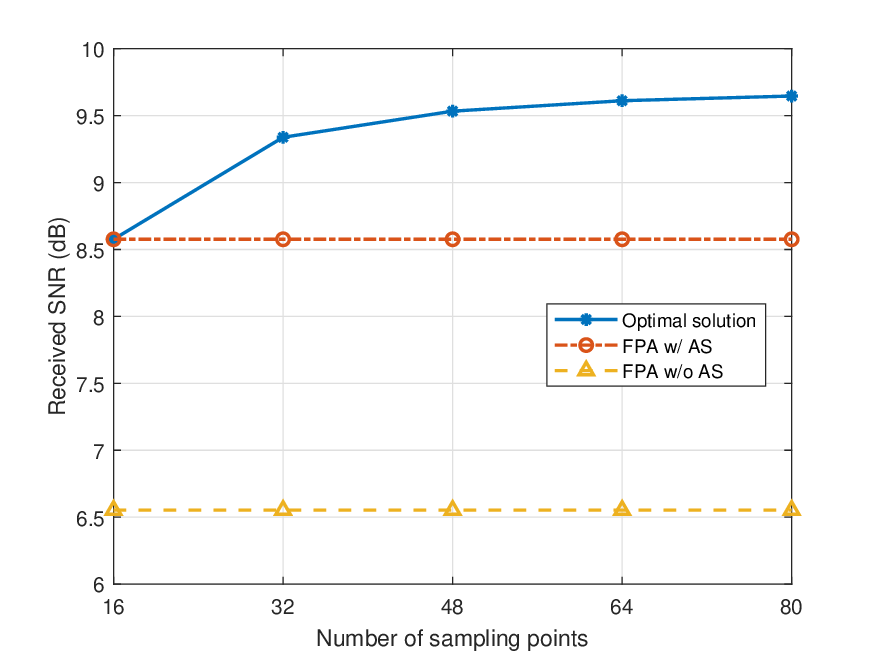}
	\DeclareGraphicsExtensions.
	\caption{Received SNR versus the number of sampling points.}\label{MAvsNumSamp}
	\vspace{-6pt}
\end{figure}
In Fig.\,\ref{MAvsNumSamp}, we show the performance of the optimal graph-based algorithm versus the number of sampling points $M$, and compare it with the performance of the FPAs with and without AS. The carrier frequency is 5 GHz and the number of transmit MAs is $N_{\mathrm{t}} = 8$. The length of the 1D antenna moving region is $A=8\lambda=0.48\;{\text{m}}$ and the number of transmit paths is $L_\mathrm{t}=7$, where the AoDs and path coefficients are the same as those in Fig. \ref{Fig_SISO_gain}. It is observed that the performance of the optimal graph-based algorithms improves with $M$ thanks to the refined sampling resolution and significantly outperforms the two FPA benchmarks. However, when $M \ge 48$ or $\delta_s \le \lambda/6$, further increasing $M$ can hardly improve the received signal-to-noise ratio (SNR). This suggests that a moderate number of sampling points suffice to achieve near-optimal performance as compared to the optimal continuous MA position searching.

\subsection{MA-Aided MIMO and Multiuser Systems}\label{section_MIMOmultiuser}
\subsubsection{MIMO Systems}
\begin{figure}[!t]
	\centering
	\includegraphics[width=89mm]{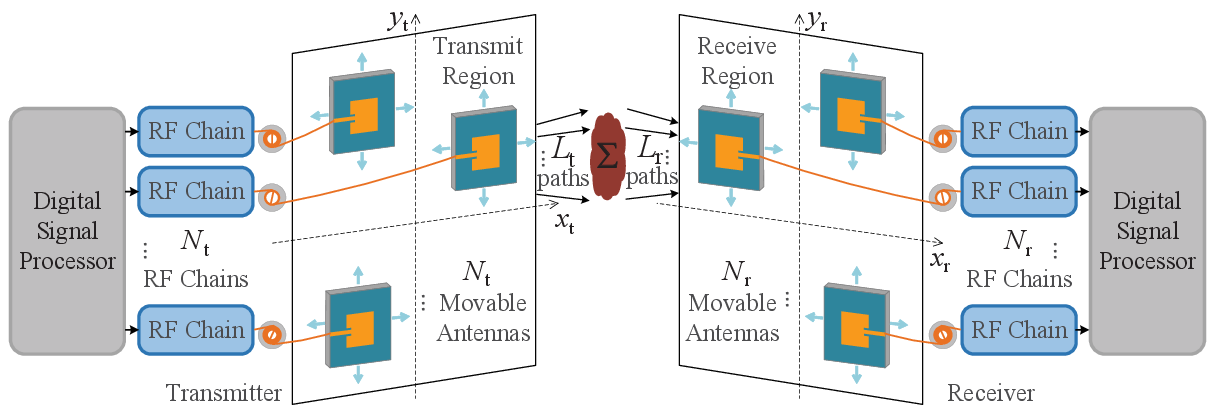}
	\caption{An MA-aided MIMO communication system.}
	\label{Fig_MIMO_model}
	\vspace{-12pt}
\end{figure}

The superiority of MAs in MIMO communication systems lies in the capability of flexibly reshaping MIMO channel matrices for enhanced spatial multiplexing performance. For example, in the low-SNR regime, the MIMO channel matrix can be reconfigured to maximize its largest singular value for improving the single-stream beamforming performance. In contrast, in the high-SNR regime, the MA position optimization can balance the singular values of the channel matrix to cater to the water-filling based optimal power allocation of multiple eigenchannels. Such performance gains can be attained by MA systems via antenna position optimization to reconfigure either instantaneous or statistical MIMO channels between the Tx and Rx. To illustrate this, we consider a MIMO communication system with $N_{\rm{t}}$ Tx-MAs and $N_{\rm{r}}$ Rx-MAs shown in Fig.~\ref{Fig_MIMO_model}, with the channel matrix given by $\mathbf{H}(\tilde{\mathbf{t}},\tilde{\mathbf{r}})=\mathbf{F}(\tilde{\mathbf{r}})^{\rm{H}}\mathbf{\Sigma}\mathbf{G}(\tilde{\mathbf{t}}) \in\mathbb{C}^{N_{\rm{r}} \times N_{\rm{t}}}$ shown in \eqref{eq_channel_MIMO}. 
%Given the PRM as $\mathbf{\Sigma}\in\mathbb{C}^{L_{\rm{r}} \times L_{\rm{t}}}$ and the FRMs for all antennas at the Tx and Rx as $\mathbf{G}(\tilde{\mathbf{t}})=[\mathbf{g}({\mathbf{t}_1}),\mathbf{g}({\mathbf{t}_2}),\ldots,\mathbf{g}({\mathbf{t}_{N_{\rm{t}}}})]\in\mathbb{C}^{L_{\rm{t}} \times N_{\rm{t}}}$ and $\mathbf{F}(\tilde{\mathbf{r}})=[\mathbf{f}({\mathbf{r}_1}),\mathbf{f}({\mathbf{r}_2}),\ldots,\mathbf{f}({\mathbf{r}_{N_{\rm{r}}}})] \in\mathbb{C}^{L_{\rm{r}} \times N_{\rm{r}}}$, respectively, with $\tilde{\mathbf{t}}=[\mathbf{t}_1^{\rm{T}},\mathbf{t}_2^{\rm{T}},\ldots,\mathbf{t}_{N_{\rm{t}}}^{\rm{T}}]^{\rm{T}}$ and $\tilde{\mathbf{r}}=[\mathbf{r}_1^{\rm{T}},\mathbf{r}_2^{\rm{T}},\ldots,\mathbf{r}_{N_{\rm{r}}}^{\rm{T}}]^{\rm{T}}$ representing the APVs for all Tx-MAs and Rx-MAs, the instantaneous channel matrix from the Tx to Rx is given by $\mathbf{H}(\tilde{\mathbf{t}},\tilde{\mathbf{r}})=\mathbf{F}(\tilde{\mathbf{r}})^{\rm{H}}\mathbf{\Sigma}\mathbf{G}(\tilde{\mathbf{t}}) \in\mathbb{C}^{N_{\rm{r}} \times N_{\rm{t}}}$, which is reconfigurable by adjusting $\tilde{\mathbf{t}}$ and $\tilde{\mathbf{r}}$ in general.

Let $\mathbf{Q} \in\mathbb{C}^{N_{\rm{t}} \times N_{\rm{t}}}$ denote the covariance matrix of the transmit signal vector, with $\mathbf{Q} \succeq \mathbf{0}$ and ${\rm{Tr}}(\mathbf{Q})\leq P$, where $P$ is the transmit power. Then, the MIMO channel capacity with instantaneous CSI can be written as
\begin{align}\label{MIMO_capacity}
	C^{\rm{ins}}(\tilde{\mathbf{t}}, \tilde{\mathbf{r}})=\max_{{\substack{\mathbf{Q}: \mathbf{Q} \succeq \mathbf{0}, \\ {\rm {Tr}}(\mathbf{Q}) \leq P}} } \bar{C}(\tilde{\mathbf{t}}, \tilde{\mathbf{r}},\mathbf{Q}),
\end{align}
where $\bar{C}(\tilde{\mathbf{t}}, \tilde{\mathbf{r}},\mathbf{Q}) \triangleq \log_{2} \det\left(\mathbf{I}_{N_{\rm{r}}}+\frac{1}{\sigma^2} \mathbf{H}(\tilde{\mathbf{t}}, \tilde{\mathbf{r}}) \mathbf{Q} \mathbf{H}(\tilde{\mathbf{t}}, \tilde{\mathbf{r}})^{\rm{H}}\right)$ and $\sigma^2$ is the average noise power.
The MIMO capacity $C^{\rm{ins}}(\tilde{\mathbf{t}}, \tilde{\mathbf{r}})$ can be improved by jointly optimizing Tx-MAs' positions $\tilde{\mathbf{t}}$, Rx-MAs' positions $\tilde{\mathbf{r}}$, and transmit covariance matrix $\mathbf{Q}$, which is formulated as the following optimization problem \cite{ma2022MAmimo} 

%In low-SNR scenarios, the optimal strategy is single-stream beamforming, where the entire transmit power is concentrated on the strongest eigenchannel of the MIMO channel matrix. In this case, optimizing  $\tilde{\mathbf{t}}$ and $\tilde{\mathbf{r}}$ can increase the largest singular value of the channel matrix. Conversely, in high-SNR scenarios, transmit power is nearly uniformly distributed across all eigenchannels. Therefore, the optimization of $\tilde{\mathbf{t}}$ and $\tilde{\mathbf{r}}$ should aim to balance the singular values of the channel matrix to maximize the overall MIMO capacity, rather than solely enhancing the power of the strongest eigenchannel.

%Accordingly, the optimization problem for maximizing the MIMO channel capacity by jointly optimizing Tx-MAs' positions $\tilde{\mathbf{t}}$, Rx-MAs' positions $\tilde{\mathbf{r}}$, and transmit covariance matrix $\mathbf{Q}$ can be formulated as
\begin{subequations}
	\begin{align}
		\textrm {(P-MIMO-ins)}~~\max_{\tilde{\mathbf{t}}, \tilde{\mathbf{r}}} \quad & C^{\rm{ins}}(\tilde{\mathbf{t}}, \tilde{\mathbf{r}}) \label{P-MIMO-a}\\
		\text{s.t.} ~\quad & \mathbf{t}_{n} \in \mathcal{C}_{\rm{t}},~1 \leq n \leq N_{\mathrm{t}}, \label{P-MIMO-b}\\
		& \mathbf{r}_{n} \in \mathcal{C}_{\rm{r}},~1 \leq n \leq N_{\mathrm{r}}, \label{P-MIMO-c}\\
		& \|\mathbf{t}_m-\mathbf{t}_n\|_2 \geq d_{\min},~~  m\neq n,\label{P-MIMO-d}\\
		& \|\mathbf{r}_m-\mathbf{r}_n\|_2 \geq d_{\min},~~  m\neq n,\label{P-MIMO-e}
	\end{align}
\end{subequations}
where $\mathcal{C}_{\rm{t}}$ and $\mathcal{C}_{\rm{r}}$ are the Tx-side and Rx-side movable regions, respectively; $d_{\min}$ is the minimum antenna spacing to avoid antenna coupling. Problem (P-MIMO-ins) is challenging to solve due to the non-concave objective function and the non-convex minimum distance constraints \eqref{P-MIMO-d} and \eqref{P-MIMO-e}. To address this challenging problem, local optimization techniques (e.g., gradient ascent and SCA) or global optimization techniques (e.g., particle swarm optimization (PSO)) can be employed to find suboptimal positions for the Tx-MAs and Rx-MAs. Moreover, each Tx-/Rx-MA's position can be alternatively optimized to reduce the computational complexity for a large number of MAs \cite{ma2022MAmimo}. In particular, to handle the non-convex constraint on inter-antenna spacing, an interesting framework was developed in \cite{Jin2024handling} by introducing auxiliary variables.

It is worth noting that the MA-MIMO system considered in (P-MIMO-ins) requires antenna positioning based on instantaneous CSI between the Tx and Rx regions. This is viable for scenarios with slowly varying channels, such as MTC. However, for wireless systems with fast-fading channels, the overhead of antenna movement may be exorbitant. Nonetheless, the MAs' positions can also be appropriately configured to improve the statistical MIMO channels between the Tx and Rx over a long time period, which thus effectively reduces the antenna movement overhead. In such cases, the ergodic MIMO channel capacity can be adopted as the performance metric, which is given by 
\begin{align}\label{statistical_MIMO_capacity}
	C^{\rm{sta}}(\tilde{\mathbf{t}}, \tilde{\mathbf{r}})= 
	\mathbb{E} \left\{ \max_{{\substack{\mathbf{Q}: \mathbf{Q} \succeq \mathbf{0}, \\ {\rm {Tr}}(\mathbf{Q}) \leq P}} } \bar{C}(\tilde{\mathbf{t}}, \tilde{\mathbf{r}},\mathbf{Q}) \right\},
\end{align}
where the expectation in \eqref{statistical_MIMO_capacity} is operated on the random channels, while the transmit covariance matrix can be designed based on the instantaneous channels because of the fast response speed of digital signal processing. Then, the optimization problem for maximizing the ergodic MIMO channel capacity based on statistical CSI can be formulated as \cite{chen2023joint,Ye2024fluidstc,Hu20242024twotimeMA}
\begin{subequations}
	\begin{align}
		\textrm {(P-MIMO-sta)}~~\max_{\tilde{\mathbf{t}}, \tilde{\mathbf{r}}, } ~ & C^{\rm{sta}}(\tilde{\mathbf{t}}, \tilde{\mathbf{r}}) \label{P2-MIMO-a}\\
		\text{s.t.} ~ & \eqref{P-MIMO-b}, \eqref{P-MIMO-c},\eqref{P-MIMO-d},\eqref{P-MIMO-e}.
	\end{align}
\end{subequations}

To characterize the statistical MIMO channel, the Rician-fading channel model, which consists of an instantaneous LoS path and random Rayleigh-fading components, was considered in \cite{chen2023joint}. However, there remains a lack of accurate and efficient statistical MIMO channel models that can capture the fundamental relationship between MAs' positions and the statistical channel information in practical environments with a finite number of scatterers. Additionally, deriving the expectation in \eqref{P2-MIMO-a} analytically is challenging. To tackle this issue, the Monte Carlo method can be employed to approximate $C^{\rm{sta}}(\tilde{\mathbf{t}}, \tilde{\mathbf{r}})$ \cite{shao20246DMA,zhu2024nearfield}. Specifically, given any statistical CSI, e.g., the statistical distributions of AoDs, AoAs, and PRMs, a large number of instantaneous channel realizations can be independently generated offline. The ergodic MIMO capacity is approximately equal to the average value of the instantaneous channel capacities if the number of instantaneous channel realizations is sufficiently large. Then, using existing optimization techniques such as SCA and PSO, the positions of Tx-MAs and Rx-MAs can be optimized to maximize the average channel capacity, where the transmit covariance matrix is used as an auxiliary variable in the offline optimization stage. Besides, some stochastic optimization techniques can also be adopted to approximate the ergodic capacity \cite{chen2023joint,shi20246DMAcellfree,chen2025MAenergy}. In practice, once the MAs have been moved to the optimized positions, the transmit precoding and receive combining matrices can be designed based on instantaneous CSI similar to that in conventional MIMO systems with FPAs.

\begin{figure}[!t]
	\centering
	\includegraphics[width=80mm]{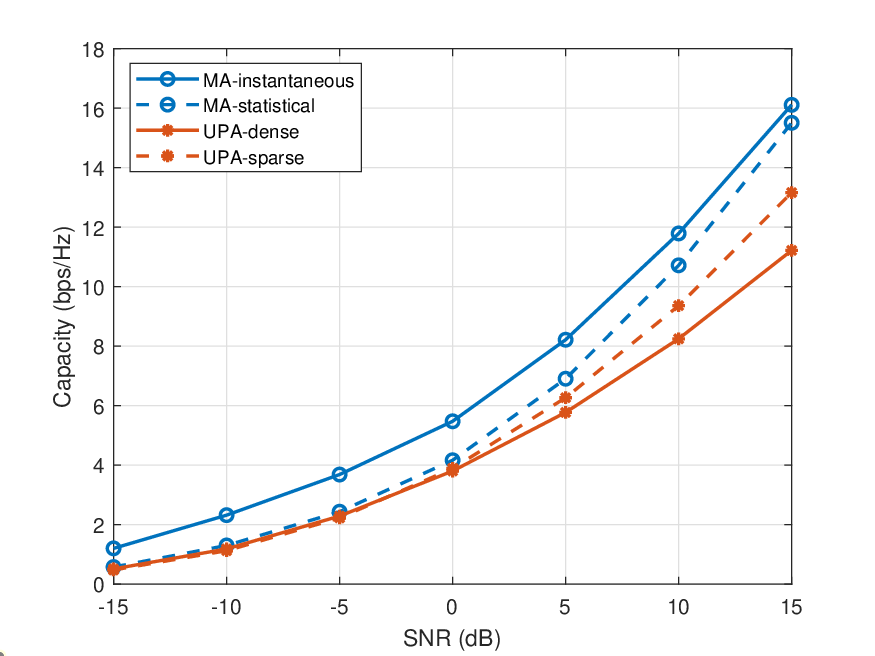}
	\caption{Comparison of MIMO capacity for MA and FPA systems versus SNR.}
	\label{Fig_MIMO_rate_SNR}
	\vspace{-12pt}
\end{figure}

To evaluate the effectiveness of MAs for improving the MIMO communication performance, we show in Fig.~\ref{Fig_MIMO_rate_SNR} the capacity of MA-MIMO and FPA-MIMO systems versus the average SNR $P/\sigma^2$. Both systems employ $N_{\rm{t}}=4$ transmit antennas and $N_{\rm{r}}=4$ receive antennas. $\mathcal{C}_{\rm{t}}$ and $\mathcal{C}_{\rm{r}}$ are set as 2D square regions with size $3\lambda\times 3\lambda$. The geometric channel model is considered, where $\mathbf{\Sigma}$ is assumed to be diagonal with $L_{\rm{t}} = L_{\rm{r}} = 6$. The diagonal elements of $\mathbf{\Sigma}$ are set as $[\mathbf{\Sigma}]_{1,1}\sim \mathcal{CN}(0, \kappa/(\kappa+1))$ and $[\mathbf{\Sigma}]_{\ell,\ell}\sim \mathcal{CN}(0, 1/((\kappa+1)(L_{\rm{t}}-1)))$ for $\ell=2,3,\ldots,L_{\rm{t}}$, where $\kappa=1$ represents the ratio of the average power between LoS and NLoS paths. The distributions of AoDs and AoAs are the same as those in Fig. \ref{Fig_SISO_gain}. The minimum spacing between MAs is set as $d_{\min}=\lambda/2$. For MA-MIMO systems with perfect instantaneous CSI, each of the $N_{\rm{t}}+N_{\rm{r}}+1$ variables $\{\mathbf{t}_n\}_{n=1}^{N_{\rm{t}}} \cup \{\mathbf{r}_m\}_{m=1}^{N_{\rm{r}}}  \cup \mathbf{Q}$ is alternately optimized via SCA to maximize the channel capacity \cite{ma2022MAmimo}. In the case of statistical CSI, we assume that the Tx and Rx undergo random movement within small local regions, which leads to constant AoAs, AoDs, and amplitude of $\mathbf{\Sigma}$ across different channel realizations, while the phase of $\mathbf{\Sigma}$ is uniformly distributed over $[0, 2\pi]$ due to the local movement of the Tx or Rx. To address the phase randomness in $\mathbf{\Sigma}$, the Monte Carlo method is employed to approximate the ergodic capacity for ease of antenna position optimization, with the AO of MAs' positions applied to all $10^3$ random channel realizations. We also evaluate two FPA-based benchmark schemes for comparison: (i) Uniform planar array (UPA)-dense: The Tx-FPAs and Rx-FPAs form two $2\times 2$ UPAs with $\lambda/2$ inter-antenna spacing; (ii) UPA-sparse: The Tx-FPAs and Rx-FPAs form two $2\times 2$ UPAs with $3\lambda$ inter-antenna spacing.

As shown in Fig.~\ref{Fig_MIMO_rate_SNR}, the MA-MIMO systems always achieve higher capacity than their FPA-MIMO counterparts. Moreover, at high SNR, the MA scheme with statistical CSI approaches the capacity performance of the scheme with instantaneous CSI. This is because for high-SNR regimes, the optimal transmit power of multiple data streams is nearly uniformly distributed across all eigenchannels for achieving the best spatial multiplexing performance. The optimal antenna positioning strategy tends to orthogonalize the array responses of multiple channel paths such that they are well separated as distinct eigenchannels. As such, the antenna position optimization is mainly determined by the AoDs and AoAs of channel paths rather than the instantaneous channel phases of their complex coefficients, which leads to a close performance between the MA-MIMO systems with instantaneous and statistical CSI. These results validate the significant potential of MAs to enhance MIMO capacity with a low antenna repositioning overhead. 

\subsubsection{Multiuser Systems}
\begin{figure}[!t]
	\centering
	\includegraphics[width=89mm]{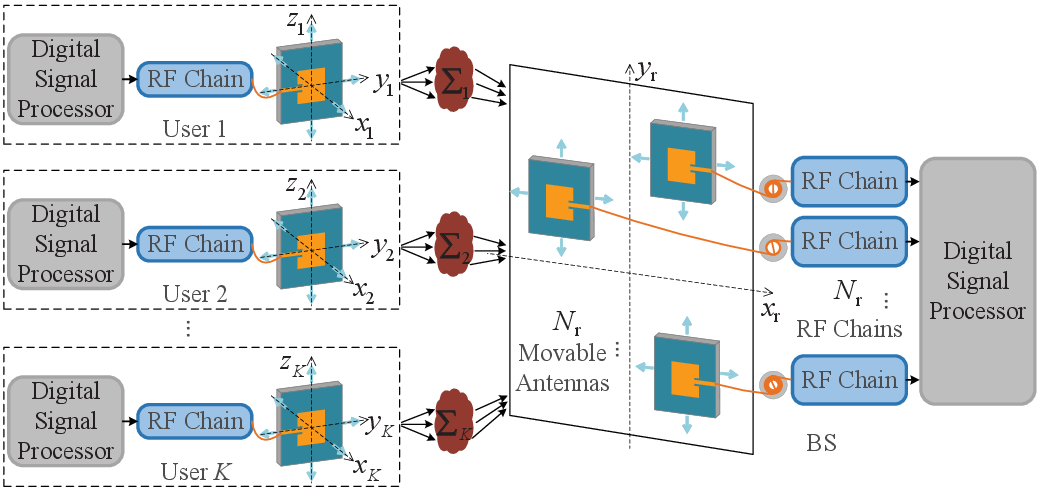}
	\caption{An MA-aided multiuser communication system.}
	\vspace{-12pt}
	\label{Fig_MU_model}
\end{figure}

The enhanced multiplexing performance provided by MAs can also be reaped in multiuser communication systems, as they are mathematically analogous to MIMO systems. However, unlike point-to-point MIMO setups with co-located antennas at both the Tx and Rx, multiuser systems involve spatially separated users, resulting in their channels typically independent. In this context, the channel correlations among multiple users can be more effectively decreased with the aid of MAs, which helps yield even higher performance gains over FPA systems. To characterize such fundamental performance gains, we consider the uplink from $K$ single-MA users to a BS equipped with $N_\mathrm{r}$ MAs shown in Fig.~\ref{Fig_MU_model}. Denote the PRM between all the Tx and Rx channel paths from the $k$-th user to the BS as $\mathbf{\Sigma}_k\in\mathbb{C}^{L_{\rm{r}}^k \times L_{\rm{t}}^k}$, and the corresponding Tx-FRV and Rx-FRM as $\mathbf{g}({\mathbf{t}_k})\in\mathbb{C}^{L_{\rm{r}}^k \times 1}$ and $\mathbf{F}_k(\tilde{\mathbf{r}})=[\mathbf{f}({\mathbf{r}_1}),\mathbf{f}({\mathbf{r}_2}),\ldots,\mathbf{f}({\mathbf{r}_{N_{\rm{r}}}})] \in\mathbb{C}^{L_{\rm{r}}^k \times N_{\rm{r}}}$, respectively, with $\tilde{\mathbf{t}}=[\mathbf{t}_1^{\rm{T}},\mathbf{t}_2^{\rm{T}},\ldots,\mathbf{t}_{K}^{\rm{T}}]^{\rm{T}}$ and $\tilde{\mathbf{r}}=[\mathbf{r}_1^{\rm{T}},\mathbf{r}_2^{\rm{T}},\ldots,\mathbf{r}_{N_{\rm{r}}}^{\rm{T}}]^{\rm{T}}$ representing the APVs for all user-side MAs and BS-side MAs. The instantaneous channel vector from the $k$-th user to the BS is given by $\mathbf{h}_k(\mathbf{t}_k,\tilde{\mathbf{r}})=\mathbf{F}_k(\tilde{\mathbf{r}})^{\rm{H}}\mathbf{\Sigma}_k\mathbf{g}({\mathbf{t}_k}) \in\mathbb{C}^{N_\mathrm{r} \times 1}$, which can be reconfigured by adjusting $\mathbf{t}_k$ and $\tilde{\mathbf{r}}$.

Let $\mathbf{W} = [\mathbf{w}_1,\mathbf{w}_2,\ldots,\mathbf{w}_K] \in\mathbb{C}^{N_\mathrm{r} \times K}$ denote the receive combining matrix of the BS, where $\mathbf{w}_k \in\mathbb{C}^{N_\mathrm{r} \times 1}$ is the combining vector for the $k$-th user. Define $\mathbf{p} = [p_1,p_2,\ldots,p_K]^{\rm{T}} \in\mathbb{R}^{K \times 1}$ with $p_k$ being the transmit power of user $k$. Then, the instantaneous achievable rate for user $k$ with perfect CSI can be written as
\begin{align}\label{MU_rate}
	R_k^{\rm{ins}}(\tilde{\mathbf{t}}, \tilde{\mathbf{r}},\mathbf{W},\mathbf{p})=\log_{2}\left( 1+ \gamma_k(\tilde{\mathbf{t}},\tilde{\mathbf{r}},\mathbf{W},\mathbf{p}) \right),
\end{align}
where $\gamma_k(\tilde{\mathbf{t}},\tilde{\mathbf{r}},\mathbf{W},\mathbf{p}) \triangleq \frac{|\mathbf{w}_k^{\rm{H}} \mathbf{h}_k(\mathbf{t}_k,\tilde{\mathbf{r}})|^2 p_k}{\sum_{q=1,q\neq k}^{K} |\mathbf{w}_k^{\rm{H}} \mathbf{h}_k(\mathbf{t}_q,\tilde{\mathbf{r}})|^2 + \|\mathbf{w}_k\|_2^2\sigma^2 }$ denotes the received SINR of the signal from user $k$, and $\sigma^2$ is the average noise power.

The achievable rate performance can be improved by jointly optimizing user-side MAs' positions $\tilde{\mathbf{t}}$, BS-side MAs' positions $\tilde{\mathbf{r}}$, receive combining matrix $\mathbf{W}$, and transmit power allocation vector $\mathbf{p}$. On one hand, channel gains can be enhanced by moving each MA to the position where the phases of all propagation paths are aligned, leading to a constructive superposition of the signals across all paths. This increases the received signal power, $|\mathbf{w}_k^{\rm{H}} \mathbf{h}(\mathbf{t}_k,\tilde{\mathbf{r}})|^2$. On the other hand, optimizing the positions of MAs can effectively reduce the correlation between channel vectors of different users, thereby contributing to suppressing the interference power, $|\mathbf{w}_k^{\rm{H}} \mathbf{h}(\mathbf{t}_q,\tilde{\mathbf{r}})|^2$, $q\neq k$, among multiple users. As such, the spatial multiplexing performance of multiple users can be significantly improved.

For the considered MA-aided multiuser communication system, there exists a basic trade-off between minimizing the transmit power and maximizing the achievable rate for multiple users. Let $f(\tilde{\mathbf{t}}, \tilde{\mathbf{r}},\mathbf{W},\mathbf{p})$ denote the achievable rate utility for multiple users, which can take the form of: (i) minimum rate, $f_1(\tilde{\mathbf{t}}, \tilde{\mathbf{r}},\mathbf{W},\mathbf{p}) = \min_{k} R_k^{\rm{ins}}(\tilde{\mathbf{t}}, \tilde{\mathbf{r}},\mathbf{W},\mathbf{p})$, or (ii) sum-rate, $f_2(\tilde{\mathbf{t}}, \tilde{\mathbf{r}},\mathbf{W},\mathbf{p}) = \sum_{k=1}^{K} R_k^{\rm{ins}}(\tilde{\mathbf{t}}, \tilde{\mathbf{r}},\mathbf{W},\mathbf{p})$. Similarly, let $g(\mathbf{p})$ denote the transmit power metric, which can take the form of: (i) maximum power, $g_1(\mathbf{p}) = \max_{k} p_k$, or (ii) sum-power, $g_2(\mathbf{p}) = \sum_{k=1}^{K} p_k$.

Accordingly, the rate-centric optimization problem for maximizing the achievable rates subject to the given constraint on the maximum transmit power $P$ by jointly optimizing user-side MAs' positions $\tilde{\mathbf{t}}$, BS-side MAs' positions $\tilde{\mathbf{r}}$, combining matrix of the BS $\mathbf{W}$, and power allocation vector $\mathbf{p}$ can be formulated as \cite{xiao2023multiuser,Feng2024MAweighted,Hu2024twouserMA,Xu2024FASmultiple,gao2024jointMA,cheng2024MAmulticast}
\begin{subequations}
	\begin{align}
		\textrm {(P-MU-r-ins)}~~\max_{\tilde{\mathbf{t}}, \tilde{\mathbf{r}},\mathbf{W},\mathbf{p}} \quad & f(\tilde{\mathbf{t}}, \tilde{\mathbf{r}},\mathbf{W},\mathbf{p}) \label{P-MU-r-a}\\
		\text{s.t.} ~~\quad & g(\mathbf{p}) \leq P, \label{P-MU-r-b}\\
		& \mathbf{t}_k \in \mathcal{C}_{\rm{t}}^k,~ 1 \leq k \leq K, \label{P-MU-r-c}\\
		& \mathbf{r}_{n} \in \mathcal{C}_{\rm{r}},~ 1 \leq n \leq N_{\mathrm{r}}, \label{P-MU-r-d}\\
		& \|\mathbf{r}_m-\mathbf{r}_n\|_2 \geq d_{\min},~  m\neq n,\label{P-MU-r-e}
	\end{align}
\end{subequations}
where $\mathcal{C}_{\rm{t}}^k$ and $\mathcal{C}_{\rm{r}}$ are the user-side and BS-side movable regions, respectively. Similarly, the dual power-centric optimization problem for minimizing the transmit power subject to the given constraint on the minimum achievable rate $\eta$ can be formulated as \cite{zhu2023MAmultiuser,hu2023power,Yang2024movable,wu2023movable}
\begin{subequations}
	\begin{align}
		\textrm {(P-MU-p-ins)}~~\min_{\tilde{\mathbf{t}}, \tilde{\mathbf{r}},\mathbf{W},\mathbf{p}} \quad & g(\mathbf{p}) \label{P-MU-p-a}\\
		\text{s.t.} ~~\quad & f(\tilde{\mathbf{t}}, \tilde{\mathbf{r}},\mathbf{W},\mathbf{p}) \geq \eta, \label{P-MU-p-b}\\
		& \eqref{P-MU-r-c}, \eqref{P-MU-r-d}, \eqref{P-MU-r-e}.
	\end{align}
\end{subequations}

To address the non-convex problems (P-MU-r-ins) and (P-MU-p-ins), the combining matrix of the BS $\mathbf{W}$ and the transmit power of users $\mathbf{p}$ can be expressed as a function of the MAs’ positions (e.g., based on minimum mean squared error (MMSE) or ZF combiner), which are then optimized by using various candidate approaches, such as gradient descent/ascent \cite{zhu2023MAmultiuser,hu2023power}, PSO \cite{xiao2023multiuser}, orthogonal matching pursuit (OMP) \cite{Yang2024movable}, and branch and bound (BnB) \cite{wu2023movable}.

If the users and their surrounding environments have slow mobility, the wireless channels between multiple users and BS exhibit slow variation. The MAs can be dynamically repositioned to track the instantaneous changes in wireless channels. For the cases with high-mobility users and fast-fading channels, an alternative way is to perform antenna movement to adapt to the statistical CSI of multiple users over a long time period. This strategy can effectively reduce the overhead of antenna movement, while the combining matrix $\mathbf{W}$ and power allocation vector $\mathbf{p}$ can be designed based on the instantaneous channels, similar to that in conventional FPA-aided systems. Then, the rate-centric and power-centric optimization problems for multiuser communications based on statistical CSI can be formulated as \cite{Hu20242024twotimeMA, zheng2024twotimeMA}
\begin{subequations}
	\begin{align}
		\textrm {(P-MU-r-sta)}~~\max_{\tilde{\mathbf{t}}, \tilde{\mathbf{r}}} \quad & \mathbb{E} \left\{ \max_{\mathbf{W},\mathbf{p}} f(\tilde{\mathbf{t}}, \tilde{\mathbf{r}},\mathbf{W},\mathbf{p}) \right\} \label{P-MU-r-sta-a}\\
		\text{s.t.} \quad & \eqref{P-MU-r-b}, \eqref{P-MU-r-c}, \eqref{P-MU-r-d}, \eqref{P-MU-r-e},
	\end{align}
\end{subequations}
and
\begin{subequations}
	\begin{align}
		\textrm {(P-MU-p-sta)}~~\min_{\tilde{\mathbf{t}}, \tilde{\mathbf{r}}} \quad & \mathbb{E} \left\{\min_{\mathbf{W},\mathbf{p}} 
		g(\mathbf{p})\right\} \label{P-MU-p-sta-a}\\
		\text{s.t.} \quad & \eqref{P-MU-p-b}, \eqref{P-MU-r-c}, \eqref{P-MU-r-d}, \eqref{P-MU-r-e},
	\end{align}
\end{subequations}
respectively, where the expectations are operated on random channels. For any given statistical CSI, the expectations in \eqref{P-MU-r-sta-a} and \eqref{P-MU-p-sta-a} can be approximated using stochastic optimization techniques, which is similar to the method used for solving problem (P-MIMO-sta) \cite{shao20246DMA,chen2023joint,shi20246DMAcellfree}.

\begin{figure}[!t]
	\centering
	\includegraphics[width=80mm]{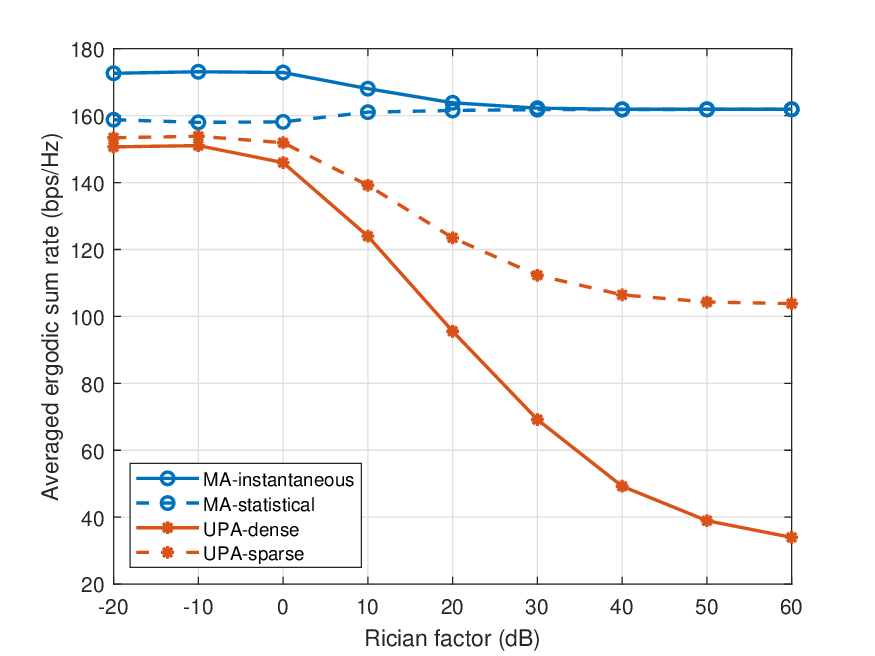}
	\caption{Achievable sum rate of multiple users versus Rician factor.}
	\vspace{-12pt}
	\label{Fig_MU_rate}
\end{figure}

To evaluate the performance gain of MAs in improving multiuser communication performance, we show in Fig.~\ref{Fig_MU_rate} the achievable sum rate of MA systems with antenna position optimization based on both instantaneous and statistical channels. 
The height of the BS’s MA array is set as $20$ m, with the number of MAs at the BS set as $N_\mathrm{r}=16$ and the minimum spacing between MAs set as $d_{\min}=\lambda/2$. The carrier frequency is $5$ GHz and $\mathcal{C}_{\rm{r}}$ is set as a square region of size $8\lambda\times 8\lambda$. The total number of users is set as $K = 12$. They are each equipped with a single FPA and randomly distributed on the ground with a distance $d_k$ from the BS uniformly distributed within the range $[20, 100]$ m. The geometric channel model is considered, where $\mathbf{\Sigma}_k$ is assumed to be diagonal with $L_{\rm{r}}^k = L_{\rm{t}}^k = 6$, $1 \leq k \leq K$. The diagonal elements of $\mathbf{\Sigma}_k$ are set as $[\mathbf{\Sigma}_k]_{1,1}\sim \mathcal{CN}\left(0, \frac{g_{k}\kappa}{\kappa+1}\right)$ and $[\mathbf{\Sigma}_k]_{\ell,\ell}\sim \mathcal{CN}\left(0, \frac{g_{k}}{(\kappa+1)(L_{\rm{t}}^k-1)}\right)$ for $2 \leq \ell \leq L_{\rm{t}}^k$, where $\kappa$ denotes the Rician factor and $g_{k}=\frac{\lambda^{2}\kappa}{16\pi^{2} d_k^{2.5}}$ denotes the total channel power gain for user $k$. The LoS path's AoDs and AoAs, for both elevation and azimuth, are determined by the physical locations of the BS and users. The AoDs and AoAs for NLoS paths are generated similarly to those in Fig. \ref{Fig_SISO_gain}. The BS's maximum transmit power is set as $P = 30$ dBm, and the noise power is set as $\sigma^2 = -90$ dBm. For statistical channels, it is assumed that the users exhibit random movement within a small local region, leading to constant AoAs, AoDs, and amplitude of $\mathbf{\Sigma}_k$ across different channel realizations, while the phase of each element in $\mathbf{\Sigma}_k$ is uniformly distributed over $[0, 2\pi]$ across different channel realizations. To address the phase randomness in $\mathbf{\Sigma}_k$, the Monte Carlo method is employed to approximate the ergodic sum rate in a similar way to that in Fig. \ref{Fig_MIMO_rate_SNR}. For comparison, we also evaluate two FPA-based benchmark schemes: (i) UPA-dense: The BS-side antennas form a $4\times 4$ UPA with $\lambda/2$ inter-antenna spacing; (ii) UPA-sparse: The BS-side antennas form a $4\times 4$ UPA with $2\lambda$ inter-antenna spacing.

As shown in Fig.~\ref{Fig_MU_rate}, the MA schemes always outperform other benchmark schemes with FPAs. Especially for larger Rician factors, the MA scheme with statistical CSI approaches the sum-rate performance of the MA scheme with instantaneous CSI. This is because the optimal MA positioning is mainly determined by the dominant LoS channels of multiple users. In practice, since the statistical CSI-based MA scheme does not require frequent antenna repositioning, it offers an effective balance between communication performance and system overhead. Additionally, as the Rician factor increases, the sum-rate advantage of MA schemes over FPA schemes becomes more pronounced, highlighting the significant potential of MA-enhanced multiuser communications in future wireless systems at high-frequency bands in which LoS channels become more likely to be dominant.

\subsection{MA-Aided Sensing and ISAC}
\subsubsection{Wireless Sensing}
Wireless sensing, which involves detecting, estimating, and extracting physical information from environmental targets, is expected to become a key service in future 6G wireless networks. Common applications of wireless sensing include radar systems for estimating target information such as distance, speed, and orientation, wireless device localization, and RF imaging for creating high-resolution images of objects in the environment \cite{Shao2022target}. By enabling antenna movement in extended regions, the angular resolution of wireless sensing can be significantly improved, without the need for massive antennas and RF chains \cite{ma2024MAsensing, wang2025MAnearsensing}. For ease of exposition, we mainly focus on MA-aided radar sensing in the sequel of this article, while similar discussions can be inferred for other wireless sensing applications due to their common design challenges and performance limitations.

\begin{figure}[!t]
	\centering
	\includegraphics[width=70mm]{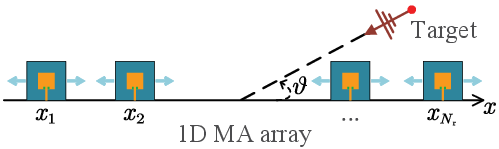}
	\caption{The 1D MA array for target angle estimation.}
	\vspace{-6pt}
	\label{Fig_sensing_1D_model}
	\vspace{-6pt}
\end{figure}

As shown in Fig.~\ref{Fig_sensing_1D_model}, we consider a radar sensing system with a 1D MA array composed of $N_\mathrm{r}$ MAs to estimate a target's AoA. The position of the $n$-th MA, $1 \leq n \leq N_\mathrm{r}$, is represented by $x_n\in[0,A]$, where $A$ denotes the length of the 1D line segment. The APV of the $N_\mathrm{r}$ MAs is denoted by $\mathbf{x} \triangleq [x_1,x_2,\ldots,x_{N_\mathrm{r}}]^{\rm{T}} \in \mathbb{R}^{N_\mathrm{r} \times 1}$, with $0 \leq x_1 < x_2 < \ldots < x_{N_\mathrm{r}} \leq A$ without loss of generality. We assume an LoS target-Rx channel, which remains static across $T_{\mathrm{s}}$ snapshots. The physical steering angle of the LoS path is denoted by $\vartheta \in [0, \pi]$, and for simplicity, the spatial AoA is defined as $u \triangleq \cos \vartheta$. The steering vector of the MA array can then be expressed as a function of the APV $\mathbf{x}$ and spatial AoA $u$, i.e., $\boldsymbol{\alpha}(\mathbf{x},u) \triangleq [ \mathrm{e}^{\mathrm{j}\frac{2\pi}{\lambda}x_1 u}, \mathrm{e}^{\mathrm{j}\frac{2\pi}{\lambda}x_2 u}, \ldots, \mathrm{e}^{\mathrm{j}\frac{2\pi}{\lambda}x_{N_\mathrm{r}} u} ]^{\rm{T}} \in{\mathbb{C}^{N_\mathrm{r}\times 1} }$, where $\lambda$ denotes the wavelength. Additionally, let $\beta$ denote the complex path coefficient from the target to the origin of the line segment. Then, the received signals over $T_{\mathrm{s}}$ snapshots are given by
\begin{equation}\label{Y}
	\mathbf{Y}=\beta\boldsymbol{\alpha}(\mathbf{x},u)\mathbf{s}^{\rm{T}} + \mathbf{N},
\end{equation}
where $\mathbf{s}\in \mathbb{C}^{T_{\mathrm{s}} \times 1}$ denotes the vector of signals reflected by the target with average power $\mathbb{E}\{|[\mathbf{s}]_t|^2\} = P$, $1 \leq t \leq T_{\mathrm{s}}$, and $\mathbf{N}\in \mathbb{C}^{N_\mathrm{r} \times T_{\mathrm{s}}}$ represents the noise matrix at the sensing Rx, where each element given by $[\mathbf{N}]_{n,t}\sim \mathcal{CN}(0,\sigma^2)$, $1 \leq n \leq N_\mathrm{r}$, with $\sigma^2$ denoting the average noise power. 

The radar sensing performance metric involves maximizing the radar sensing SNR  \cite{kuang2024movableISAC, khalili2024advanced, WuHS_MA_RIS_ISAC, hao2024fluid, zhang2024efficient} or SINR \cite{xiu2024movable, lyu2024flexibleISAC, wang2024multiuser, peng2024jointISAC} with given target AoA(s), and minimizing the lower-bound on the AoA estimation mean square error (MSE), i.e., the CRB, either for the given AoA \cite{qin2024cramer} or without any prior AoA information \cite{ma2024MAsensing}. For any given APV $\mathbf{x}$, we can adopt classical methods, such as multiple signal classification (MUSIC), to estimate the target's spatial AoA $u$ \cite{ma2024MAsensing}. Let $\hat{u}$ denote the estimation of $u$. Then, the AoA estimation MSE can be expressed as ${\rm{MSE}}(u)\triangleq\mathbb{E}\{|u-\hat{u}|^2\}$, and the CRB of ${\rm{MSE}}(u)$ is given by \cite{ma2024MAsensing}
\begin{align}\label{CRB1D}
	{\rm{MSE}}(u)\geq{\rm{CRB}}_u(\mathbf{x}) =  \frac{\sigma^2\lambda^2}{8\pi^2T_{\mathrm{s}}PN_\mathrm{r}|\beta|^2}\frac{1}{{\rm{var}}(\mathbf{x})},
\end{align}
where the variance function is defined as ${\rm{var}}(\mathbf{x})\triangleq\frac{1}{N_\mathrm{r}}\sum_{n=1}^{N_\mathrm{r}}x_n^2 - \mu(\mathbf{x})^2$ with $\mu(\mathbf{x})=\frac{1}{N_\mathrm{r}}\sum_{n=1}^{N_\mathrm{r}}x_n$ being the mean of vector $\mathbf{x}$. The result in \eqref{CRB1D} shows that ${\rm{CRB}}_u(\mathbf{x})$ decreases with the increase of ${\rm{var}}(\mathbf{x})$. Therefore, to minimize the CRB, the optimization problem for $\mathbf{x}$ can be formulated as
\begin{subequations}
	\begin{align}
		\textrm {(P-sen-1D)}~~\max_{\mathbf{x}} \quad & {\rm{var}}(\mathbf{x})   \label{P1a}\\
		\text{s.t.} \quad & x_1\geq0,~ x_{N_\mathrm{r}}\leq A, \label{P1b}\\
		& x_n-x_{n-1} \geq d_{\min},~ 2 \leq n \leq N_\mathrm{r}, \label{P1c}
	\end{align}
\end{subequations}
where $d_{\min}$ denotes the minimum antenna spacing to avoid antenna coupling. Although problem (P-sen-1D) is non-convex, an optimal solution for problem (P-sen-1D) is given by \cite{ma2024MAsensing}
\begin{align}\label{1Doptimal}
	&x^\star_n=\left\{
	\begin{array}{ll}
		(n-1)d_{\min},    & 1 \leq n \leq \lfloor N_\mathrm{r}/2 \rfloor;\\
		A-(N_\mathrm{r}-n)d_{\min},  & \lfloor N_\mathrm{r}/2 \rfloor+1 \leq n \leq N_\mathrm{r}.
	\end{array} \right.
\end{align}
The result in \eqref{1Doptimal} indicates that to achieve the minimum CRB for AoA estimation MSE along the 1D line segment, the MAs should be divided into two groups. Specifically, one half of the MAs should be placed at the leftmost end of the segment, while the other half should be placed at the rightmost end, both with the minimum inter-antenna spacing, $d_{\min}$. Such an APV can effectively maximize the array aperture and thus achieve the best sensing performance.

\begin{figure}[!t]
	\centering
	\includegraphics[width=75mm]{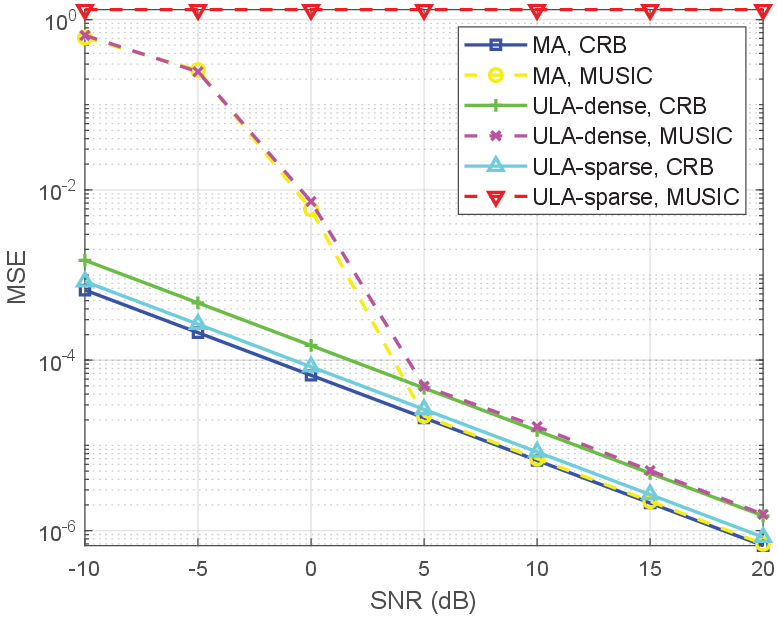}
	\vspace{-6pt}
	\caption{MSE versus SNR in the case of 1D array \cite{ma2024MAsensing}.}
	\vspace{-6pt}
	\label{Fig_1D_SNR}
\end{figure}

To evaluate the performance gain of MAs in improving the target's AoA estimation, we show in Fig.~\ref{Fig_1D_SNR} the AoA estimation MSE (including both the actual MSE via MUSIC algorithm and its CRB) versus SNR. We set $N_\mathrm{r}=16$, $A=10\lambda$, $d_{\min}=\lambda/2$, $\vartheta=45^{\circ}$, and thus $u=\cos \vartheta=0.71$. The average received SNR is defined as $P|\beta|^2/\sigma^2$. The number of snapshots is set as $T_{\mathrm{s}}=1$. We also evaluate two FPA-based benchmark schemes for comparison: (i) Uniform linear array (ULA)-dense: a ULA with $\lambda/2$ inter-antenna spacing; (ii) ULA-sparse: a ULA with $A/(N_\mathrm{r}-1)$ inter-antenna spacing. It is observed that the AoA estimation MSE curves obtained via the MUSIC algorithm can closely approach the CRB for both the optimal antennas' positions in \eqref{1Doptimal} and ULA-dense in the high-SNR regime. Moreover, the proposed optimal antennas' positions in \eqref{1Doptimal} achieve a significantly lower MSE compared to the benchmark schemes. This is because the optimized MA array achieves a significantly larger aperture in the spatial domain compared to the dense ULA, thereby enhancing angular resolution. In comparison, although the sparse ULA also offers a large aperture, it suffers from obstinate grating lobes in the angular beam patterns, similar to that shown in Fig. \ref{beampattern}(b)\footnote{Grating lobes may benefit wireless communication systems by enabling multiple beams to cover users in different directions. However, for radar systems, grating lobes are highly detrimental as they introduce angular ambiguity and degrade sensing accuracy. This stark contrast highlights a fundamental conflict in system requirements: the optimal MA positions for communication and sensing can lead to contradictory outcomes. Consequently, achieving an effective balance requires deliberate trade-offs in MA position optimization to accommodate both functionalities in ISAC systems.}. These grating lobes introduce ambiguity in the angular domain, resulting in worse MSE performance for the ULA-sparse scheme. In summary, the MA-aided sensing systems can effectively enlarge the array aperture while mitigating grating lobes, which ensures their superior performance in improving the accuracy of AoA estimation.

\begin{figure}[!t]
	\centering
	\includegraphics[width=65mm]{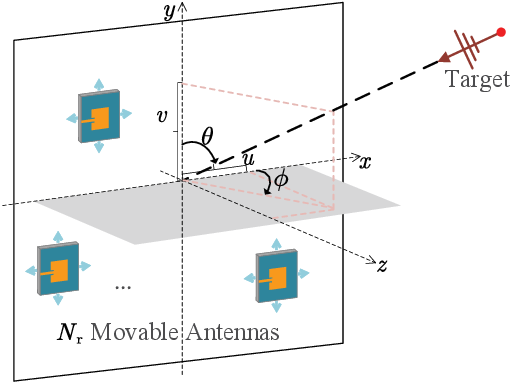}
	\caption{The 2D MA array for target angle estimation.}
	\vspace{-12pt}
	\label{Fig_sensing_2D_model}
\end{figure}

Next, we consider a more general radar sensing system where $N_\mathrm{r}$ MAs are deployed on a 2D plane $\mathcal{C}_{\rm{r}}$ to estimate the target's AoAs along both the $x$ and $y$ axes shown in Fig.~\ref{Fig_sensing_2D_model}. The 2D coordinate of the $n$-th MA, $1 \leq n \leq N_\mathrm{r}$, is denoted as $\mathbf{r}_n = [x_n, y_n]^{\rm{T}} \in \mathcal{C}_{\rm{r}}$. Denote the collection of $N_\mathrm{r}$ MAs' APV by $\tilde{\mathbf{r}}=\left[\mathbf{r}_1^{\rm{T}}, \mathbf{r}_2^{\rm{T}}, \ldots, \mathbf{r}_{N_\mathrm{r}}^{\rm{T}}\right] \in{\mathbb{R}^{2{N_\mathrm{r}} \times 1}}$. The physical elevation and azimuth AoAs of the target-Rx LoS path are denoted by $\theta \in [0, \pi]$ and $\phi \in [0, \pi]$, respectively. For convenience, the spatial AoAs are defined as $u \triangleq \sin \theta \cos \phi$ and $v \triangleq \cos \theta$. The steering vector of the 2D MA array can then be expressed as a function of the APV and the two spatial AoAs, i.e., $\boldsymbol{\alpha}(\tilde{\mathbf{r}},u,v) \triangleq [ \mathrm{e}^{\mathrm{j}\frac{2\pi}{\lambda}(x_1 u + y_1 v)}, \mathrm{e}^{\mathrm{j}\frac{2\pi}{\lambda}(x_2 u + y_2 v)}, \ldots, \mathrm{e}^{\mathrm{j}\frac{2\pi}{\lambda}(x_{N_\mathrm{r}} u + y_{N_\mathrm{r}} v)} ]^{\rm{T}} \in{\mathbb{C}^{N_\mathrm{r}\times 1}}$. Similar to the AoA estimation procedure in the 1D MA array scenario, for any given APV, parameters $u$ and $v$ can be jointly estimated using classical approaches, such as the MUSIC algorithm. Accordingly, the CRB of the AoA estimation MSE is given by \cite{ma2024MAsensing}
\begin{align}\label{CRBr}
	{\rm{MSE}}(u)\geq{\rm{CRB}}_u(\tilde{\mathbf{r}}) &=  \frac{\sigma^2\lambda^2}{8\pi^2T_{\mathrm{s}}PN_\mathrm{r}|\beta|^2}\frac{1}{{\rm{var}}(\mathbf{x})-\frac{{\rm{cov}}(\mathbf{x},\mathbf{y})^2}{{\rm{var}}(\mathbf{y})}}, \notag\\
	{\rm{MSE}}(v)\geq{\rm{CRB}}_v(\tilde{\mathbf{r}}) &=  \frac{\sigma^2\lambda^2}{8\pi^2T_{\mathrm{s}}PN_\mathrm{r}|\beta|^2}\frac{1}{{\rm{var}}(\mathbf{y})-\frac{{\rm{cov}}(\mathbf{x},\mathbf{y})^2}{{\rm{var}}(\mathbf{x})}},
\end{align}
where $\mathbf{x}\triangleq[x_1,x_2,\ldots,x_{N_\mathrm{r}}]^{\rm{T}}\in \mathbb{R}^{N_\mathrm{r} \times 1}$ and $\mathbf{y}\triangleq[y_1,y_2,\ldots,y_{N_\mathrm{r}}]^{\rm{T}}\in \mathbb{R}^{N_\mathrm{r} \times 1}$. The covariance function is defined as ${\rm{cov}}(\mathbf{x},\mathbf{y})\triangleq \frac{1}{N_\mathrm{r}}\sum_{n=1}^{N_\mathrm{r}}x_n y_n - \mu(\mathbf{x})\mu(\mathbf{y})$. Let $c(\tilde{\mathbf{r}})$ denote the CRB metric, which can take the form of: (i) maximum CRB, $c_1(\tilde{\mathbf{r}}) = \max \{{\rm{CRB}}_u(\tilde{\mathbf{r}}), {\rm{CRB}}_v(\tilde{\mathbf{r}})\}$, or (ii) sum-CRB, $c_2(\tilde{\mathbf{r}}) = {\rm{CRB}}_u(\tilde{\mathbf{r}}) + {\rm{CRB}}_v(\tilde{\mathbf{r}})$. Therefore, the optimization problem of minimizing the CRB metric by optimizing $\tilde{\mathbf{r}}$ can be formulated as
\begin{subequations}
	\begin{align}
		\textrm {(P-sen-2D)}~~\min_{\tilde{\mathbf{r}}} \quad & c(\tilde{\mathbf{r}}) \label{P2a}\\
		\text{s.t.} \quad &  \mathbf{r}_{n} \in \mathcal{C}_{\rm{r}},~1 \leq n \leq N_{\mathrm{r}}, \label{P2d}\\
		& \|\mathbf{r}_m-\mathbf{r}_n\|_2 \geq d_{\min},~~  m\neq n.\label{P2e}
	\end{align}
\end{subequations}
Denoting $A^\textrm{cir}$ as the radius of the minimum circumcircle of $\mathcal{C}_{\rm{r}}$, it can be proven that a lower bound on $c_1(\tilde{\mathbf{r}})$ is given by \cite{ma2024MAsensing}
\begin{align}\label{CRBlowerbound}
	\min_{\tilde{\mathbf{r}}}~ c_1(\tilde{\mathbf{r}}) \geq \frac{\sigma^2\lambda^2}{4\pi^2T_{\mathrm{s}}PN_\mathrm{r}|\beta|^2(A^\textrm{cir})^2}.
\end{align}
Despite having this lower bound, it is still challenging to optimally solve the non-convex problem (P-sen-2D). To address this issue, the AO between $\mathbf{x}$ and $\mathbf{y}$ can be employed to obtain locally optimal solutions for problem (P-sen-2D), where the SCA technique can be applied for each subproblem of optimizing one of $\{\mathbf{x}, \mathbf{y}\}$ with the other fixed \cite{ma2024MAsensing}.

\begin{figure}[!t]
	\centering
	\includegraphics[width=75mm]{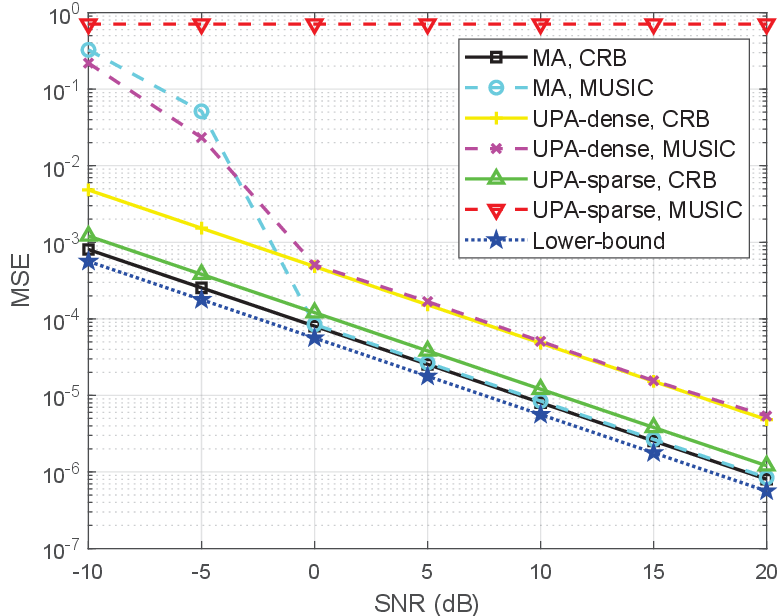}
	\vspace{-6pt}
	\caption{${\rm{MSE}}(u)$ versus SNR in the case of 2D array \cite{ma2024MAsensing}.}
	\vspace{-12pt}
	\label{Fig_2D_SNR}
\end{figure}

To validate the effectiveness of the 2D MA array for AoA estimation, we consider the maximum CRB metric and show in Fig.~\ref{Fig_2D_SNR} the AoA estimation MSE (including both the actual MSE via MUSIC algorithm and its CRB) versus SNR. Since ${\rm{MSE}}(u)$ and ${\rm{MSE}}(v)$ have similar performance, we only show ${\rm{MSE}}(u)$ for convenience. We consider the 2D square region $\mathcal{C}_{\rm{r}}$ with size $A\times A$. We set $N_\mathrm{r}=36$, $A=5\lambda$, $\theta=45^{\circ}$, $\phi=60^{\circ}$, and thus $u=\sin \theta \cos \phi=0.35$ and $v=\cos \theta=0.71$. We also evaluate two FPA-based benchmark schemes for comparison: (i) UPA-dense: a $6\times 6$ UPA with $\lambda/2$ inter-antenna spacing; (ii) UPA-sparse: a $6\times 6$ UPA with $A/(\sqrt{N_\mathrm{r}}-1)$ inter-antenna spacing.
It can be observed that the curves depicting AoA estimation MSE via the MUSIC algorithm can approach the CRB for both the proposed MA scheme and the UPA-dense scheme in the high-SNR regime. In addition, the proposed MA scheme achieves significantly lower MSE compared to the UPA-dense and UPA-sparse schemes and its MSE is also close to the CRB lower-bound in \eqref{CRBlowerbound}. These results demonstrate the advantages of MAs in improving wireless sensing performance via antenna position optimization.

\subsubsection{ISAC}
\begin{figure}[!t]
	\centering
	\includegraphics[width=65mm]{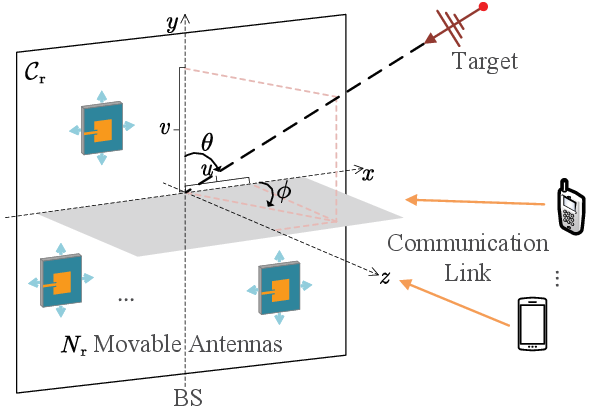}
	\caption{The MA-aided ISAC system.}
	\vspace{-12pt}
	\label{Fig_ISAC_model}
\end{figure}

There is a growing interest in the new paradigm of ISAC in 6G wireless networks, which is able to provide both sensing and communication services by sharing the hardware and/or radio resources \cite{Shao2022target}. Considering the superiority of MAs in enhancing the spatial multiplexing and beamforming performance in wireless communication systems and improving the spatial resolution in wireless sensing systems, they are promising to be employed in ISAC applications to increase design flexibility and achieve combined gains. First, antenna movement effectively enlarges the array aperture, enhancing angular resolution for wireless sensing and boosting spatial multiplexing performance for wireless communication. Additionally, the reconfigurable geometry of an MA array can reduce the correlation between steering vectors across different directions, thereby decreasing angle estimation ambiguity and mitigating interference in multiuser communication scenarios. Moreover, the dynamic adjustability of MA positions enables ISAC systems to adapt to time-varying environmental conditions and diverse communication/sensing tasks. In practice, the geometry of an MA array can be either pre-configured for specific communication/sensing applications or dynamically adjusted in real time to meet varying ISAC performance requirements. 

To examine the potential of MAs in ISAC systems, we consider a BS equipped with $N_\mathrm{r}$ MAs on a 2D plane $\mathcal{C}_{\rm{r}}$ to estimate a target's AoA, while simultaneously providing communication services. It is assumed that the sensing and communication subsystems share the same hardware but operate on separate time or frequency resource blocks, thereby avoiding interference between the two tasks. For the considered sensing system, the CRB of the AoA estimation's MSE is given in \eqref{CRBr}, while the sensing metric is given in \eqref{P2a}. The communication system can be the MA-aided MIMO or multiuser systems with instantaneous/statistical channels in Section \ref{section_MIMOmultiuser}. Since the speed of antenna movement is limited, it is practically difficult to conduct frequent antenna movement if the time of duration for sensing/communication signals is short. As such, the sensing and communication subsystems should share the MAs with the same layout in practice. Due to the distinct design objectives, there exists a trade-off between the communication and sensing performance when optimizing the positions of MAs. For example, for the MA-aided MIMO systems with statistical channels, the communication-centric optimization problem for maximizing the ergodic MIMO capacity can be formulated by imposing an additional CRB constraint on problem (P-MIMO-sta) as
\begin{subequations}
	\begin{align}
		\textrm {(P-ISAC-com)}~~\max_{\tilde{\mathbf{t}}, \tilde{\mathbf{r}}, } ~ & C^{\rm{sta}}(\tilde{\mathbf{t}}, \tilde{\mathbf{r}}) \label{P-ISAC-com-a}\\
		\text{s.t.} \quad & c(\tilde{\mathbf{r}}) \leq \epsilon_{\rm{s}}, \label{P-ISAC-com-b}\\
		&\eqref{P-MIMO-b}, \eqref{P-MIMO-c},\eqref{P-MIMO-d},\eqref{P-MIMO-e},
	\end{align}
\end{subequations}
where $\epsilon_{\rm{s}}$ is a given CRB threshold. Similarly, the sensing-centric optimization problem can be formulated as
\begin{subequations}
	\begin{align}
		\textrm {(P-ISAC-sen)}~~\min_{\tilde{\mathbf{t}}, \tilde{\mathbf{r}}} \quad & c(\tilde{\mathbf{r}}) \label{P-ISAC-sen-a}\\
		\text{s.t.} \quad &  C^{\rm{sta}}(\tilde{\mathbf{t}}, \tilde{\mathbf{r}}) \geq \epsilon_{\rm{c}}, \label{P-ISAC-sen-b}\\
		&\eqref{P-MIMO-b}, \eqref{P-MIMO-c},\eqref{P-MIMO-d},\eqref{P-MIMO-e},
	\end{align}
\end{subequations}
where $\epsilon_{\rm{c}}$ is the given capacity threshold.
To address the non-convex problem (P-ISAC-com) or (P-ISAC-sen), the algorithms for solving the MIMO capacity or multiuser achievable rate maximization problems with instantaneous or statistical channels in Section \ref{section_MIMOmultiuser} can be applied similarly, where the sensing CRB metric can be relaxed via the SCA technique \cite{ma2024MAsensing,ma2025MAISAC}. 

Note that sensing and communication operating on the same or different time/frequency resource blocks represent two distinct technical approaches for ISAC systems. In ISAC systems that utilize different time/frequency resource blocks, the transmit signals or waveforms for the communication and sensing subsystems can be independently designed, which significantly simplifies the overall system design and is practically more viable. However, this approach underutilizes available time/frequency resources. In comparison, ISAC systems that operate on the same time/frequency resource blocks maximize resource efficiency. However, this approach introduces additional challenges such as pronounced interference and complicated signal processing. Nonetheless, exploiting the additional spatial DoF, the MA technology can be effectively applied to both approaches to enhance overall performance and improve the desired trade-offs \cite{ma2025MAISAC,kuang2024movableISAC, khalili2024advanced, WuHS_MA_RIS_ISAC, hao2024fluid, zhang2024efficient, xiu2024movable, lyu2024flexibleISAC, wang2024multiuser, peng2024jointISAC, guo2024movable, MaY2024movableISAC,zhou2024fluidISAC,xiu2025movableISAC,yang2025MAISACjoint,chen2025MAISACopt}.

\begin{figure}[!t]
	\centering
	\includegraphics[width=80mm]{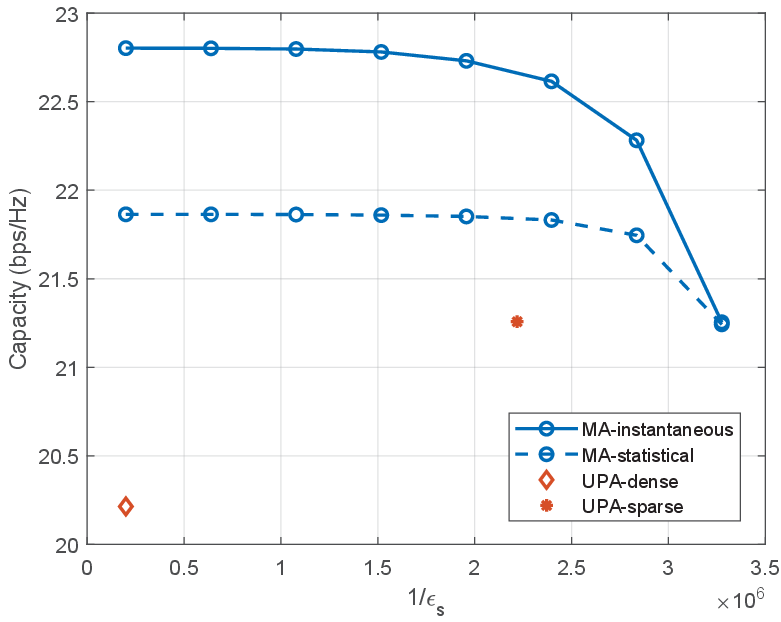}
	\caption{Capacity versus the reciprocal of the CRB threshold.}
	\vspace{-12pt}
	\label{Fig_ISAC_rate_CRB}
\end{figure}

To characterize the trade-off between communication and sensing performance, we consider an MA-aided MIMO communication system and use the acceptable maximum CRB as the sensing metric. We solve the communication-centric optimization problem (P-ISAC-com), and show in Fig.~\ref{Fig_ISAC_rate_CRB} the capacity of the MA-MIMO system versus the reciprocal of the CRB threshold, $1/\epsilon_{\rm{s}}$, under both instantaneous and statistical channels. We set $N_\mathrm{r}=16$, $A=5\lambda$, $P/\sigma^2=15$ dB, and $|\beta|=1$. The number of snapshots for sensing is set as $T_{\mathrm{s}}=16$. For the MIMO communication system, the simulation parameters are the same as those in Fig.~\ref{Fig_MIMO_rate_SNR}, except that the Tx is equipped with a dense UPA, and thus only the Rx-side MAs and the transmit covariance matrix are alternately optimized to maximize the channel capacity. The two FPA-based benchmark schemes are the same as those in Fig.~\ref{Fig_MIMO_rate_SNR}.
As shown in Fig.~\ref{Fig_ISAC_rate_CRB}, the MA schemes always outperform other benchmark schemes with FPAs in terms of minimizing sensing CRB and/or maximizing MIMO communication capacity. To achieve homogenous sensing performance across all directions in the angular domain, the BS should generate an omnidirectional beam pattern to scan targets uniformly in all possible directions. Consequently, the covariance matrix of the transmit probing signals is set as an identity matrix. As indicated in \eqref{CRBr}, the CRB of MSE in AoA estimation depends on the positions of the MAs. Once the antenna positions are fixed, the sensing CRB becomes constant. In this context, the two FPA-based benchmark schemes correspond to specific points in the capacity versus reciprocal CRB threshold region. In contrast, the MA-aided ISAC system enables flexible trade-offs between sensing and communication performance through adaptive antenna positioning. Note that for a sufficiently large threshold of $1/\epsilon_{\rm{s}}$, the positions of MAs should be reconfigured to minimize the sensing CRB, and thus both MA-instantaneous and MA-statistical schemes yield the same array geometry with the same communication performance. This flexibility highlights the superiority of MAs in enhancing the overall performance of ISAC systems.

While MAs offer significant theoretical advantages, their integration into ISAC systems necessitates the development of new standardization frameworks, protocols, and practical deployment strategies. Ensuring compatibility with existing infrastructures and addressing challenges such as maintenance and reliability are critical for real-world implementation. The additional DoFs introduced by MAs' positions require joint optimization with other system parameters such as transmit beamforming and power allocation. These optimization problems are typically high-dimensional and non-convex, demanding the design of efficient and scalable algorithms. Furthermore, MAs enable real-time position adjustment, which is essential in dynamic environments where the locations of communication users or sensing targets frequently change. In this context, leveraging statistical channel knowledge for antenna movement optimization emerges as a promising yet challenging approach for realizing the potential of MA-aided ISAC systems.

\subsection{MA-Enabled Sparse Array}
MA provides an effective method to realize sparse arrays, for which the spacing between adjacent array elements can be larger than the typical value of half-wavelength \cite{li2024sparse}. Depending on whether all adjacent antenna elements have an equal distance, sparse arrays can be classified into the uniform sparse array (USA) and non-uniform sparse array (NUSA). Compared to conventional dense arrays with inter-antenna spacing $d_0=\lambda/2$, USA has one additional design DoF to optimize the array sparsity $\eta$, which is defined as $\eta=d_\mathrm{us}/d_0$, with $d_\mathrm{us}$ being the adjacent antenna separation of the USA. In comparison, NUSA is an even more flexible architecture with higher design DoFs than the USA. For example, for a linear NUSA with $N$ array elements, the design DoF is $N-1$ since the distances of all the $N-1$ pairs of adjacent elements can be freely designed. Over the years, various architectures for NUSA have been designed based on different criteria, such as minimum redundant array (MRA) \cite{chun2008minimum}, modular array (MoA) \cite{li2022near}, nested array (NA) \cite{pal2010nested}, and co-prime array (CPA) \cite{yang2019estimation}. 

%However, all these arrays have a fixed geometry once fabricated which cannot adapt to varying requirements of wireless communications and/or sensing. In this context, MA provides an effective method to realize flexible switches between different dense/sparse array geometries. In addition, compared to systems with independent MA elements, the MA-enabled sparse array significantly reduces the hardware complexity by using the necessary DoFs in antenna movement for geometry reconfiguration only.

Compared to conventional MIMO systems with uniform arrays of half-wavelength antenna separation, sparse arrays-enabled MIMO systems encompass several appealing advantages. First, given the same number of array elements, sparse arrays can achieve a larger array aperture than conventional dense arrays. This results in sharper beams and finer spatial resolution, which is desirable for interference rejection for both wireless communications and sensing. Besides, the enlarged array aperture leads to a larger near-field region, which can be exploited to enhance communication and sensing performance. Second, by carefully designing the physical geometry of antenna arrays, sparse arrays enable the formation of virtual MIMO by forming difference or sum co-arrays that have a much larger number of virtual elements. This has been extensively exploited for wireless localization and sensing, for which a sensing DoF in the order of $N^2$ can be achieved by using only $N$ elements. Third, thanks to the wider and more flexible array spacing, sparse arrays are expected to reduce the mutual coupling effect and enable more flexible deployment to achieve conformity with the mounting structure.

By exploiting the above appealing features, sparse arrays have recently received growing attention, not only for wireless localization and sensing alone but also for communication and ISAC. For example, in \cite{wang2023can1}, it was revealed that given the same number of array elements, USA significantly outperforms the conventional dense array for multiuser communications, especially for closely located users. This can be intuitively explained as follows. While sparse arrays generate undesired grating lobes, it achieves a narrower main lobe. The fact that the users' spatial AoD/AoA differences are typically non-uniformly distributed provides a natural filtering for mitigating the effect of undesired grating lobes \cite{wang2023can1}. In \cite{wang2023can}, by taking into account the near-field effects, the closed-form expression of the spatial multiplexing gain in terms of effective DoF (EDoF) for uniform sparse arrays was derived, mathematically quantifying that the near-field region can be enlarged by sparse arrays. In addition, multiuser communication with MoAs was studied in \cite{li2023multi}. To mitigate grating lobes, an efficient user grouping method was proposed to avoid allocating the users located within each other's grating lobes to the same time-frequency resource block. By this means, it was shown that the sparse arrays outperform the conventional dense arrays for multiuser communications.

Sparse arrays have also received growing attention in ISAC systems. In \cite{li2024sparse}, the authors provided an overview of the architectures, opportunities, challenges, and design issues for sparse array-based ISAC. In \cite{min2024}, the authors conducted a beam pattern analysis of NAs and derived the closed-form expressions for the beam-pattern metrics. The results revealed that compared to conventional uniform arrays, NAs can achieve better communication performance for densely located users, while they maintain the advantage of sensing thanks to the virtual aperture. In \cite{Gao2023Integrated}, the USA architecture was used for radar Rx to enhance the angular resolution, where the OMP algorithm was used to alleviate angular ambiguity. To exploit the extra virtual aperture of the sparse array in ISAC systems, a unified antenna structure and a new virtual aperture scheme were proposed in \cite{liu2024virtual}. Utilizing virtual MIMO, the authors in \cite{sankar2024sparse} investigated the joint AS and transmit precoder design for ISAC systems to meet communication users' requirements while being capable of identifying a certain number of sources.

\begin{figure}[t]
	\centering
	\includegraphics[width=0.7\columnwidth]{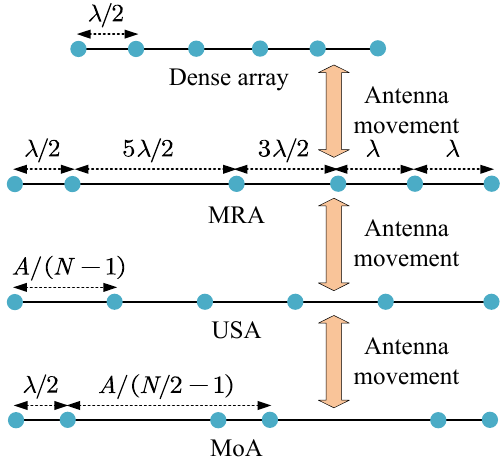}
	\caption{Schematic diagrams of dense array and exemplar sparse arrays, including MRA, USA, and MoA.}
	\vspace{-12pt}
	\label{Fig_MA_meets_sparse}
\end{figure}

\begin{figure}[t]
	\centering
	\centerline{\includegraphics[width=3.5in,height=2.625in]{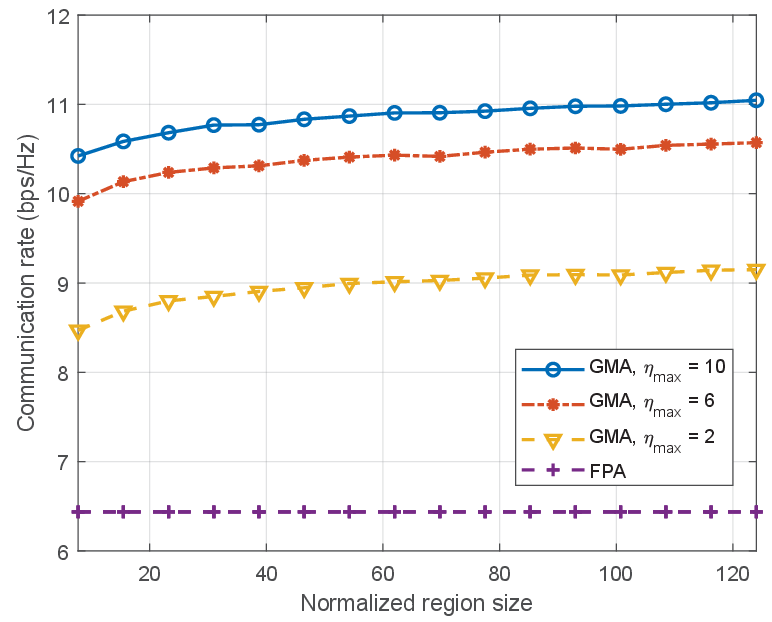}}
	\caption{Uplink communication rate versus the normalized movable region size for GMA-based sparse arrays \cite{lu2024group}.}
	\vspace{-12pt}
	\label{Fig_sparse_sum}
\end{figure}

To realize sparse arrays, most prior studies are based on the conventional FPA architecture, where the desired geometry of the sparse array is predetermined and cannot be altered once the array is deployed. As shown in Fig.~\ref{Fig_MA_meets_sparse}, we consider a 1D antenna movement region with a length of $A$. The advancement of MA technology brings a paradigm shift from the conventional FPA-based sparse array to MA-based sparse arrays, for which the geometry of the dense/sparse array can be dynamically switched based on the scenarios and the distribution of communication users and/or sensing targets. For instance, for ISAC systems with MA-based sparse arrays, by applying the time-switching principle, the NUSA can be formed for sensing to increase spatial resolution, while the USA can be formed in orthogonal time for communication. Besides, for multiuser communications employing MA-based USA at the BS, higher array sparsity can be used if the users are densely located. To this end, a novel architecture of group MA (GMA) with flexible sparsity was proposed in \cite{lu2024group}, which significantly eases the antenna movement overhead. In particular, it aims to jointly optimize the array sparsity $\eta$ (realized by AS over a large array) and the 1D position of the GMA $x$ (realized by array movement), and the associated optimization problem is formulated as
\begin{subequations}
	\begin{align}
		{\text{(P-MA-sp)}}\;\underset{x, \eta}{\max}~ & R(x, \eta) \label{eq:Obj}\\
		\text{s.t.}~ & 1 \leq \eta \leq \eta_{\max},  \\
		& 0 \leq x \leq A,\label{eq:contr2}
	\end{align} 
\end{subequations}
where $\eta_{\max}$ denotes the maximum possible sparsity level and $R(x, \eta)$ is the communications rate of the multiple access channel, which is a function of $x$ and $\eta$ given by
\begin{equation}\label{eq:R}
	R(x,\eta)\!=\! \sum_{k=1}^K \log_2\left(1+ \bar p_k \mathbf h_k(x,\eta)^{\mathrm{H}} \mathbf C_k^{-1}(x,\eta)\mathbf h_k(x,\eta) \right), 
\end{equation}
where $\bar p_k$ is the transmit power of user $k$ normalized by the noise power, $\mathbf h_k(x,\eta)$ is the channel vector of user $k$ that depends on the location and sparsity level of the GMA, and $\mathbf C_k(x,\eta)=\mathbf I_{N} + \sum_{i=1, i\neq k}^K \bar p_i \mathbf h_i(x,\eta) \mathbf h_i (x, \eta)^{\mathrm{H}}$ is the interference-plus-noise covariance matrix of user $k$. Problem (P-MA-sp) is non-convex as the objective function $R(x,\eta)$ is a non-concave and intricate function in $x$ and $\eta$. In \cite{lu2024group}, an AO algorithm was adopted to solve (P-MA-sp) by optimizing $x$ and $\eta$ iteratively. Fig.~\ref{Fig_sparse_sum} shows the uplink communication rate achieved by the GMA-based sparse array versus the normalized movable region size under different $\eta_{\max}$'s. The simulation setup and parameter settings are the same as those in \cite{lu2024group}. Specifically, a linear GMA is employed at the BS with $4$ antennas activated. The carrier frequency is $28$ GHz. The number of users is $K=5$, which are uniformly distributed around the BS with the maximum distance of $50$ m. The channel for each user is composed of $1$ LoS and $5$ NLoS paths, with their AoAs and coefficients randomly generated following similar distributions to those in Fig. \ref{Fig_MU_rate}. In particular, the Rician factor is set as $\kappa=10$ dB. The transmit power of each user is $10$ dBm and the noise power is $-94$ dBm. It is observed that compared to the conventional FPA-based dense array, the GMA-based sparse array achieves considerable performance gains, thanks to the flexible array geometry with optimized sparsity level $\eta$ and array position $x$.

The research on MA-enabled sparse arrays is still in its infancy with many important problems remaining to be resolved. For example, in addition to the GMA-enabled uniform sparse arrays studied in \cite{lu2024group}, the general problem of jointly optimizing the antenna movement and/or sparse array geometry is yet to be studied. Note that compared to the antenna movement optimization problems studied in the previous subsections, the MA-enabled sparse array is expected to have confined optimization space and hence lower complexity, thanks to its array structures. Besides, the design of low-complexity architectures for realizing MA-enabled sparse arrays and the joint optimization of antenna movement/array geometry and resource allocation are important problems deserving further investigation.

\begin{table*}[htp]
	\centering
	\scriptsize 
	\caption{MA Position Optimization under Different System Setups.}
	\begin{tabular}{|c|c|c|c|c|c|c|c|}
		\hline
		\textbf{Usage} &
		\textbf{Ref.} &
		\textbf{\makecell{System setup}} &
		\textbf{\makecell{Antenna\\ setup}} &
		\textbf{CSI level} &
		\textbf{Channel model} &
		\textbf{Objective function} &
		\textbf{\makecell{MA position \\optimization algorithm}} \\ \hline
		\multirow{16}{*}{\rotatebox{90}{Communications}} &
		\cite{zhu2022MAmodel} &
		Single-user, narrowband &
		SISO &
		Instantaneous &
		Field-response &
		\makecell{Received SNR\\ maximization} &
		\makecell{Closed-form solutions \\under certain conditions} \\ \cline{2-8} 
		&
		\cite{zeng2024csi} &
		Single-user, narrowband &
		SISO &
		CSI-free &
		Field-response &
		\makecell{Received SNR\\ maximization} &
		\makecell{Zeroth-order \\gradient approximation} \\ \cline{2-8} 
		&
		\cite{zhu2024wideband} &
		Single-user, wideband &
		SISO &
		Instantaneous &
		Field-response &
		\makecell{Achievable rate\\ maximization} &
		\makecell{Parallel greedy \\ascent algorithm} \\ \cline{2-8} 
		&
		\cite{zhu2023MAarray} &
		\makecell{Beam nulling,\\ narrowband} &
		MISO &
		AoD/AoA &
		LoS only &
		\makecell{Beam gain\\ maximization} &
		\makecell{Closed-form solutions \\under certain conditions} \\ \cline{2-8} 
		&
		\cite{ma2024multi} &
		\makecell{Multi-beam forming,\\ narrowband} &
		MISO &
		AoD/AoA &
		LoS only&
		\makecell{Minimum beam gain \\maximization } &
		AO   and SCA \\ \cline{2-8} 
		&
		\cite{wang2024flexible} &
		\makecell{Wide-beam coverage,\\ narrowband} &
		MISO &
		AoD/AoA &
		LoS only&
		\makecell{Minimum beam gain \\maximization } &
		AO and SCA \\ \cline{2-8} 
		&
		\cite{mei2024movable} &
		Single-user, narrowband &
		MISO &
		Instantaneous &
		N/A &
		\makecell{Received signal \\power maximization} &
		\makecell{Discrete sampling and\\ graph-based method} \\ \cline{2-8} 
		&
		\cite{ma2022MAmimo} &
		Single-user, narrowband &
		MIMO &
		Instantaneous &
		Field-response &
		\makecell{MIMO capacity\\ maximization} &
		AO and SCA \\ \cline{2-8} 
		&
		\cite{chen2023joint} &
		Single-user, narrowband &
		MIMO &
		Statistical &
		Field-response &
		\makecell{MIMO capacity\\ maximization} &
		\makecell{Constrained stochastic SCA} \\ \cline{2-8} 
		&
		\cite{shao20246DMA} &
		Multiuser, narrowband &
		MISO &
		Statistical &
		Field-response &
		\makecell{Sum-rate \\maximization} &
		\makecell{AO and conditional\\ gradient algorithm} \\ \cline{2-8} 
		&
		\cite{xiao2023multiuser} &
		Multiuser, narrowband &
		MISO &
		Instantaneous &
		Field-response &
		\makecell{Minimum-rate \\maximization} &
		PSO \\ \cline{2-8} 
		&
		\cite{zhu2023MAmultiuser} &
		Multiuser, narrowband &
		MISO &
		Instantaneous &
		Field-response &
		\makecell{Sum-rate \\maximization} &
		\makecell{Multi-directional \\gradient descent algorithm} \\ \cline{2-8} 
		&
		\cite{hu2023power} &
		Multiuser, narrowband &
		MISO &
		AoD/AoA &
		LoS only&
		\makecell{Transmit power \\minimization} &
		Gradient descent algorithm \\ \cline{2-8} 
		&
		\cite{Yang2024movable} &
		Multiuser, narrowband &
		MISO &
		Instantaneous &
		Field-response &
		\makecell{Sum-rate \\maximization} &
		\makecell{Regularized LS-based OMP} \\ \cline{2-8} 
		&
		\cite{wu2023movable} &
		Multiuser, narrowband &
		MISO &
		Instantaneous &
		N/A &
		\makecell{Transmit power \\minimization} &
		\makecell{Discrete sampling \\and generalized\\ Bender's decomposition} \\ \cline{2-8} 
		&
		\cite{zhu2024nearfield} &
		Multiuser, narrowband &
		MISO &
		\makecell{Instantaneous \\ or statistical} &
		Field-response &
		\makecell{Minimum SINR/SNR \\maximization} &
		\makecell{Gardient ascent and AO} \\ \hline
		\multirow{11}{*}{\rotatebox{90}{Sensing/ISAC}} &
		\cite{kuang2024movableISAC} &
		\makecell{Multiuser, single-target,\\ narrowband} &
		MISO &
		\makecell{Instantaneous \\and AoD/AoA} &
		\makecell{Field-response\\ and LoS} &
		\makecell{Sum-rate \\maximization} &
		PSO \\ \cline{2-8} 
		&
		\cite{khalili2024advanced} &
		\makecell{Multiuser, single-target,\\ narrowband} &
		MISO &
		\makecell{Instantaneous \\and AoD/AoA} &
		\makecell{Field-response\\ and LoS} &
		\makecell{Transmit power \\minimization} &
		AO \\ \cline{2-8} 
		&
		\cite{WuHS_MA_RIS_ISAC} &
		\makecell{Multiuser, multi-target,\\ narrowband} &
		MISO &
		\makecell{Instantaneous \\and AoD/AoA} &
		\makecell{Field-response\\ and LoS} &
		\makecell{Minimum beampattern \\gain maximization} &
		AO and SCA \\ \cline{2-8} 
		&
		\cite{hao2024fluid} &
		\makecell{Multiuser, single-target,\\ narrowband} &
		MISO &
		\makecell{Instantaneous \\and AoD/AoA} &
		\makecell{Field-response\\ and LoS} &
		\makecell{Minimum sensing\\ SNR maximization} &
		AO and SCA \\ \cline{2-8} 
		&
		\cite{zhang2024efficient} &
		\makecell{Multiuser, single-target,\\ narrowband} &
		MISO &
		\makecell{Instantaneous \\and AoD/AoA} &
		\makecell{Field-response\\ and LoS} &
		\makecell{Sum-rate \\maximization} &
		\makecell{Extrapolated projected \\gradient algorithm} \\ \cline{2-8} 
		&
		\cite{xiu2024movable} &
		\makecell{Single-user, single-target,\\ narrowband} &
		MISO &
		\makecell{Instantaneous \\and AoD/AoA} &
		LoS only&
		\makecell{Communication SNR\\ maximization} &
		BCD and SCA \\ \cline{2-8} 
		&
		\cite{lyu2024flexibleISAC} &
		\makecell{Multiuser, single-target,\\ narrowband} &
		MISO &
		\makecell{Instantaneous \\and AoD/AoA} &
		\makecell{Field-response\\ and LoS} &
		\makecell{Weighted sum of \\communication rate and \\sensing mutual information} &
		\makecell{Search-based projected \\gradient ascent algorithm} \\ \cline{2-8} 
		&
		\cite{wang2024multiuser} &
		\makecell{Multiuser, single-target,\\ narrowband} &
		MIMO &
		AoD/AoA &
		Spatial-correlation &
		\makecell{Sum-rate \\maximization} &
		Deep reinforcement learning \\ \cline{2-8} 
		&
		\cite{peng2024jointISAC} &
		\makecell{Multiuser, single-target,\\ narrowband} &
		MISO &
		\makecell{Instantaneous \\and AoD/AoA} &
		\makecell{Field-response\\ and LoS} &
		\makecell{Weighted sum of \\communication rate and \\sensing mutual information} &
		Gradient projection method \\ \cline{2-8} 
		&
		\cite{qin2024cramer} &
		\makecell{Multiuser, single-target,\\ narrowband} &
		MISO &
		\makecell{Instantaneous \\and AoD/AoA} &
		\makecell{Field-response\\ and LoS} &
		CRB minimization &
		AO and SCA \\ \cline{2-8} 
		&
		\cite{ma2024MAsensing} &
		Single-target, narrowband &
		MISO &
		AoD/AoA &
		LoS only&
		CRB minimization &
		AO and SCA \\ \hline
		\multirow{3}{*}{\rotatebox{90}{Sparse array}} &
		\cite{pal2010nested} &
		Single-target, narrowband &
		MISO &
		AoD/AoA &
		LoS only&
		\makecell{Maximum DoF\\ maximization} &
		Closed-form solutions \\ \cline{2-8} 
		&
		\cite{sankar2024sparse} &
		\makecell{Multiuser, multi-target,\\ narrowband} &
		MISO &
		AoD/AoA &
		N/A &
		\makecell{Transmit power \\minimization} &
		$l_1$-norm minimization \\ \cline{2-8} 
		&
		\cite{lu2024group} &
		Multiuser, narrowband &
		MISO &
		Instantaneous &
		Field-response &
		\makecell{Sum-rate \\maximization} &
		AO and SCA \\ \hline
	\end{tabular}%
	\label{OptAlg}
\end{table*}

\subsection{Lessons Learned and Future Directions}
In Table \ref{OptAlg}, we summarize the existing works on MA position optimization under different basic system setups, including SISO, MISO/SIMO, and MIMO with single or multiple users, as well as setups involving sensing, ISAC, and sparse arrays. It is observed that most existing works focus on optimizing MA positions for wireless communications, with a particular emphasis on narrowband system setups, likely due to the complexity involved in wideband channel modeling and performance optimization. From an optimization perspective, the most desirable strategy is to derive globally optimal solutions for MA positions in closed form, as this approach ensures the best system performance with minimal computational complexity. However, such solutions are typically feasible only for systems with simple configurations or specific optimization structures, such as single-MA setups \cite{zhu2022MAmodel} or pure LoS channels \cite{zhu2023MAarray}. For most practical scenarios, the highly nonlinear relationship between antenna positions and channel responses precludes closed-form globally optimal solutions. To address this, locally optimal optimization techniques are widely adopted in the literature. Examples include gradient-based ascent/descent methods \cite{zhu2023MAmultiuser,hu2023power,zhu2024wideband,peng2024jointISAC,shao20246DMA}, SCA \cite{ma2022MAmimo,ma2024multi,wang2024flexible,WuHS_MA_RIS_ISAC}, OMP \cite{Yang2024movable}, AO \cite{hao2024fluid,xiu2024movable,qin2024cramer,lu2024group,hu2024comp}, and penalty-based method \cite{qin2024antenna,xiao2023multiuser}. While these approaches are computationally efficient, they are prone to being trapped in local optima. In contrast, swarm intelligence-based optimization tools offer stronger global search capabilities, at the cost of higher computational complexities. Examples include PSO \cite{xiao2023multiuser,kuang2024movableISAC,DingJZ_MA_FD_secure_1}, hippopotamus optimization \cite{xiao2024MANOMA}, and the firefly algorithm \cite{Hoang2024firefly}. Additionally, some studies have reformulated continuous position optimization problems into discrete versions by discretizing the MA movement region. Techniques such as graph theory \cite{mei2024movable,mei2024movable_secure} and BnB methods \cite{wu2023movable,wu2024globallyMA} enable globally optimal solutions over discrete positions. However, this approach is computationally expensive, particularly for higher-resolution discretization or larger movable regions, with worst-case complexities that grow exponentially. Beyond traditional methods, AI-based optimization has emerged as a promising avenue for MA system design \cite{zhu2023MAMag,WangC_FAS_RIS_AI_survey}. Techniques such as deep learning \cite{Kang2024DeepMA,tang2024deepMA}, reinforcement learning \cite{weng2024learnMA,Waqar2024leaningfluid}, deep reinforcement learning \cite{wang2024multiuser,bai2024Deepmovable}, and federated learning \cite{zhao2024FLmovable,ahmadzadeh2024enhancement} have demonstrated superior performance compared to conventional optimization approaches. However, these methods rely heavily on extensive training data and significant computational resources. Looking forward, advanced AI techniques such as generative adversarial networks \cite{Creswell2018GAN} and large language models \cite{jiang2024LLM} are expected to further enhance the optimization and configuration of MA-aided wireless networks, paving the way for more intelligent and efficient designs.

The optimization methods discussed above, including gradient-based optimization, swarm intelligence-based optimization, discrete position optimization, and AI-powered optimization, are general approaches applicable to SISO, MISO, SIMO, and MIMO systems in communication, sensing, and ISAC designs. These methods typically involve different trade-offs between optimality and complexity, making it essential to select the most appropriate approach based on specific system configurations and design objectives. Nevertheless, there remains a need for more efficient algorithms for MA position optimization to further reduce computational complexity while enhancing solution quality. In addition, there is significant room for improvement in areas such as CSI accuracy, near- versus far-field propagation, and MA architecture, which are highlighted as potential future research directions in the following.

\subsubsection{Robust MA Position Optimization}
Most of the above works have assumed perfect instantaneous CSI to optimize the MA positions. However, due to various practical factors such as channel aging, limited training/feedback overhead, and noise/interference effect, it is difficult to acquire perfect CSI in practice. As such, further works are needed to look into practical MA position optimization and beamforming design that accounts for imperfect CSI \cite{wu2024globallyMA}. In particular, under the field-response channel model, there may exist estimation errors in both the angle and amplitude information on each spatial path. To model such CSI errors, there are two widely applied models, namely, the deterministic model and the stochastic model. The former model assumes that the norms of the CSI errors are upper-bounded by a set of deterministic values. Hence, the MA positions and beamforming are optimized to ensure the worst-case performance of a given utility function among all possible CSI, subject to the given maximum norm of CSI errors. In contrast, the latter model assumes that the CSI errors are random variables (usually following the Gaussian distribution). Due to the randomly distributed CSI error, the MA positions and beamforming are generally optimized to ensure the non-outage performance of a given utility target. However, due to the highly nonlinear channel expressions in terms of the angles and amplitudes of the spatial paths, both models render the associated robust MA position optimization problems challenging to be optimally solved, calling for more efficient algorithms. Furthermore, the hardware impairments for communication modules and positioning errors for movement modules further aggravate the challenges of robust MA position optimization \cite{yao2024rethinking}, especially in conjunction with the effects of imperfect CSI estimation.

\subsubsection{MA Position Optimization under Near-Field Channels}
In practice, to enable more flexible movements of multiple MAs for communication performance enhancement, MA-aided wireless communication systems typically feature larger aperture sizes compared to their conventional FPA-based counterparts. Moreover, the current wireless communication systems are expected to migrate to higher frequency bands (e.g., mmWave and THz bands) in the future to achieve broader bandwidth. The combination of larger antenna apertures and higher frequency bands, however, increases the Rayleigh distance, rendering the far-field plane-wave channel model inaccurate and necessitates the use of the near-field spherical-wave model. However, most of the above works optimize the MA positions based on the far-field assumption, and research on MA position optimization under near-field conditions remains in its early stages. In particular, more efficient optimization algorithms are needed to handle the more complex channel expressions involving both angle and distance information for each spatial path. Some recent works \cite{zhu2024nearfield,ding2024near,chen2024joint} developed a variety of optimization algorithms to jointly optimize the transmit beamforming and MA positions under near-field channel models. The authors in \cite{WeiX_MA_RIS} analyzed the performance advantages of MAs over FPAs at the BS in an IRS-aided wireless system, under a near-field BS-IRS LoS channel. However, further research is expected to explore the effectiveness of MAs in flexible beamforming (e.g., beam nulling and coverage), enhancing wireless communication performance under different system setups (e.g., wideband and MIMO), wireless sensing/localization applications, sparse-array formation, etc.

\subsubsection{Movement Overhead Management}
MAs are usually implemented with mechanical drivers such as motors, which can introduce additional energy consumption. Furthermore, jointly optimizing antenna movement with communication resource allocation increases computational complexity and signal processing overhead. However, the overhead associated with antenna movement diminishes as the movement frequency decreases. In practical applications, antenna positions can be reconfigured more frequently to enhance communication or sensing performance, or less frequently to minimize overall overhead \cite{Vahid2024movable, Vahid2024MAplan}. In this context, how to properly schedule the movement of MAs within a given region is a crucial but largely unresolved challenge. The authors in \cite{li2024minimizing} have made an initial attempt to tackle this issue from the perspective of MA trajectory optimization. In particular, given the desired destination positions of multiple MAs, they aimed to jointly optimize their associations with the initial MA positions and the trajectories for moving them from their respective initial to destination positions within a given 2D region, with a goal to minimize the delay of antenna movement subject to the inter-MA minimum distance constraints in the movement. Future research could extend this work in several directions. For instance, it would be interesting to explore triggering mechanisms for antenna movement, such as initiating adjustments when system performance falls below a certain threshold or when channel conditions/statistics undergo significant changes. Another important avenue is to determine the optimal achievable MA positions within movement period/delay/energy constraints, thereby balancing the trade-off between movement overhead and communication/sensing performance \cite{wang2024MAdelay}.

\subsubsection{MA Position Optimization under Other Architectures}
Most of the current works assume that each MA element’s position can be individually and globally adjusted within the transmit/receive moving region. However, this architecture may lead to excessively high energy consumption and position-tuning delays, thus challenging the practical implementation of MAs. Furthermore, such an architecture requires a fundamental paradigm shift from existing FPA arrays, potentially resulting in high network infrastructure costs. Due to these practical considerations, there is also a need to investigate the MA position optimization under other more cost-effective MA architectures to better balance communication performance and implementation complexity. Note that the complexity of the MA architecture can be moderated by varying several key aspects such as the moving unit (from individual elements to arrays)\cite{ning2024movable,lu2024group, yichi2024movable}, moving range (from global to local)\cite{ning2024movable,chen2023joint, mei2024movable_secure}, proportion of MAs (from full to partial) \cite{ning2024movable, shi2024capacity}, and so on. These variations can be further jointly applied with the sliding, turning, and/or folding of the antenna elements/arrays for specific communication scenarios, as presented in Section \ref{MA_arch}. As such, it is necessary to delve into the fundamental trade-offs between the beamforming flexibility or communication/sensing performance and the implementation complexity in these architectures, which is worth pursuing in future work.

\section{MA Channel Acquisition} \label{Sec_Channel}
To perform antenna position optimization and fully leverage the spatial DoFs for MA-aided communication systems, it is essential to acquire instantaneous/statistical CSI between any two points within the Tx and Rx antenna moving regions in MA-aided wireless communication systems. This allows for the construction of a complete channel mapping between the Tx and Rx antenna moving regions, i.e., $h(\mathbf{t}, \mathbf{r})$ in \eqref{eq_channel_basic}. Unlike large-scale channel knowledge maps (CKMs) that capture the CSI between the Tx/Rx and its surrounding environments over a range of a few meters to several hundred meters \cite{zeng2024CKMtut}, the channel mapping for MA-aided systems mainly focuses on the Tx and Rx antenna moving regions in the wavelength scale. In fact, such channel mapping includes infinite small-scale CKMs at the Rx by changing the position of the Tx-MA. 

For relatively small Tx/Rx antenna moving regions, their channel mapping can be obtained through exhaustive antenna movement, where the CSI is measured for all possible pairs of antenna positions within the spatial regions. However, when the antenna moving regions are large, exhaustive antenna movement can lead to prohibitively high training overhead and excessive energy consumption due to the need for repositioning the antennas to all potential locations for CSI measurement. To address this challenge as well as to reduce the overhead of channel measurement, two main categories of MA channel acquisition approaches can be adopted, i.e., model-based and model-free methods, which are discussed as follows.

\subsection{Model-Based Methods}
Model-based methods rely on the field-response channel model for MA systems. As shown in \eqref{eq_channel_basic}, given the MAs' positions, the channel between the Tx-MA and Rx-MA is determined by the field response information (FRI), which includes the wave vectors (or AoDs/AoAs) of all Tx and Rx paths, as well as the PRM. Compared to the continuous mapping between arbitrary antenna positions within the Tx and Rx regions, there are only a finite number of unknown parameters in the FRI. Moreover, given the limited size of the antenna moving regions, the angular resolution of channel paths is also limited, where the paths with similar AoDs/AoAs can be equivalently regarded as a single effective channel path w.r.t. the Tx/Rx region. Consequently, the channel of an MA system exhibits inherent sparsity in the angular domain \cite{ma2023MAestimation,xiao2023channel,Xu2024estimation}, while the channel mapping between the Tx and Rx antenna moving regions can be reconstructed by estimating the FRI. These model-based approaches significantly reduce the time overhead for channel acquisition, as it mainly depends on the number of Tx and Rx paths rather than the physical size of the antenna moving regions. Therefore, it can be applied to larger regions with considerably lower complexity compared to direct channel measurement via exhaustive antenna movement. Next, we introduce two efficient methods to achieve this goal based on compressed sensing and tensor decomposition, respectively, following the model-based approach.

\subsubsection{Compressed Sensing-based Method}
Due to the inherent sparsity of finite channel paths in the angular domain, compressed sensing-based methods can be applied for MA channel acquisition \cite{ma2023MAestimation,xiao2023channel}. In general, the compressed sensing framework consists of three key phases: (i) Measuring channel responses at a limited number of selected MA locations; (ii) Estimating the FRI, including the wave vectors (or AoDs/AoAs) and PRM; and (iii) Reconstructing the complete channel mapping between Tx and Rx regions. For the FRI estimation phase, the wave vectors are usually approximated by uniformly quantizing each entry of the Tx and Rx wave vectors (within the range of $-1$ to $1$) into $G$ discrete grids. In general, there is a trade-off between the accuracy of FRI estimation and computational complexity. In particular, joint estimation of FRI components \cite{xiao2023channel} offers greater accuracy compared to the successive estimation of FRI components \cite{ma2023MAestimation}. However, the computational complexity of joint estimation scales quadratically with that of successive estimation due to the need to search across all grids.

For ease of exposition, we consider an MA-aided communication system equipping a single MA at both the Tx and Rx within 2D antenna moving regions. Without loss of generality, we assume $z_{\mathrm{t}}=z_{\mathrm{r}}=0$ and let $\mathbf{t}$ and $\mathbf{r}$ denote the 2D positions of the Tx-MA and Rx-MA in their LCSs, respectively. In the successive FRI estimation scheme \cite{ma2023MAestimation}, there are three steps to estimate FRI components: (i) Estimate Tx wave vectors by moving the Tx-MA while fixing the Rx-MA, and solve the Tx-side low-dimensional sparse signal recovery problem; (ii) Estimate Rx wave vectors by moving the Rx-MA while fixing the Tx-MA, and solve the Rx-side low-dimensional sparse signal recovery problem; and (iii) Based on the estimated Tx and Rx wave vectors, estimate the PRM via least square (LS). Specifically, in step (i), let $M_{\rm{t}}$ denote the total number of measurement positions for the Tx-MA, which are denoted by $\tilde{\mathbf{t}}=[\mathbf{t}_1^{\rm{T}},\mathbf{t}_2^{\rm{T}},\ldots,\mathbf{t}_{M_{\rm{t}}}^{\rm{T}}]^{\rm{T}} \in\mathbb{R}^{2M_{\rm{t}} \times 1}$. The Rx-MA's position is fixed as $\mathbf{r}_1 \in\mathbb{R}^{2\times 1}$. We assume that one pilot signal is transmitted by the Tx-MA at each measurement position. Without loss of generality, we adopt the normalized pilot signal $s=1$ throughout this section. Thus, in this step, the received signals are given by
\begin{align}
	\left(\mathbf{y}_{\rm{t}}\right)^{\rm{T}} &= \sqrt{P}\mathbf{f}(\mathbf{r}_1)^{\rm{H}} \mathbf{\Sigma} \mathbf{G}(\tilde{\mathbf{t}})  + \left(\mathbf{z}_{\rm{t}}\right)^{\rm{T}} \\
	&\triangleq \left(\mathbf{x}_{\rm{t}}\right)^{\rm{H}} \mathbf{G}(\tilde{\mathbf{t}})  + \left(\mathbf{z}_{\rm{t}}\right)^{\rm{T}} \in \mathbb{C}^{1 \times M_{\rm{t}}}, \notag
\end{align}
where $\mathbf{x}_{\rm{t}} \triangleq \sqrt{P}\mathbf{\Sigma}^{\rm{H}}\mathbf{f}(\mathbf{r}_1) \in\mathbb{C}^{L_{\rm{t}} \times 1}$,  $\mathbf{G}(\tilde{\mathbf{t}})=[\mathbf{g}({\mathbf{t}_1}),\mathbf{g}({\mathbf{t}_2}),\ldots,\mathbf{g}({\mathbf{t}_{M_{\rm{t}}}})]\in\mathbb{C}^{L_{\rm{t}} \times M_{\rm{t}}}$, and $\mathbf{z}_{\rm{t}}\sim \mathcal{CN}(0,\sigma^2\mathbf{I}_{M_{\rm{t}}})$ is the additive white Gaussian noise (AWGN) vector, with $\sigma^2$ denoting the average noise power.

The 2D wave vector of the $j$-th Tx path is defined as $\tilde{\mathbf{k}}_{{\rm{t}},j}=[\cos\theta_{{\rm{t}},j}\cos\phi_{{\rm{t}},j},\cos\theta_{{\rm{t}},j}\sin\phi_{{\rm{t}},j}]^{\rm{T}} \triangleq [u_{{\rm{t}},j},v_{{\rm{t}},j}]^{\rm{T}}$. By uniformly discretizing $u\in[-1,1]$ and $v\in[-1,1]$ into $G$ grids with $G\gg L_{\rm{t}}$, we have $\left(\mathbf{x}_{\rm{t}}\right)^{\rm{H}} \mathbf{G}(\tilde{\mathbf{t}}) \approx \left(\bar{\mathbf{x}}_{\rm{t}}\right)^{\rm{H}} \bar{\mathbf{G}}(\tilde{\mathbf{t}})$, where $\bar{\mathbf{G}}(\tilde{\mathbf{t}}) \in{\mathbb{C}^{G^2\times{M_{\rm{t}}}}}$
is an over-complete matrix and its $(g_1+(g_2-1)G)$-th row is in the form of $\left[ e^{{\rm{j}}\frac{2\pi}{\lambda}\tilde{\mathbf{k}}_{g_1,g_2}^{\rm{T}}\mathbf{t}_1}, e^{{\rm{j}}\frac{2\pi}{\lambda}\tilde{\mathbf{k}}_{g_1,g_2}^{\rm{T}}\mathbf{t}_2},\ldots,e^{{\rm{j}}\frac{2\pi}{\lambda}\tilde{\mathbf{k}}_{g_1,g_2}^{\rm{T}}\mathbf{t}_{M_{\rm{t}}}} \right] \in\mathbb{C}^{1\times M_{\rm{t}}}$ with $\tilde{\mathbf{k}}_{g_1,g_2}\triangleq[-1+2g_1/G,-1+2g_2/G]^{\rm{T}}$, $1 \leq g_1, g_2 \leq G$, and $\bar{\mathbf{x}}_{\rm{t}} \in{\mathbb{C}^{G^2\times 1}}$ is a sparse vector with $L_{\rm{t}}$ nonzero elements corresponding to $\mathbf{x}_{\rm{t}}$. If $G\gg L_{\rm{t}}$, the Tx wave vector estimation problem can be transformed into a sparse signal recovery problem, i.e., finding a sparse $\bar{\mathbf{x}}_{\rm{t}}$ to minimize $\|\left(\mathbf{y}_{\rm{t}}\right)^{\rm{T}}-\left(\bar{\mathbf{x}}_{\rm{t}}\right)^{\rm{H}} \bar{\mathbf{G}}(\tilde{\mathbf{t}})\|_2$. Classical compressed sensing algorithms, such as the OMP \cite{lee2016channel}, can be employed to estimate $\{u_{{\rm{t}},j},v_{{\rm{t}},j}\}_{j=1}^{L_{\rm{t}}}$ corresponding to the rows of $\bar{\mathbf{G}}(\tilde{\mathbf{t}})$ with non-zero coefficients in $\bar{\mathbf{x}}_{\rm{t}}$. 

In step (ii), let $M_{\rm{r}}$ denote the total number of measurement positions for the Rx-MA, which are denoted by $\tilde{\mathbf{r}}=[\mathbf{r}_1^{\rm{T}},\mathbf{r}_2^{\rm{T}},\ldots,\mathbf{r}_{M_{\rm{r}}}^{\rm{T}}]^{\rm{T}} \in\mathbb{R}^{2M_{\rm{r}}\times 1}$. The Tx-MA's position is fixed as $\mathbf{t}_1 \in\mathbb{R}^{2 \times 1}$. Thus, in this step, the received signals are given by
\begin{align}
	\mathbf{y}_{\rm{r}} &= \sqrt{P}\mathbf{F}(\tilde{\mathbf{r}})^{\rm{H}} \mathbf{\Sigma}  \mathbf{g}(\mathbf{t}_1) + \mathbf{z}_{\rm{r}} \\
	&\triangleq \mathbf{F}(\tilde{\mathbf{r}})^{\rm{H}} \mathbf{x}_{\rm{r}}  + \mathbf{z}_{\rm{r}}
	\in \mathbb{C}^{M_{\rm{r}} \times 1}, \notag
\end{align}
where $\mathbf{x}_{\rm{r}} \triangleq \sqrt{P}\mathbf{\Sigma}  \mathbf{g}(\mathbf{t}_1) \in\mathbb{C}^{L_{\rm{r}} \times 1}$,  $\mathbf{F}(\tilde{\mathbf{r}})=[\mathbf{f}({\mathbf{r}_1}),\mathbf{f}({\mathbf{r}_2}),\ldots,\mathbf{f}({\mathbf{r}_{M_{\rm{r}}}})]\in\mathbb{C}^{L_{\rm{r}} \times M_{\rm{r}}}$, and $\mathbf{z}_{\rm{r}}\sim \mathcal{CN}(0,\sigma^2\mathbf{I}_{M_{\rm{r}}})$ is the AWGN vector. Similar to the Tx wave vector estimation procedure, the Rx wave vectors can be estimated via the compressed sensing algorithms. Finally, in step (iii), the PRM can be estimated via LS based on $\mathbf{y}_{\rm{t}}$, $\mathbf{y}_{\rm{r}}$, as well as the estimation of $\mathbf{G}(\tilde{\mathbf{t}})$ and $\mathbf{F}(\tilde{\mathbf{r}})$ \cite{ma2023MAestimation}.

To further improve the FRI estimation accuracy, the joint estimation of FRI components was considered in \cite{xiao2023channel}. Specifically, let $M$ denote the total number of measurement position pairs for the Tx-MA and Rx-MA, which are denoted by $\tilde{\mathbf{t}}=[\mathbf{t}_1^{\rm{T}},\mathbf{t}_2^{\rm{T}},\ldots,\mathbf{t}_{M}^{\rm{T}}]^{\rm{T}} \in\mathbb{R}^{2M\times 1}$ and $\tilde{\mathbf{r}}=[\mathbf{r}_1^{\rm{T}},\mathbf{r}_2^{\rm{T}},\ldots,\mathbf{r}_{M}^{\rm{T}}]^{\rm{T}} \in\mathbb{R}^{2M\times 1}$, respectively. We assume that one pilot signal is transmitted by the Tx-MA and received by the Rx-MA at each measurement position pair. Thus, the received signals are given by
\begin{align}
	\mathbf{y} &= \sqrt{P}\begin{bmatrix}
		\mathbf{f}(\mathbf{r}_1)^{\rm{H}} \mathbf{\Sigma}  \mathbf{g}(\mathbf{t}_1) \\
		\vdots \\
		\mathbf{f}(\mathbf{r}_M)^{\rm{H}} \mathbf{\Sigma}  \mathbf{g}(\mathbf{t}_M)
	\end{bmatrix} + \mathbf{z} \\
	&=\sqrt{P}\begin{bmatrix}
		\mathbf{g}(\mathbf{t}_1)^{\rm{T}} \otimes \mathbf{f}(\mathbf{r}_1)^{\rm{H}} \\
		\vdots \\
		\mathbf{g}(\mathbf{t}_M)^{\rm{T}} \otimes \mathbf{f}(\mathbf{r}_M)^{\rm{H}}
	\end{bmatrix} \mathbf{u} + \mathbf{z} \notag\\
	&\triangleq\sqrt{P}\mathbf{\Phi}\mathbf{u} + \mathbf{z} \in \mathbb{C}^{M \times 1}, \notag
\end{align}
where $\mathbf{u}\triangleq {\rm{vec}}(\mathbf{\Sigma}) \in\mathbb{C}^{L_{\rm{r}} L_{\rm{t}}\times 1}$, $\mathbf{z}\sim \mathcal{CN}(0,\sigma^2\mathbf{I}_{M})$ is the AWGN vector. By uniformly discretizing all entries of Tx and Rx wave vectors into $G$ grids with $G\gg L_{\rm{t}}L_{\rm{r}}$, we have $\mathbf{\Phi}\mathbf{u} \approx \bar{\mathbf{\Phi}}\bar{\mathbf{u}}$, where $\bar{\mathbf{\Phi}} \in{\mathbb{C}^{M\times{G^4}}}$
is a measurement matrix and its $(g_1+(g_2-1)G+(g_3-1)G^2+(g_4-1)G^3)$-th column is in the form of $\left[ e^{{\rm{j}}\frac{2\pi}{\lambda}(\mathbf{k}_{g_1,g_2}^{\rm{T}}\mathbf{t}_1+\mathbf{k}_{g_3,g_4}^{\rm{T}}\mathbf{r}_1)}, \ldots,e^{{\rm{j}}\frac{2\pi}{\lambda}(\mathbf{k}_{g_1,g_2}^{\rm{T}}\mathbf{t}_M+\mathbf{k}_{g_3,g_4}^{\rm{T}}\mathbf{r}_M)} \right]^{\rm{T}} \in\mathbb{C}^{M \times 1}$, and $\bar{\mathbf{u}} \in{\mathbb{C}^{G^4\times 1}}$ is a sparse vector with $L_{\rm{t}}L_{\rm{r}}$ nonzero elements corresponding to $\mathbf{u}$. If $G\gg L_{\rm{t}}L_{\rm{r}}$, the FRI estimation problem can be transformed into a sparse signal recovery problem, i.e., finding a sparse $\bar{\mathbf{u}}$ to minimize $\|\mathbf{y}-\sqrt{P}\bar{\mathbf{\Phi}}\bar{\mathbf{u}}\|_2$. Classical compressed sensing algorithms, such as the OMP \cite{lee2016channel}, can be employed to estimate Tx and Rx wave vectors corresponding to the columns of $\bar{\mathbf{\Phi}}$ with non-zero coefficients in $\bar{\mathbf{u}}$. Although the computational complexity of joint FRI estimation is much higher than that of successive estimation, as it involves searching across $G^4$ grids, it offers improved estimation accuracy due to the joint estimation of all unknown FRI components. Moreover, the measurement locations of MAs determine the measurement matrix of the compressed sensing algorithm and thus significantly influence the estimation accuracy. Therefore, it is essential to strategically design the measurement locations of MAs to enhance the estimation accuracy. The design criteria and specific implementations can be referred to \cite{xiao2023channel}.

Despite the MA-SISO systems considered in \cite{xiao2023channel,ma2023MAestimation}, the compressed sensing-based methods are also applicable to MISO/SIMO/MIMO systems. In particular, multiple MAs at the Tx/Rx can conduct channel measurements at different locations simultaneously, which can help reduce the antenna movement overhead. The MA systems with 1D movement regions and those with Tx-MA or Rx-MA only can be regarded as special cases of the joint 2D movement at Tx and Rx. Besides, using the FRI acquired by 2D movement, the channel mapping between 3D Tx and Rx regions can also be constructed according to \eqref{eq_channel_basic} because for each wave vector $\mathbf{k}_{\mathrm{t},j}$ (or $\mathbf{k}_{\mathrm{r}, i}$), the third entry can be determined by the other two estimated elements in $\tilde{\mathbf{k}}_{\mathrm{t},j}$ (or $\tilde{\mathbf{k}}_{\mathrm{r}, i}$) according to its unit-norm property. Moreover, the compressed sensing-based channel acquisition method has also been extended to MA-aided wideband communication systems in \cite{xiao2024channelwide} by jointly estimating the Tx-Rx wave vectors, delays, and path-response coefficients.

\begin{table*}[t]
	\caption{A summary of representative channel acquisition methods for MA systems.} \label{Tab_channel_CE}
	\centering 
	\small
	\begin{tabular}{|>{\centering\arraybackslash}m{1.1cm}|>{\centering\arraybackslash}m{1.7cm}|>{\centering\arraybackslash}m{1.8cm}|>{\centering\arraybackslash}m{0.5cm}|>{\centering\arraybackslash}m{1.5cm}|>{\centering\arraybackslash}m{1.6cm}|>{\centering\arraybackslash}m{1.9cm}|>{\centering\arraybackslash}m{1.9cm}|>{\centering\arraybackslash}m{1.9cm}|}
		\hline
		& \textbf{Advantages}    & \textbf{Limitations}         & \textbf{Ref.} & \textbf{System Setup}        & \textbf{Channel Dimension}      & \textbf{Estimation Method}  & \textbf{Estimation Accuracy}  & \textbf{Computational Complexity} \\ \hline
		\multirow{3}{*}{\makecell[c] {\\ \\  \\ \\ Model-\\based\\ methods} }     & \multirow{3}{*}{\begin{tabular}[c]{@{}c@{}}\makecell[c]{1) Higher \\estimation \\accuracy \\ for fewer \\channel \\ paths; }\\ \makecell[c]{2) Relatively\\ low channel \\measurement \\ overhead\\ for large \\regions} \end{tabular}} & \multirow{3}{*}{\begin{tabular}[c]{@{}c@{}}\makecell[c]{\\ 1) Rely on \\accurate \\channel \\models;}\\ \makecell[c]{2) Sensitive \\to channel \\measurement \\errors}\end{tabular}}                                                              
		& \cite{ma2023MAestimation}       & SISO with Tx-MA and Rx-MA    & 2D continuous Tx/Rx regions          & Successive FRI estimation via compressed sensing            & Proportional to number of discrete grids of Tx/Rx wave vector & Proportional to number of discrete grids of Tx/Rx wave vector       \\ \cline{4-9} 
		&  & & \cite{xiao2023channel}        & SISO with Tx-MA and Rx-MA    & 2D continuous Tx/Rx regions            & Joint FRI estimation via compressed sensing     & Proportional to number of discrete grids of Tx and Rx wave vectors & Proportional to number of discrete grids of Tx and Rx wave vectors            \\ \cline{4-9} 
		&  & & \cite{zhang2024TensorCE}      & MIMO with Tx-MA and Rx-MA    & 2D continuous Tx/Rx regions              & Successive FRI estimation via tensor decomposition     & Super-resolution   & Proportional to number of channel measurements       \\ \hline
		\multirow{3}{*}{\makecell[c]{\\ \\  \\   Model-\\free\\ methods}} & \multirow{3}{*}{\begin{tabular}[c]{@{}c@{}}\makecell[c]{1) Robust to \\channel \\modeling \\errors;}\\ \makecell[c]{  2) Relatively\\ low channel\\ measurement \\ overhead\\ for small \\ regions} \end{tabular}}                               & \multirow{3}{*}{\begin{tabular}[c]{@{}c@{}}\makecell[c]{ 1) Much high \\ overhead\\ for large\\ and high-\\dimensional \\ regions;}\\ \makecell[c]{ 2) Reconstruct \\ channels for \\discrete ports \\only}\end{tabular}} 
		& \cite{Skouroumounis2023fluidCE}      & SISO with Rx-MA only         & 1D discrete region                & Unmeasured channels are assumed equal to those at nearby measured positions       & Proportional to number of channel measurements & Proportional to number of channel measurements  \\ \cline{4-9} 
		&  &  & \cite{zhang2023successive} & SIMO with Rx-MA only         & 1D discrete region              & Bayesian linear regression & Proportional to number of channel measurements & Proportional to number of channel measurements      \\ \cline{4-9} 
		&  &  &  \cite{ji2024correlation}    & SISO with Rx-MA only         & 1D discrete region              & Machine learning     & Proportional to number of channel measurements & Proportional to number of channel measurements             \\ \hline
	\end{tabular}
\end{table*}

\subsubsection{Tensor Decomposition-based Method}
The compressed sensing-based methods rely on discrete wave vector quantization for sparse representation of channels, which inevitably limits the FRI estimation accuracy. For one thing, since the computational complexity is proportional to the number of discrete grids used for discretizing wave vectors, searching across all grids for high accuracy can be computationally intensive. For another, the estimation accuracy is limited by the grid resolution, preventing super-resolution in high-SNR regimes. To overcome these limitations, a tensor decomposition-based method was proposed in \cite{zhang2024TensorCE}, which models the received pilot signals as a third-order tensor. The factor matrices of the tensor are obtained via canonical polyadic (CP) decomposition, allowing for grid-free estimation of wave vectors based on designed antenna movement positions, and then the PRM is estimated via LS.

We consider an MA-MIMO communication system with $N_{\rm{t}}$ Tx-MAs and $N_{\rm{r}}$ Rx-MAs at both the 2D Tx and Rx antenna moving regions. In the tensor decomposition-based method, there are three steps to estimate FRI components: (i) Solve the Tx-side factor matrices estimation problem by moving the Tx-MAs while fixing the Rx-MAs, and estimate Tx wave vectors via the estimation of signal parameters via rotational invariance technique (ESPRIT) algorithm; (ii) Solve the Rx-side factor matrices estimation problem by moving the Rx-MAs while fixing the Tx-MAs, and estimate Rx wave vectors via the ESPRIT algorithm; and (iii) Based on the estimated Tx and Rx wave vectors, estimate the PRM via LS. Specifically, in step (i), let $M$ denote the total number of measurement positions for one Tx-MA (with the other Tx-MAs all switched off), which are denoted by $\tilde{\mathbf{t}}=[\mathbf{t}_1^{\rm{T}},\mathbf{t}_2^{\rm{T}},\ldots,\mathbf{t}_{M}^{\rm{T}}]^{\rm{T}} \in\mathbb{R}^{2M \times 1}$. The Rx-MAs' positions are fixed as $\tilde{\mathbf{r}}_0=[\mathbf{r}_1^{\rm{T}},\mathbf{r}_2^{\rm{T}},\ldots,\mathbf{r}_{N_{\rm{r}}}^{\rm{T}}]^{\rm{T}} \in\mathbb{R}^{2N_{\rm{r}}\times 1}$. It is assumed that one pilot signal is transmitted by the Tx-MA at each measurement position and received by all Rx-MAs. Thus, in this step, the received signals are given by
\begin{align}
	\mathbf{Y}_{\rm{t}} &= \sqrt{P}\mathbf{F}(\tilde{\mathbf{r}}_0)^{\rm{H}} \mathbf{\Sigma} \mathbf{G}(\tilde{\mathbf{t}})  + \mathbf{Z}_{\rm{t}} \in \mathbb{C}^{N_{\rm{r}} \times M},
\end{align}
where $\mathbf{Z}_{\rm{t}}\in \mathbb{C}^{N_{\rm{r}} \times M}$ represents the AWGN matrix at the receiver.

The ESPRIT algorithm can achieve grid-free estimation of wave vectors. Denote $\mathbf{r}_n = [x_{\rm{r}}^n,y_{\rm{r}}^n]^{\rm{T}}$ and $\mathbf{t}_m = [x_{\rm{t}}^m,y_{\rm{t}}^m]^{\rm{T}}$, $1 \leq n \leq N_{\rm{r}}$, $1 \leq m \leq M$. To apply the ESPRIT algorithm, $\tilde{\mathbf{t}}$ should form the UPA shape with $M=M_{\rm{x}}M_{\rm{y}}$, such that $\mathbf{G}(\tilde{\mathbf{t}})$ can be decomposed as $\mathbf{G}(\tilde{\mathbf{t}})^{\rm{H}} = \mathbf{A}_{\rm{x}} \odot \mathbf{A}_{\rm{y}}$, where $\mathbf{A}_{\rm{x}}=[\mathbf{a}_{\rm{x}}(u_{{\rm{t}},1}),\ldots,\mathbf{a}_{\rm{x}}(u_{{\rm{t}},L_{\rm{t}}})]\in\mathbb{C}^{M_{\rm{x}} \times L_{\rm{t}}}$ and $\mathbf{A}_{\rm{y}}=[\mathbf{a}_{\rm{y}}(v_{{\rm{t}},1}),\ldots,\mathbf{a}_{\rm{y}}(v_{{\rm{t}},L_{\rm{t}}})]\in\mathbb{C}^{M_{\rm{y}} \times L_{\rm{t}}}$, with $\mathbf{a}_{\rm{x}}(u)\triangleq\left[e^{{\rm{j}}\frac{2\pi}{\lambda}x_{\rm{t}}^1 u},\ldots,e^{{\rm{j}}\frac{2\pi}{\lambda}x_{\rm{t}}^{M_{\rm{x}}} u}\right]^{\rm{T}}\in\mathbb{C}^{M_{\rm{x}}\times 1}$ and $\mathbf{a}_{\rm{y}}(u)\triangleq\left[e^{{\rm{j}}\frac{2\pi}{\lambda}y_{\rm{t}}^1 u},\ldots,e^{{\rm{j}}\frac{2\pi}{\lambda}y_{\rm{t}}^{M_{\rm{y}}} u}\right]^{\rm{T}}\in\mathbb{C}^{M_{\rm{y}} \times 1}$. Then, the received signal matrix can be rewritten as
\begin{align}\label{Yttensor}
	\left(\mathbf{Y}_{\rm{t}}\right)^{\rm{H}} &= \sqrt{P}\mathbf{G}(\tilde{\mathbf{t}})^{\rm{H}} \mathbf{\Sigma}^{\rm{H}} \mathbf{F}(\tilde{\mathbf{r}}_0)  + \left(\mathbf{Z}_{\rm{t}}\right)^{\rm{H}}, \\
	&\triangleq (\mathbf{A}_{\rm{x}} \odot \mathbf{A}_{\rm{y}}) \mathbf{D}_{\rm{t}} + \left(\mathbf{Z}_{\rm{t}}\right)^{\rm{H}}, \notag
\end{align}
where $\mathbf{D}_{\rm{t}} \triangleq \sqrt{P}\mathbf{\Sigma}^{\rm{H}} \mathbf{F}(\tilde{\mathbf{r}}_0)\in \mathbb{C}^{L_{\rm{t}} \times N_{\rm{r}}}$.

Based on the definition of CP decomposition, $\left(\mathbf{Y}_{\rm{t}}\right)^{\rm{H}}$ can form an $M_{\rm{x}}\times M_{\rm{y}}\times N_{\rm{r}}$ tensor with three factor matrices $\{\mathbf{A}_{\rm{x}}, \mathbf{A}_{\rm{y}}, \mathbf{D}_{\rm{t}}\}$. Then, $\{\mathbf{A}_{\rm{x}}, \mathbf{A}_{\rm{y}}, \mathbf{D}_{\rm{t}}\}$ can be estimated via the alternating least square (ALS) based on $\mathbf{Y}_{\rm{t}}$ and the structure of \eqref{Yttensor}. Since the factor matrices $\mathbf{A}_{\rm{x}}$ and $\mathbf{A}_{\rm{y}}$ exhibit the inherent Vandermonde structure with $\tilde{\mathbf{t}}$ forming the UPA shape, the Tx wave vectors can be estimated via the ESPRIT algorithm based on the estimation of $\{\mathbf{A}_{\rm{x}}, \mathbf{A}_{\rm{y}}\}$.

Similarly, in step (ii), the Tx-MAs are fixed while the Rx-MAs are moved to form a UPA shape. The Rx wave vectors can then be estimated in the same manner as in step (i). Finally, in step (iii), the PRM can be estimated via LS based on the received signals and the previously estimated $\mathbf{G}(\tilde{\mathbf{t}})$ and $\mathbf{F}(\tilde{\mathbf{r}})$.

The tensor decomposition-based method can achieve super-resolution FRI estimation in high-SNR regimes by utilizing the ESPRIT algorithm. However, this method is limited to MIMO systems satisfying $\min\{M_{\rm{x}},L_{\rm{t}}\} + \min\{M_{\rm{y}},L_{\rm{t}}\} + \min\{N_{\rm{r}},L_{\rm{t}}\} \geq 2L_{\rm{t}}+2$ and $\min\{M_{\rm{x}},L_{\rm{r}}\} + \min\{M_{\rm{y}},L_{\rm{r}}\} + \min\{N_{\rm{t}},L_{\rm{r}}\} \geq 2L_{\rm{r}}+2$, making it unsuitable for MA systems with fewer antennas. Additionally, the ESPRIT algorithm necessitates that the MAs' positions for channel measurement form a UPA shape, which limits the channel acquisition performance for antenna moving regions with irregular shapes.

\subsection{Model-Free Methods}
Model-based methods depend on a specific channel model for MA systems, where key parameters such as the wave vectors for all Tx and Rx paths, as well as the corresponding PRM, are estimated to reconstruct the channel mapping. However, these methods may experience performance degradation if discrepancies exist between the assumed and the actual channel models. In contrast, model-free methods do not rely on any channel model assumptions. Instead, the MA measures the channel at certain positions, and the channels at unmeasured positions are reconstructed by assuming correlation to those at nearby measured positions \cite{Skouroumounis2023fluidCE,new2024CE}, by Bayesian linear regression \cite{zhang2023successive,cui2024near}, or by machine learning \cite{ji2024correlation,zhang2024MLCE}. In general, the model-free methods consist of two steps: (i) Measuring channel responses at a number of selected MA positions; and (ii) Estimating the channel for the unmeasured positions based on the measured channel responses. 

For ease of exposition, we consider an MA communication system in which a single MA is positioned within the Rx antenna moving region, while the Tx is equipped with an FPA. In step (i), let $M$ denote the total number of measurement positions for the MA, which are denoted by $\tilde{\mathbf{r}}=[\mathbf{r}_1^{\rm{T}},\mathbf{r}_2^{\rm{T}},\ldots,\mathbf{r}_{M}^{\rm{T}}]^{\rm{T}}$. We assume that one pilot signal is received by the MA at each measurement position. Thus, the received signals are given by
\begin{align}
	\mathbf{y}(\tilde{\mathbf{r}}) &= \sqrt{P} \mathbf{h}(\tilde{\mathbf{r}})  + \mathbf{z} \in\mathbb{C}^{M \times 1},
\end{align}
where $\mathbf{h}(\tilde{\mathbf{r}})=[h(\mathbf{r}_1),h(\mathbf{r}_2),\ldots,h(\mathbf{r}_M)]^{\rm{T}} \in\mathbb{C}^{M \times 1}$ denotes the channel vector from the Tx-FPA to the Rx-MA at positions $\tilde{\mathbf{r}}$, and $\mathbf{z}\sim \mathcal{CN}(0,\sigma^2\mathbf{I}_{M})$ is the AWGN vector. Then, in step (ii), the channel at unmeasured position $\mathbf{r}$ can be estimated by
\begin{align}
	h(\mathbf{r}) = U(\mathbf{y}(\tilde{\mathbf{r}})),
\end{align}
where $U(\mathbf{y}(\tilde{\mathbf{r}}))$ is the estimation function of the received signal vector $\mathbf{y}(\tilde{\mathbf{r}})$.

The key challenge for model-free methods lies in the efficient design of the estimation function $U(\cdot)$. A straightforward approach is to assume that the channels at unmeasured locations are the same as those at nearby measured locations \cite{Skouroumounis2023fluidCE,new2024CE}. However, this approach depends on the assumption of spatially slow-varying channels. To overcome this limitation, Bayesian linear regression \cite{zhang2023successive,cui2024near} and machine learning \cite{ji2024correlation,zhang2024MLCE} techniques can be employed to estimate $U(\cdot)$ based on the received signal vector $\mathbf{y}(\tilde{\mathbf{r}})$. The Bayesian linear regression method models the channel response as a stochastic process or random field with respect to the MA’s position, where each instantaneous channel realization in a given propagation environment is treated as a sample of the Gaussian random process/field. Based on this framework, the instantaneous channel can be estimated from measurements at different MA positions. For instance, in \cite{zhang2023successive}, the posterior distribution of the instantaneous channel is computed using a set of channel measurements, which then determines the selection of the next antenna position for measurement. It has been demonstrated that the uncertainty of the channel, defined as the posterior variance, decreases asymptotically as the number of measurements increases. Once this uncertainty falls below a predefined threshold, the posterior mean serves as the Bayesian estimator of the instantaneous channel. On the other hand, machine learning-based methods assume a mapping between measured and unmeasured positions within the antenna movement region \cite{ji2024correlation,zhang2024MLCE}. Typically, the input to the learning network consists of the channel responses at measured positions, while the output predicts the channel responses at all possible positions within the movement region. The network is initially trained offline using a data set that includes CSI for both measured and unmeasured positions. During online deployment, the trained network leverages the measured channel data to predict the channel responses at all unmeasured positions. It is noteworthy that machine learning is not limited to model-free channel acquisition, while it can also enhance model-based approaches. Through careful design and training, machine learning techniques can effectively extract and utilize channel structures and features, significantly improving channel acquisition performance. This highlights a promising direction for future research to further explore the potential of machine learning in MA channel acquisition. Nevertheless, for Bayesian linear regression to compute the joint probability distribution of the entire channel mapping \cite{zhang2023successive,cui2024near}, or for machine learning to create a training dataset for the neural network \cite{ji2024correlation,zhang2024MLCE}, the antenna moving region must consist of discrete ports rather than a continuous region. Furthermore, most of these studies focus solely on Rx-side channel acquisition with 1D antenna movement, leaving the reconstruction of high-dimensional channel mapping involving both Tx-MAs and Rx-MAs moving within 2D/3D regions unexplored.

\subsection{Performance Comparison}
Representative works on channel acquisition for MA systems as well as their respective advantages and limitations are summarized and compared in Table \ref{Tab_channel_CE}. In general, model-based methods can yield relatively low channel measurement overhead and high channel acquisition accuracy if the antenna moving regions have a high dimension or large size, as they depend on the number of Tx and Rx paths rather than the physical size of the antenna moving regions \cite{ma2023MAestimation, xiao2023channel, zhang2024TensorCE}. However, these methods rely on accurate channel structures between the Tx and Rx regions and are sensitive to channel measurement errors in practice. In comparison, model-free methods are robust to modeling and/or measurement errors since they make no assumptions on the channel model, which can achieve a high channel acquisition accuracy for antenna moving regions with low dimensions and small sizes \cite{Skouroumounis2023fluidCE,zhang2023successive,ji2024correlation}. However, they may incur much higher overheads for larger regions, as the number of required channel measurements scales with the size of the antenna moving region. Especially for joint Tx-Rx antenna movement in 2D/3D regions, the feasibility of these model-free approaches and their required channel measurement overhead are still open problems deserving further investigation.

\begin{figure}[!t]
	\centering
	\includegraphics[width=80mm]{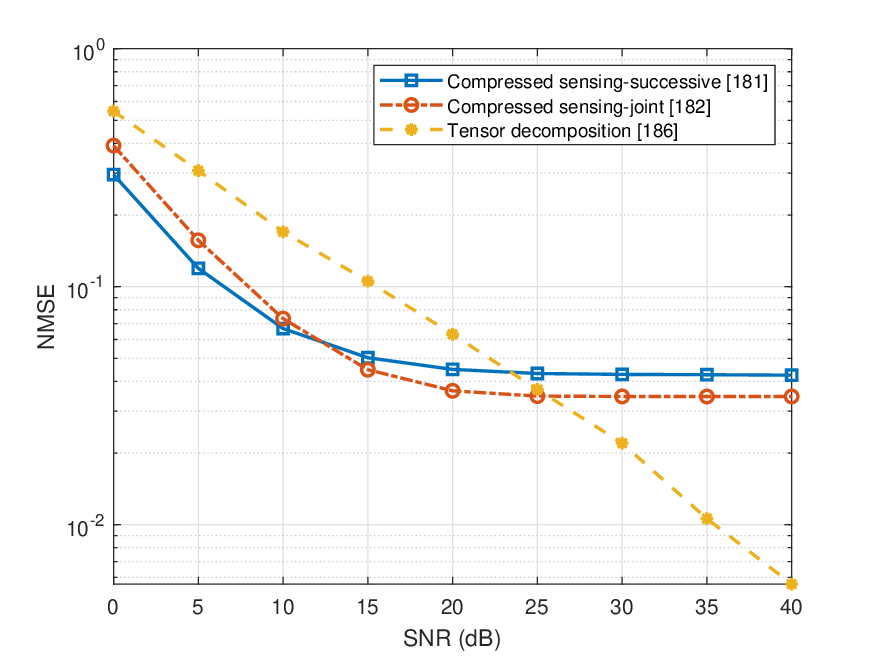}
	\caption{NMSE versus SNR for model-based methods.}
	\label{Fig_CE_model_based}
	\vspace{-12pt}
\end{figure}

To evaluate the performance of model-based channel acquisition methods, we show in Fig.~\ref{Fig_CE_model_based} the normalized mean square error (NMSE) of compressed sensing-based and tensor decomposition-based methods versus the average SNR $P/\sigma^2$. The simulation parameters are the same as that in Fig.~\ref{Fig_MIMO_rate_SNR}, except that $L_{\rm{t}} = L_{\rm{r}} = 3$ and $\kappa=0.5$. We set $N_{\rm{t}}=N_{\rm{r}}=1$ for the compressed sensing-based methods \cite{ma2023MAestimation, xiao2023channel} and $N_{\rm{t}}=N_{\rm{r}}=4$ for the tensor decomposition-based method \cite{zhang2024TensorCE}. The total number of channel measurements is set as $256$ for all methods. To assess channel reconstruction performance, the Tx and Rx antenna movement regions are uniformly divided into multiple grids, with the distance between adjacent grids set as $\lambda/5$. Let $\mathbf{H}$ and $\hat{\mathbf{H}}$ denote the ground-truth and reconstructed channel matrices from the centers of all transmit grids to those of all receive grids, respectively. The NMSE for channel reconstruction is defined as $\textrm{NMSE} = \mathbb{E}\{ \|\mathbf{H}-\hat{\mathbf{H}}\|_{\mathrm{F}}^2/\|\mathbf{H}\|_{\mathrm{F}}^2 \}$, which is obtained by averaging over $10^4$ random channel realizations. As shown in Fig.~\ref{Fig_CE_model_based}, the tensor decomposition-based method achieves a lower NMSE than the compressed sensing-based methods for SNR $>25$ dB. This improvement is due to the limited estimation accuracy of FRI via compressed sensing, which depends on the number of discrete grids of Tx/Rx wave vectors, whereas the tensor decomposition-based method can achieve super-resolution FRI estimation in high-SNR regimes by utilizing the ESPRIT algorithm. Additionally, joint estimation of FRI components via compressed sensing achieves a lower NMSE than the successive estimation method for SNR $>12$ dB; however, the computational complexity of joint estimation scales quadratically with that of successive estimation, as it requires searching across all grids. Thus, there is a fundamental trade-off between computational complexity and channel reconstruction accuracy for model-based methods across different SNR regimes.

\begin{figure}[!t]
	\centering
	\includegraphics[width=80mm]{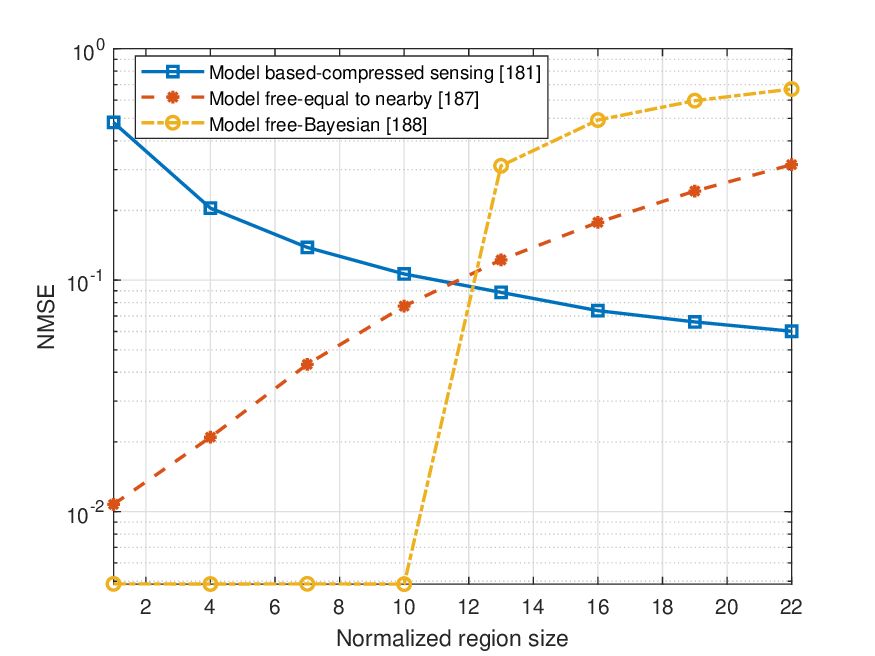}
	\caption{NMSE versus normalized region size for model-based and model-free methods.}
	\label{Fig_CE_model_free}
	\vspace{-12pt}
\end{figure}

Moreover, we show in Fig.~\ref{Fig_CE_model_free} the NMSE of model-based and model-free methods versus the normalized region size. A SISO system is considered, where the Tx is equipped with an FPA and the Rx is equipped with an MA that can be positioned freely along a 1D line segment of length $A$. The total number of channel measurements and SNR are set as $50$ and $20$ dB, respectively, for all methods. The remaining simulation parameters are the same as those in Fig. \ref{Fig_CE_model_based}, except that $L_{\rm{t}} = L_{\rm{r}} = 2$ and $\kappa=1$. We select the compressed sensing with successive FRI estimation as the model-based method \cite{ma2023MAestimation}, and the assumption of unmeasured positions being equal to nearby measured positions \cite{Skouroumounis2023fluidCE} and Bayesian linear regression \cite{zhang2023successive} as two model-free methods. As shown in Fig.~\ref{Fig_CE_model_free}, the model-free methods achieve a lower NMSE compared to the model-based method for smaller antenna movement regions. However, the model-based method outperforms the model-free methods for larger regions. This is because model-free methods rely heavily on adjacent measured positions for reconstructing the channel at unmeasured positions, necessitating a greater number of measurements to densely cover larger and higher-dimensional regions. In contrast, the model-based method can achieve a higher angular resolution as the size of the antenna moving region increases, which thus yields a higher channel acquisition accuracy. In summary, a fundamental trade-off exists between time/measurement overhead and channel reconstruction accuracy for model-based and model-free methods across different region sizes.

% Considering the respective advantages of both the model-based and model-free methods, interested readers are suggested to select appropriate methods under different application scenarios and design constraints. In addition, to overcome their respective limitations, more research efforts are expected for channel acquisition methods to bridge the gap between the model-based and model-free methods such that robust and accurate channel acquisition for MA-aided wireless systems can be efficiently implemented.

\subsection{Lessons Learned and Future Directions}
In this section, we have presented channel acquisition methods for MA systems, focusing on two primary approaches namely model-based and model-free. Model-based approaches leverage the channel structure to recover FRI in the angular domain, offering superior performance for high-dimensional regions with large sizes. In contrast, model-free approaches utilize channel statistical correlations to estimate coefficients in the spatial domain, making them more effective for low-dimensional regions with smaller sizes. It is observed from Table \ref{Tab_channel_CE} that most existing works for MA system channel estimation focus on narrowband systems with point-to-point communication setups. Significant knowledge gaps still exist in areas such as channel acquisition for the general MA system setups with antenna rotation, multiuser channel acquisition, statistical channel acquisition, etc., which are highlighted as potential directions for future research in the following.

\subsubsection{Channel Acquisition with Antenna Rotation}
Most current studies on MA channel acquisition assume that the orientations of antennas are fixed. In the more general MA model with antenna rotation, i.e., 6DMA systems, the PRM is influenced by the AOMs of both Tx-MA and Rx-MA, while the wave vectors of the channel paths couple solely with the MAs' positions. This indicates that the successive FRI estimation could be a promising strategy. Specifically, the wave vectors could first be estimated using existing channel acquisition methods, followed by estimating the PRM with a designed rotation sampling scheme based on prior knowledge of the wave vectors. This strategy could reduce the overall channel acquisition overhead. Developing more efficient channel acquisition algorithms for MA/6DMA systems with antenna rotation remains an important challenge for future research.

% \subsubsection{Performance Analysis}
% Most of the above works focus on the channel acquisition algorithms, where the channel acquisition accuracy is not analyzed. For both model-based and model-free methods, designing the optimal MAs' positions for channel measurement is a crucial research area for minimizing the time overhead of channel measurement. Moreover, investigating how channel acquisition accuracy scales with measurement density and the number of measurements is an important direction for future research.

% \subsubsection{Efficient Channel Acquisition Methods for Obtaining Favorable Channels}
% Most of the above works aim to reconstruct the channel at all discrete/continuous MAs' positions. However, the goal of MA channel acquisition is to find the positions where MAs can be relocated to achieve favorable communication performance, rather than reconstructing the entire channel mapping. Therefore, future research should focus on developing more efficient channel acquisition methods that prioritize finding these optimal positions. The new channel acquisition metric should focus on efficiently finding positions to achieve favorable communication performance, rather than reconstructing the entire channel mapping between the Tx and Rx regions.

\subsubsection{Multiuser Channel Acquisition}
Most existing works on MA channel acquisition focus on MA-aided point-to-point communications. In contrast, channel acquisition in MA-aided multiuser systems poses new challenges in reconstructing all channel mappings for all users. A promising approach to reduce the training overhead is via downlink channel acquisition, where the BS sends common pilots to all users, allowing multiple users to estimate their channels simultaneously and then feed them back to the BS. For uplink channel acquisition in multiuser systems, time/frequency/code division techniques are necessary to mitigate pilot interference from multiple users. However, this results in time overhead and computational complexity for channel acquisition increasing with the number of users. Additionally, the positions of the BS-side MAs should be carefully optimized to balance channel acquisition performance across all users. Investigating efficient channel acquisition methods for MA-aided multiuser systems remains a crucial topic for future research.

\subsubsection{Statistical Channel Acquisition}
So far this section has reviewed existing studies on estimating the instantaneous channel mapping for MA systems. In contrast, statistical channel acquisition for MA systems remains largely unexplored, which presents a significant challenge in accurately modeling the MA statistical channels and efficiently estimating the corresponding statistical information. A straightforward approach involves measuring the instantaneous channels of MA-aided systems over a long time period and averaging the resulting samples to approximate the statistical channel. However, this method incurs substantial time overhead for channel measurement. Hence, future research should prioritize the development of more efficient statistical channel acquisition methods for MA systems, such as estimating the distribution of the AoDs/AoAs and PRMs between the transceivers under specific signal propagation environments.

It is worth noting that the ultimate goal of MA channel acquisition is to help find optimal antenna positions for improving communication performance. In practice, blindly pursuing higher accuracy for channel acquisition may entail excessive overhead in antenna movement and energy consumption. In practice, an advisable way is to develop more efficient channel acquisition algorithms that prioritize increasing channel accuracy at candidate positions with potentially favorable channel conditions. Therein, only necessary CSI is extracted to enable antenna movement optimization, rather than reconstructing the entire channel mapping between Tx and Rx regions. Based on this idea, an initial work \cite{zeng2024csi} attempted to realize antenna position optimization by leveraging the received SNR measurements during antenna movement, without the need for complete CSI. However, only the simple setup of a single Rx-MA was considered. Future works may consider more general system setups under different practical environments by developing customized protocols and algorithms to achieve the best trade-off between channel acquisition and communication performance.

\section{Prototypes, Experiments, and Standardization} \label{Sec_Prototype}
This section provides an overview of the existing prototypes for MA-enabled wireless sensing and/or communication systems, as well as their experimental results to validate the performance gains practically achievable by antenna movement.

\subsection{Wireless Localization and Sensing}
Some earlier works were reported to implement MAs or similar concepts for wireless localization and sensing \cite{li2022using,zhuravlev2015experimental, hinske2008using, 2011doppler}.
%applications other than wireless communications, such as radar \cite{zhuravlev2015experimental}, navigation \cite{li2022using}, localization \cite{zhuravlev2015experimental}, etc.
The works in \cite{li2022using,zhuravlev2015experimental,hinske2008using} utilized the mechanical slide-based MA to develop the prototypes of wireless localization and sensing systems.
Specifically,
the authors in \cite{li2022using} proposed an MA-based integrated inertial navigation system (INS)/global navigation satellite system (GNSS) method, which can improve the heading angle estimation accuracy by more than 50\% compared to the conventional FA system.
The authors in \cite{zhuravlev2015experimental} showed a prototype of MA-aided multi-static radar, where the Tx and Rx antennas can be moved in a line segment with linear drives, such that the antenna configuration can be fine-tuned to balance the trade-off between system complexity and radar imaging quality.
In addition, in \cite{hinske2008using}, a prototype of MA-based wireless localization system using RF identification (RFID) technologies was developed, which can achieve a high positioning accuracy in the order of millimeters.
Different from the linear slides in \cite{li2022using,zhuravlev2015experimental,hinske2008using}, the authors in \cite{2011doppler} used a motor-driven rotation-type MA to implement indoor positioning.
The results showed that their developed MA-based wireless position prototype can achieve decimeter-level positioning accuracy \cite{2011doppler}.

%The authors in \cite{zhuravlev2015experimental} utilized the developed MA prototype to realize an electronically switched sparse antenna array for acquiring radar signals.

\begin{figure}[t]
	\centering
	\includegraphics[width=0.95\columnwidth]{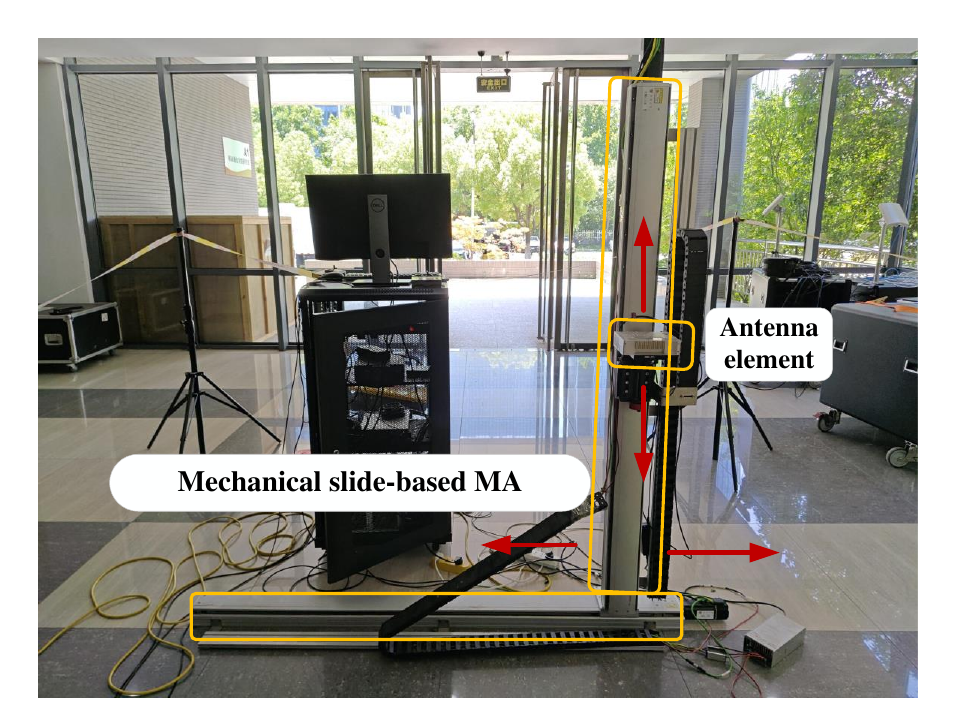}
	\caption{Mechanical slide-based MA communication prototype developed by Southeast University in \cite{dong2024movable}.}
	\vspace{-12pt}
	\label{Fig_prototype}
\end{figure}

\begin{figure}[t]
	\centering
	\includegraphics[width=0.95\columnwidth]{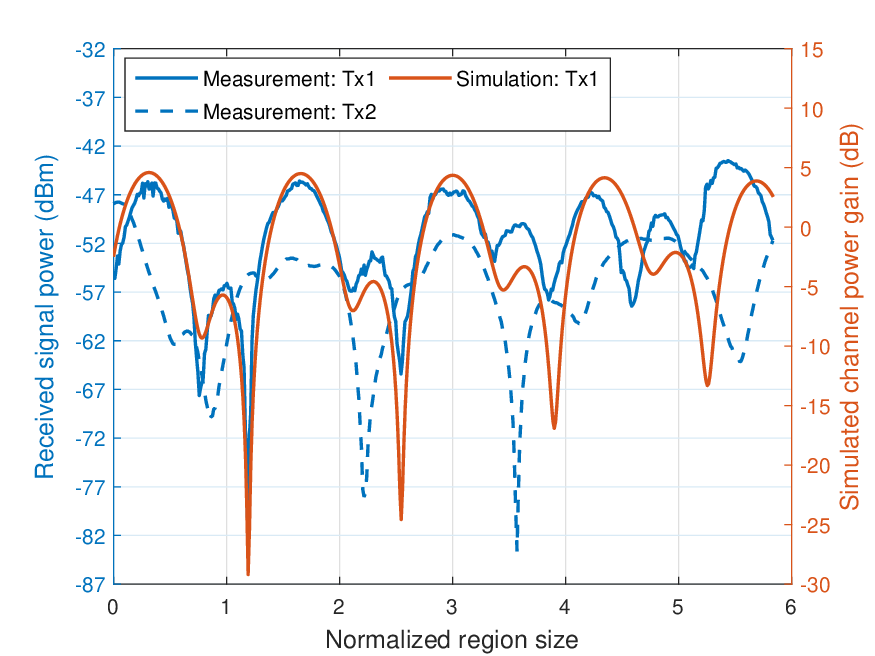}
	\caption{The measured signal power (in dBm) and the simulated channel power gain (in dB) at the carrier frequency of 3.5 GHz \cite{dong2024movable}.}
	\vspace{-12pt}
	\label{Fig_1D_gain}
\end{figure}

\begin{figure}[t]
	\centering
	\subfigure[Measurement results]{
		\begin{minipage}[c]{0.45\textwidth}
			\centering
			\includegraphics[width=\textwidth]{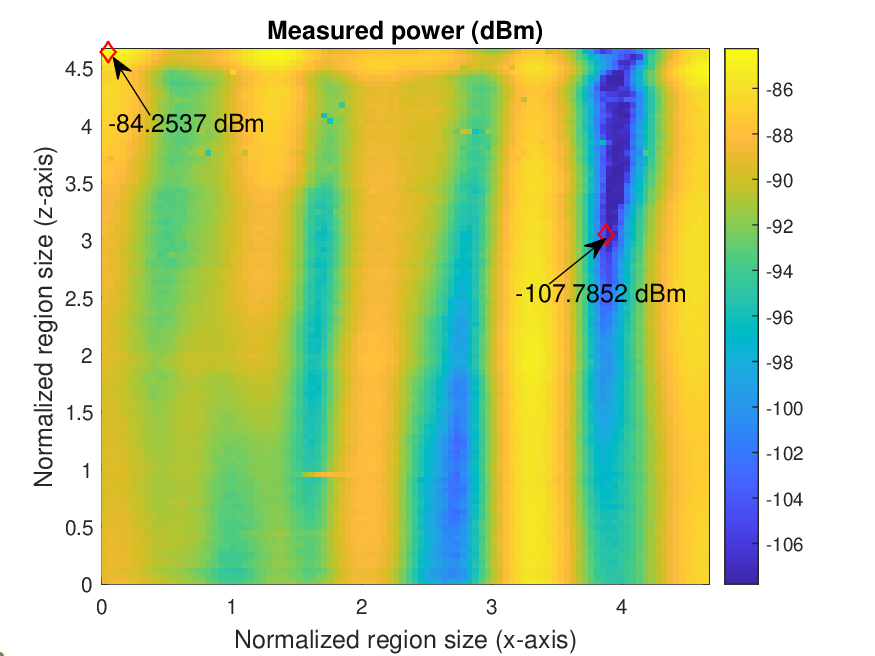}
	\end{minipage}}
	\subfigure[Simulation results]{
		\begin{minipage}[c]{0.45\textwidth}
			\centering
			\includegraphics[width=\textwidth]{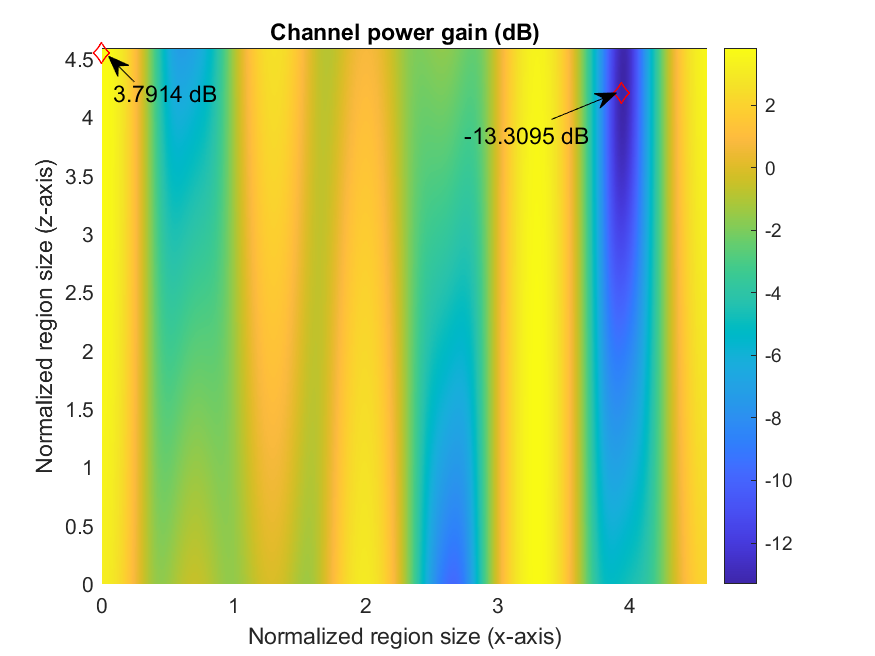}
	\end{minipage}}
	\caption{The measured received signal power (in dBm) and the simulated channel power gain (in dB) at the carrier frequency of 27.5 GHz \cite{dong2024movable}.}
	\vspace{-12pt}
	\label{Fig_2D_gain}
\end{figure}

\subsection{Wireless Communication}
%Despite extensive theoretical studies on MA-based wireless communications based on mechanical slide, the promised performance gain of mechanical slide-based MA has not been validated in practical systems yet.
Despite extensive theoretical research on MAs for wireless communications, the performance gains of MA-based communication systems were only validated experimentally recently \cite{dong2024movable,  wang2024movable, shen2024design,  zhang2024pixel}.
In particular, the authors in \cite{dong2024movable} developed a prototype of mechanical slide-based MA communication system capable of ultra-accurate movement control, with a positioning accuracy of $0.05$ millimeters, as illustrated in Fig. \ref{Fig_prototype}. Specifically, the Tx and Rx are equipped with an FPA and an MA, respectively, where the MA can be flexibly moved within a given 1D/2D region. The received signal power of the MA-aided SISO communication system was measured at $3.5$ GHz and $27.5$ GHz, where the MA moved along a 1D horizontal line with the step size of $0.01\lambda$ and within a 2D square region with the step size of $0.05\lambda$, respectively. The measurement results were compared with simulation results, which utilized the estimated path state information (PSI) obtained from the measurement data, including the number of Rx paths, their delays, elevation and azimuth AoAs, as well as the power ratio of each path. 
For carrier frequency of $3.5$ GHz, the maximum variation of the measured signal power reaches over $40$ dB within the $6\lambda$ range, as illustrated in Fig. \ref{Fig_1D_gain}, where the maximum and minimum values are $-43.7$ dBm and $-84.1$ dBm, respectively. For carrier frequency of $27.5$ GHz, the maximum variation of the measured signal power is about $23$ dB, as shown in Fig. \ref{Fig_2D_gain}(a), where the maximum and minimum power are $-84.3$ dBm and $-107.8$ dBm, respectively.
Furthermore, by comparing the measurement results and simulation results at both $3.5$ GHz and $27.5$ GHz, it was found that the simulation results match the measurement results well, while the minor difference between them may be caused by the PSI estimation error.
These experimental results in \cite{dong2024movable} revealed that the MA technology has a great potential in improving wireless communication performance via local antenna movement and the corresponding performance gain highly relies on the estimated PSI.

\begin{table*}[t]\small 
	\caption{Prototypes of movable and reconfigurable antennas for wireless sensing and communications.}\label{Tab_Prototypes}
	\newcommand{\tabincell}[2]{\begin{tabular}{@{}#1@{}}#2\end{tabular}}
	\centering
	\begin{tabular}{|c|c|c|c|}
		\hline
		\textbf{Objective} &  \textbf{Ref.}  &           \textbf{Implementation}        &   \textbf{Key findings}\\
		\hline
		
		\multirow{7}{*}{\makecell{Localization \\ and sensing}}  &
		\cite{zhuravlev2015experimental} & \multirow{7}{*}{Motor} & \tabincell{c}{Balancing trade-off between system
			\\ complexity and radar imaging quality}\\
		\cline{2-2}\cline{4-4}
		~ & \cite{li2022using}  & ~ & \tabincell{c}{Improving the heading angle estimation  \\  accuracy by more than 50\%} \\
		\cline{2-2}\cline{4-4}
		~ & \cite{hinske2008using}  & ~ &  \tabincell{c}{ Achieving a high positioning accuracy \\  in the order of millimeters}\\
		\cline{2-2}\cline{4-4}
		~ & \cite{2011doppler}  &   & \tabincell{c}{Achieving decimeter-level positioning accuracy } \\
		\hline

		\multirow{7}{*}{\makecell{Wireless \\communication}}  & \cite{dong2024movable} & \multirow{3}{*}{Motor}  & \tabincell{c}{Demonstrating maximum variation
			\\of received signal power up to 40 dB \\ at 3.5 GHz and 23 dB at 27.5 GHz}\\
		\cline{2-2}\cline{4-4}
		~ & \cite{wang2024movable}  & ~  & \tabincell{c}{Coping with multi-path fading issues;\\ Enhancing
			spectral efficiency up to 10\%} \\
		\cline{2-4}
		~ & \cite{shen2024design}  &  Fluid  &  \tabincell{c}{Reducing outage probability by 57\%; \\ Increasing multiplexing gain  to 2.27}\\
		\cline{2-4}
		~ & \cite{zhang2024pixel}  &  Pixel & \tabincell{c}{Achieving around 30 dB signal power variation; \\ Validating feasibility of multiple access} \\
		\hline
	\end{tabular}
\end{table*}

\begin{figure}[t]
	\centering
	\includegraphics[width=0.88\columnwidth]{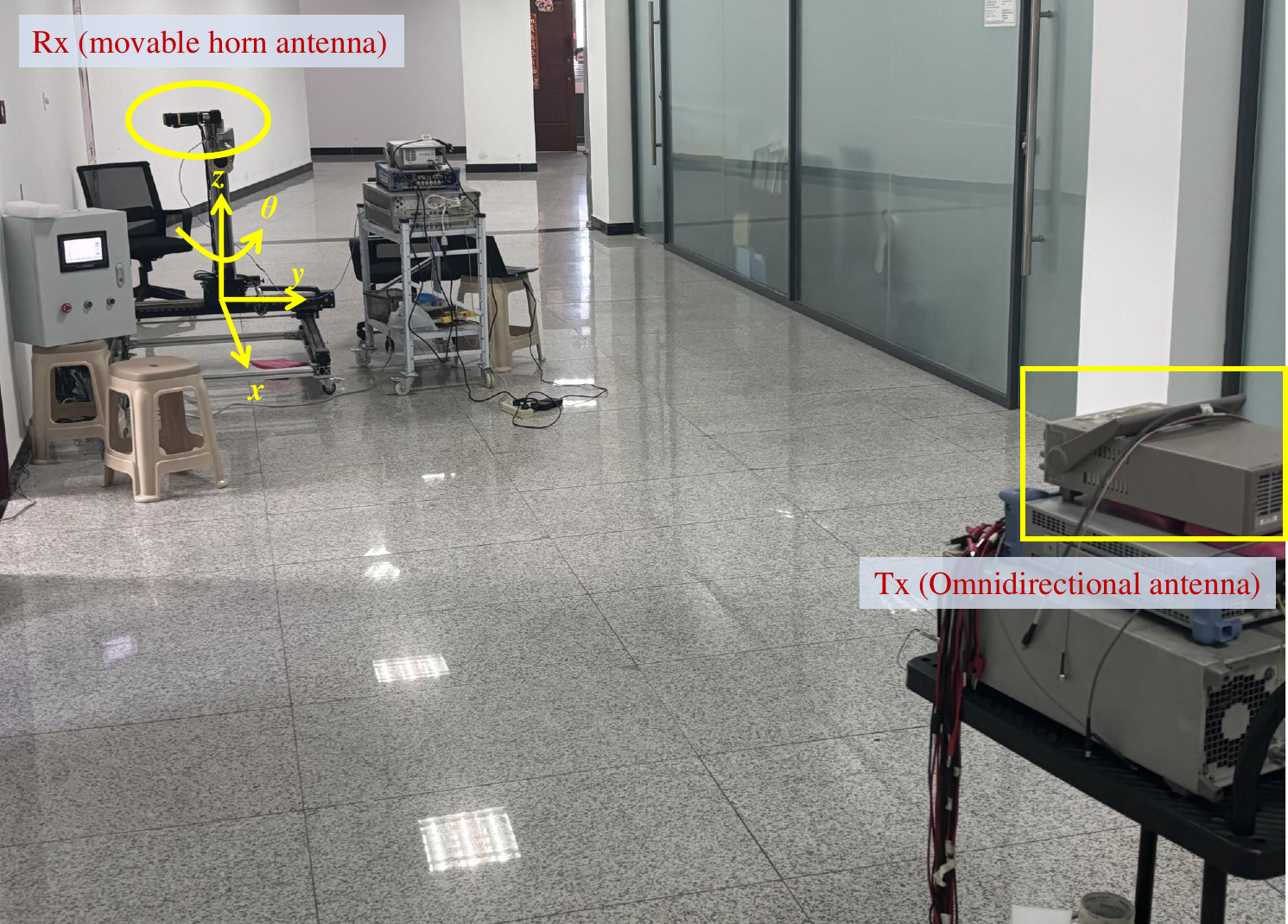}
	\caption{Mechanically-driven 6DMA communication prototype developed by UESTC.}
	\vspace{-12pt}
	\label{4DExperiment}
\end{figure}
\begin{figure}[t]
	\centering
	\includegraphics[width=0.95\columnwidth]{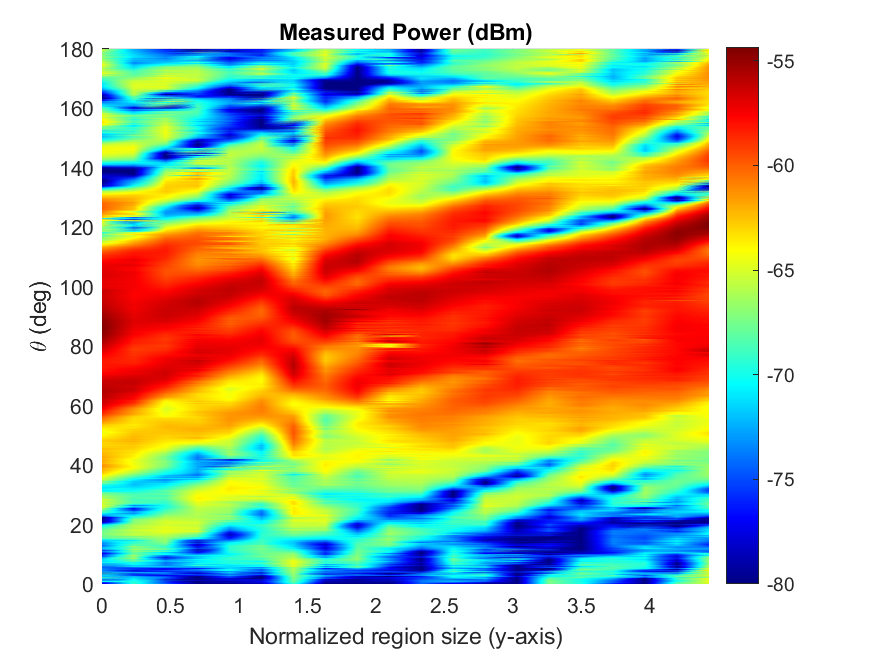}
	\caption{The measured signal power (in dBm) with the 6DMA prototype at the carrier frequency of 3.5 GHz.}
	\vspace{-12pt}
	\label{4DMA_result}
\end{figure}
Furthermore, Yafei Wu and Weidong Mei et al. from the University of Electronic Science and Technology of China (UESTC) have recently developed a mechanically-driven 6DMA prototype for 6DMA systems \cite{shao20246DMA,shao2024discrete,shao2024Mag6DMA}, as depicted in Fig.\,\ref{4DExperiment}. In this prototype, the 3D position (axes $x$, $y$, and $z$) and 1D orientation ($\theta$) of a horn antenna can be altered mechanically, allowing its main lobe to align with different spatial directions. Fig.\,\ref{4DMA_result} presents the experimental results for this 6DMA prototype in an indoor environment as in Fig.\,\ref{4DExperiment}, showing the received signal power (in dBm) for different values of $y$ and $\theta$. The horn antenna is used as an Rx with the main-lobe beam gain of $12$ dBi. The transmit power is $0$ dBm, and the carrier frequency is $3.5$ GHz. It is observed that for a fixed antenna position along the $y$-axis, rotating the horn antenna (i.e., changing $\theta$) results in fluctuations in the received signal power with a maximum difference exceeding $60$ dB. The peak signal power occurs when $\theta$ is between $60^{\circ}$ and $100^{\circ}$. These on-site measurement results suggest that optimizing antenna orientation, in addition to position, provides additional DoFs to improve channel conditions.

In addition, it was shown in \cite{wang2024movable} that the MA-based communications system outperforms the FA system in coping with multi-path deep fading, with an improvement in spectral efficiency by up to $10$\% at the frequency carrier of 300 GHz. In addition, the prototypes of wireless communication systems based on fluid antennas \cite{shen2024design} and pixel antennas \cite{zhang2024pixel} have also been developed. For a 4-user system \cite{shen2024design}, experimental results showed that the double-channel fluid antenna can significantly reduce the outage probability by $57$\% and increase the multiplexing gain to $2.27$, compared to a static omnidirectional antenna.
For the pixel-based reconfigurable antenna communication system prototype \cite{zhang2024pixel}, measurement results demonstrated that up to around $30$ dB signal power boost at $2.5$ GHz can be achieved, which validated the feasibility of a 2-user multiple access system aided by pixel antennas. In summary, different prototypes of MAs/6DMAs and reconfigurable antennas as well as their key findings are provided in Table \ref{Tab_Prototypes}.

\subsection{Standardization Issues}
The transition of a technology from theoretical research to prototype development and eventually to the formulation of standards typically spans over a decade. As the MA technology is an emerging field, there are currently no established standards dedicated to MA-enabled systems. To incorporate MA-based systems into 3GPP standards in the 6G era, such as Release 21 and beyond, it will be essential to address standardization of key technical aspects. These include MA-enabled channel modeling, movement management, channel estimation, and other critical functionalities.

\subsubsection{Standardization of Channel Modeling}
The existing 3GPP TR 38.901 (see Section 7.1.3) has already accounted for the effects of array pose changes, such as bearing, downtilt, and slant, on the channel model, with simulation parameters for the downtilt angle detailed in \cite[Table 7.8-1]{3gpp2019studyo}. For MA/6DMA-enabled systems, the antenna's position distribution, orientation, and polarization direction become highly flexible, introducing additional DoFs. Therefore, it is essential to extend the existing model to develop an MA-based channel model with higher-dimensional parameters, e.g., the field-response channel model. Such an advanced model will be instrumental in guiding link-level and system-level simulations, offering critical insights for network design and deployment.

\subsubsection{Standardization of Antenna Movement Management}
For MA-enabled systems, the antenna movement management will require thorough discussions regarding the number of permissible movement modes, the minimum time unit for antenna movement, and the periodicity of such movements. For instance, the existing 3GPP TS 38.214 (see Section 5.2.2.2) specifies the number of supported ports, i.e., 4, 8, 12, 16, 24, and 32 (as of Release 18), for gNodeB during downlink channel measurements \cite{3gpp2024Physicallayer}. Similarly, an MA-enabled system will need to define the allowable number of movement modes and their associated  identifiers. Additionally, the current 3GPP TS 38.211 (see Section 4.3.2) establishes frames, subframes, and slots as time units. An MA-enabled system will need to define a minimum time unit and allocate time resources for antenna movements, akin to Physical Random Access Channel occasions (see TS 38.211, Table 6.3.3.2-3) \cite{3gpp2020Physicalchannels}. Regarding the periodic behavior of antenna movements, these could be categorized as `periodic', `aperiodic', or `semi-persistent' (see TS 38.331, Section 6.3.2). For aperiodic or semi-persistent behaviors, a Medium Access Control Control Element or Downlink Control Information could be configured to trigger antenna movements at the gNodeB or user, while the user might feedback the control signal to the gNodeB to adjust its antenna movement modes through Uplink Control Information \cite{3gpp2022Radio}. Defining these parameters will be crucial to ensuring effective and efficient movement management within standardized frameworks.

\subsubsection{Standardization of Channel Estimation}
The standardization of channel estimation for MA-enabled systems could draw inspiration from the 5G NR beam-based channel estimation framework. According to 3GPP TS 38.331 (see Section 6.3.2), gNodeB specifies the channel measurement behavior via ``CSI-ResourceConfig'' on the Radio Resource Control, where each resource facilitates a channel measurement \cite{3gpp2022Radio}. In this framework, the BS sequentially scans through analog beams, configuring a CSI-Reference Signal resource for each beam. The user then reports back the preferred analog beam along with measured metrics such as Channel Quality Indicator, Rank Indicator, and Precoding Matrix Indicator (as detailed in TS 38.212, Table 6.3.1.1.2) \cite{3gpp2020Multiplexing}. Similarly, channel estimation for MA-enabled systems would require a tailored ``ResourceSetList'' to enable measurements under various antenna movement modes. The user would measure the channel under different modes and feed back either instantaneous or statistical information. Additionally, the feedback to the BS should include mode-specific beamforming information and details about the antenna's mode behavior. This extended framework would accommodate the unique challenges and opportunities presented by the dynamic antenna mode configurations in MA-enabled systems, ensuring robust channel estimation for both communication and sensing applications.

\subsection{Lessons Learned}
In this section, we have reviewed existing prototypes of MA/6DMA systems, focusing on their channel models and performance gains validated through real-world experiments. As summarized in Table \ref{Tab_Prototypes}, most current prototypes implement a single antenna with reconfigurable position and/or orientation. These additional DoFs in antenna movement have demonstrated significant improvements in both sensing and communication performance. Future research should explore the development of multi-MA systems to validate their advantages in flexible beamforming and spatial multiplexing. Additionally, efforts should be directed toward developing cost-effective system architectures and establishing general and standardized frameworks to facilitate the practical implementation and widespread commercialization of this technology.

\section{Extensions and Future Directions} \label{Sec_Extension}
In this section, we explore several key extensions of MA-aided wireless systems to unlock other potential applications and discuss their associated design issues to inspire future studies in this area. 
\subsection{Space-Air-Ground Integrated Networks}
MA arrays can play a significant role in space-air-ground integrated networks by enabling more flexible beamforming for various applications. For instance, due to the long propagation distances and high operating frequency bands of satellite-ground links, LEO satellites are typically equipped with high-gain antenna arrays to compensate for significant path losses. However, conventional FPA arrays usually face challenges such as severe coupling effects and heat dissipation in dense arrays, as well as undesirable sidelobes and interference leakage in sparse arrays. Additionally, as an LEO satellite orbits the Earth, the directions of its coverage and interference areas relative to the antenna array change over time. Since traditional FPA arrays cannot adjust the correlation between steering vectors over coverage and interference directions, only the antenna weights can be reconfigured to meet the varying coverage requirements of terrestrial users, which limits the DoF in beamforming. To address this issue, the authors in \cite{ZhuLP_satellite_MA} proposed using an MA array mounted on LEO satellites. This approach aims to minimize the average signal leakage power to interference areas while maintaining the minimum beamforming gain over coverage areas by optimizing MAs' positions jointly with their beamforming weights over time. Numerical results showed that compared to FPA arrays, the MA array can significantly reduce interference leakage and increase the average signal-to-leakage power ratio of satellite-ground links. This performance improvement can be further enhanced by incorporating additional orientation optimization of the MA array, particularly when antenna positions are not frequently adjusted. Thereafter, the authors in \cite{lin2024power} investigated the MA array-aided full-duplex satellite communications. The results demonstrated that the joint optimization of MAs' positions at both the Tx and Rx can effectively mitigate self-interference and multiuser interference, thereby saving the transmit power of the satellite. 

With the advancement in unmanned aerial vehicle (UAV) payload capacity and the miniaturization of communication devices, UAVs can also be integrated into terrestrial cellular networks as aerial communication platforms (e.g., BSs or relays \cite{zeng2019accessing, zhu2020UAVRelay}) or aerial users (known as cellular-connected UAVs \cite{mei2021aerial,zeng2019cellular}). By leveraging the 3D mobility of UAVs, their positions can be flexibly adjusted within broad areas to enhance communication performance. This has spurred the interest in UAV-mounted MAs, combining the large- and small-scale channel reconfigurations offered by UAVs and MAs, respectively. In complex environments such as urban areas, UAVs can address signal blockages caused by dense obstacles through changing their large-scale positions, while MAs can achieve more efficient interference mitigation by reconfiguring the correlation of channel vectors via small-scale movement. Motivated by these advantages, the authors in \cite{tang2024uav} explored the multi-beam forming problem for UAV-mounted MAs subject to interference power constraints over multiple directions, where the altitude of the UAV and the positions of MAs were jointly optimized for achieving more flexible beamforming. Besides, the authors in \cite{liu2024uav} addressed a more challenging problem by jointly optimizing the MAs' positions and UAV trajectory over time to maximize the sum rate of ground users. Their results demonstrated the advantages of dynamic antenna position optimization during the flight. Moreover, the authors in \cite{ren20246DMAUAV} studied the cellular-connected UAV communications aided by 6DMA arrays. By jointly optimizing the antenna positions, array orientation, and receive beamforming at the UAV, the SINR can be significantly improved as compared to conventional FPA-aided systems. A UAV-enabled passive 6DMA system was investigated in \cite{liu2024uav6DMA} by exploiting the angle-dependent reflection of an IRS and the controllable mobility of a UAV to enhance the communication performance. Despite the promising benefits of UAV-mounted MAs, several new challenges also arise. For example, the performance of MA-aided UAV communication systems may be affected by wind disturbances, leading to inaccurate antenna positions/orientations and thus more robust designs need to be investigated in future work.

\subsection{Over-the-Air/Mobile Edge Computing}
Over-the-air computation (AirComp) has emerged in recent years to meet the growing demand for distributed data collection from massive IoT devices \cite{zhu2021AirComp}. By enabling simultaneous data transmission from multiple devices and utilizing the superposition property of electromagnetic waves, AirComp allows for the rapid computation of a large amount of data at the sink node. In this context, MAs can be adopted to enhance wireless data aggregation by flexibly reshaping wireless channels between IoT devices and the sink node \cite{cheng2023movableAirComp}. In light of this, the authors in \cite{zhang2024AriComp} studied the MA array-aided AirComp systems with 1D antenna movement, demonstrating that MAs can significantly reduce the computation mean square error (CMSE) compared to conventional FPAs. Therein, an AO-based algorithm was developed to obtain suboptimal solutions for the positions of MAs. Building on this, a more recent study \cite{li2024over} further investigated the MA array-enhanced AirComp system with 2D antenna movement. To fully exploit the DoF in antenna position optimization, a PSO-based two-loop iterative algorithm was proposed, which can outperform the AO algorithm in minimizing the CMSE.

In addition, mobile edge computing (MEC) is a distributed computing technology that conducts data processing and storage at nodes close to the data source, thereby reducing network latency and bandwidth usage. Instead of relying on centralized cloud servers, MEC performs tasks at the edge of the network and thus enables faster responses in real-time applications. In \cite{xiu2024delayMAMEC, xiu2024latencyMAMEC, zuo2024fluid}, the authors showed that thanks to the capability of improving wireless channels, MA-aided MEC systems can significantly reduce the overall delay compared to those based on FPAs. Furthermore, the authors in \cite{ChenPC_MA_WPT_MEC} investigated a sophisticated wirelessly powered MEC system, where an MA-enabled hybrid access point first wirelessly charges devices and then receives their offloaded tasks. In this work, three types of MA configurations were proposed to balance the computational performance and implementation complexity, by allowing different degrees of flexibility for antenna movement in the power transfer and task offloading phases. Both studies \cite{zuo2024fluid, ChenPC_MA_WPT_MEC} underscore the potential of MAs to improve MEC performance in terms of either reducing system delays or boosting computation rates.

\subsection{Physical Layer Security}
Physical layer security (PLS) has been increasingly recognized as a crucial approach for secure wireless transmissions between source and destination. Given the inherent vulnerabilities of wireless signals to secure threats, a primary challenge for PLS lies in ensuring a high secrecy rate for the legitimate Rx while preventing the eavesdropper (Eve) from intercepting confidential information. However, due to the fixed feature of traditional FPA systems, they cannot fully utilize the spatial DoF, resulting in weakened secure transmission performance. In contrast, MAs have appealing benefits in exploiting the spatial variation of wireless channels through movement reconfiguration, which brings new opportunities for PLS communications. In the following, we discuss the MA-aided secure communication systems under different CSI availabilities of Eves.

If the Eves remain stationary and active, obtaining the CSI of these external devices is possible at the legitimate Tx. In such cases, security issues can be alleviated by leveraging the inherent flexibility of MAs and optimizing resource allocation. For instance, by appropriately adjusting the antennas' positions, an MA array can enhance the channel quality between the Tx and the legitimate Rx while simultaneously degrading the link to the Eves, thus significantly improving the secrecy rate. Even if the Eves and the legitimate Rx are located in close directions w.r.t. the Tx, their channel correlation can be effectively reduced by optimizing the positions of MAs \cite{hu2024secure,cheng2024secure,tang2024secure,DingJZ_MA_FD_secure_1,mao2024MAcovert,wang2024MAcovert,Xie2024MARIScovert,Liu2024MAcovert}. In particular, the authors in \cite{mei2024movable_secure} proposed a discrete sampling approach for a MISO secure communication system and obtained an optimal solution for antenna positioning in a recursive manner. To further ensure the security of the transmission, an effective approach is transmitting artificial noise (AN) to prevent the interception of Eves. Through joint optimization of the AN beamformers and the positions of MAs, the interference to unauthorized Eves can be intensified, thereby reducing the leakage of confidential information \cite{tang2024secure,DingJZ_MA_FD_secure_1,DingJZ_MA_FD_secure_2,liao2024MAjamming}. In addition, recent studies have demonstrated that by exploiting the unique advantages of MAs, the covert communication performance can be significantly enhanced, thus achieving higher-level PLS \cite{mao2024MAcovert,wang2024MAcovert,Xie2024MARIScovert,Liu2024MAcovert,cheng2024MAfrequency}.

On the other hand, if the Eves are passive or moving in certain areas, obtaining their perfect CSI is challenging. In this context, secrecy outage probability can be employed as a performance metric by utilizing the statistical CSI, and it can be significantly decreased by jointly optimizing the transmit beamforming and the positions of MAs at the Tx \cite{HuGJ_secure_CSIfree_MA}. Additionally, considering the mobility of the Eve, the authors in \cite{FengZY_MA_secure} proposed a framework for modeling the imperfect CSI of the Eve using the concept of a virtual MA, which captures all potential positions of the Eve by leveraging the flexibility of the MA. Similar idea has also been adopted in \cite{cheng2024MAnoCSI} to minimize the transmit power of confidential signals. Despite the aforementioned advantages, there are still many technical issues remaining to be solved. For instance, existing literature primarily focused on scenarios where both legitimate users and Eves are equipped with a single FPA or only one user/Eve is equipped with multiple FPAs. However, the design of MA-aided MIMO secure communication systems with multiple legitimate users and Evas have yet to be explored, which requires more sophisticated system modeling and design. Additionally, antenna movement optimization at the network level with multiple distributed access points and multiple eavesdroppers is also a challenging problem that deserves further investigation.

In addition, the introduction of MAs to wireless communication systems may also increase the risk of information leakage if they are misused by malicious individuals. A recent study has exposed this concern, where the MA-enhanced jamming system can destroy the legitimate links dramatically \cite{maghrebi2024movable}. On the other hand, the capability of Eves can also be enhanced with the aid of MAs. In this regard, effective regulations and countermeasures are required to guarantee the secure performance of wireless systems. To this end, advanced transceivers based on MAs can be developed to detect the locations of Eves/jammers and then perform physical attack to destroy hostile nodes. Alternatively, flexible beamforming enabled by MA arrays can be employed to minimize the signal leakage to the directions of Eves or the interference from the directions of jammers.

\subsection{Wireless Information and Power Transfer}
Wireless power transfer (WPT) facilitates perpetual energy supplies for massive low-power devices in the forthcoming IoT networks \cite{xu2014power}. MA serves as a promising technology to enhance the energy transmission for WPT due to its capability of channel condition improvement \cite{ZhangL_FAS_WPT,ZhouLS_FAS_SWIPT,XiaoJH_MA_RIS_WPT,Farshad_FAS_NOMA_WPT}. Specifically, by simply positioning the MA where the channel correlation among IoT devices is highest, the optimal transmission in a standalone WPT scenario can be easily achieved. Furthermore, with the advancement of RF devices, WPT in multipath propagation environments has transitioned from the conceptual stage to practice \cite{HoangTQV_WPT_hardware}, making it possible for MA to transmit energy signals leveraging the characteristics of multipath channels.

However, the objective of WPT usually conflicts with that of information transmission, making their coordination and trade-off a challenging problem when utilizing MA in simultaneous wireless information and power transfer (SWIPT) \cite{xu2014power}. Specifically, under the condition of high channel correlation among devices, strong interference signals contribute to higher received power for energy harvesting but deteriorate communication performance. Thanks to the new spatial DoF offered by MA, it provides novel solutions distinct from traditional beamforming and power allocation, e.g., by moving a single MA to a designated position to balance the trade-off between SINR for information reception and received power for energy harvesting \cite{Psomas_WPT_survey}. Moreover, multiple MAs can be moved to different locations to optimize each objective separately, with some antennas positioned to maximize SINR and others to maximize received power \cite{WongKK_FAS2}. In the aforementioned solutions, determining the optimal positions for MAs is essential, which has spurred a variety of recent investigations \cite{Christodoulos_FAS_SWIPT_1,LaiXZ_FAS_WPT,Christodoulos_FAS_SWIPT_2}. Additionally, the authors in \cite{LinX_FAS_WPT_1} and \cite{LinX_FAS_WPT_2} respectively designed time-switching and power-splitting schemes for SWIPT aided by MA, and extended the previous works to more sophisticated multi-BS networks.

Despite the current advancements in MA-aided WPT/SWIPT, several challenges and open problems still remain. First, since the mechanical movement of antennas introduces additional power consumption, it is imperative to account for the overall energy budget of the MA-assisted WPT/SWIPT system to improve energy efficiency. Thus, given the difficulties in deploying motor or servo-driven MA in low-power devices, it is crucial to consider electronically reconfigurable antenna schemes at the Rx, e.g., dual-mode antennas. Next, the theoretical performance limits of MA in SWIPT systems have not been fully revealed, and more in-depth investigation is thus needed. Furthermore, the design and analysis of SWIPT in MA-assisted MIMO systems is also a critical topic, which has yet to be thoroughly explored.

\subsection{Synergy of MA and IRS}
IRS possesses the capability to reconfigure wireless propagation environments and enhance signal coverage \cite{wu2024ISfor6G, WuQQ_IRS_magazine, mei2022intelligent, zheng2022survey}. It plays a pivotal role in repairing and extending direct links required for future wireless systems operating at higher frequencies, thereby reducing the need for costly BS densification. However, existing IRS-aided systems only exploit the potential of channel variations in signal propagation, with insufficient utilization of multipath channel characteristics at the transceivers. Furthermore, the doubly fading of cascaded channels in IRS-aided systems leads to significant power attenuation of the desired signal and its subsequent performance degradation. The introduction of MA to IRS-aided systems, which can reshape channel conditions, effectively overcomes the above issues by significantly enhancing the desired signal. Additionally, MA systems also benefit from the deployment of IRS, as IRS can be used as artificial scatterers to recreate more favorable multipath scattering conditions, allowing MA to utilize more channel paths and their spatially-varying superpositions \cite{wong2022bruce,Arman_FAS_RIS_survey,WongKK_FAS3_RIS}.

To characterize the theoretical performance of MA-IRS-aided wireless networks, several existing works have initially established the model and analyzed its performance \cite{Farshad_FAS_RIS,LaiXZ_FAS_RIS,Alireza_FAS_RIS_IT,yao2024fasRIS,zheng2024paving}. From the perspective of MA-IRS-aided system design, the coupling and interaction between MA positions and IRS phase shifts make their joint design an intractable problem \cite{SunYN_MA_RIS,MengKT_FAS_RIS_ISAC_survey,WuHS_MA_RIS_ISAC,WangC_FAS_RIS_AI_survey,Farshad_FAS_RIS_secure}, which demands for more efficient algorithms to jointly optimize a myriad of parameters. Generally, the above studies based on instantaneous CSI between MA and IRS, and between IRS and users, encounter significant implementation challenges and complexities, highlighting the imperative need for finding more advanced channel estimation methods, which remains to be addressed in the future. Fortunately, since the full array gain can be achieved with tunable main-lobe and grating-lobe provided by MA arrays, a closed-form solution for joint MA position and beamforming design under statistical CSI is possible \cite{ChenJG_FAS_RIS}. Another simplified scheme adopts random phase shifts and treats IRSs as ordinary scatterers in the propagation environments, enabling system analysis and optimization without CSI \cite{YaoJT_FAS_RIS_throughput}. However, this approach requires a large number of IRS elements \cite{wong2022bruce}. Given that both MA and IRS can improve channel conditions, one may wonder which has a greater impact on the performance of the system they compose. A recent study discussed this issue in a system where MA collaborates with active IRS \cite{YaoJT_FAS_ARIS}. Results showed that MA affects the system performance more in simple environments with fewer IRS reflecting elements or limited channel paths, while active IRS becomes more critical in the opposite case. Nevertheless, the performance gain of MAs in IRS-aided systems lacks in-depth theoretical analysis. A recent study \cite{WeiX_MA_RIS} has made an initial attempt to address this issue under the conditions of a single user and an LoS channel between the BS and IRS. The findings indicated that the performance gain of MAs over FPAs may diminish in scenarios with a single-MA BS or under the far-field BS-IRS propagation. Nonetheless, more in-depth and comprehensive research is needed to explore more general system setups for IRS-MA integrated design.

It is worth mentioning that another line of research exists, which is different from the aforementioned synergy of MA and IRS. These works endow the mobility characteristic of MA to IRS, allowing each element of IRS \cite{Hu2024IRSmoveable,ZhangY_MIS,zhao2024movableSTARS} or the entire IRS \cite{Ibrahim_MIS, cheng2022RISrotate} to move freely in a continuous spatial field. Furthermore, considering IRS as rotatable in future designs introduces new DoF and potential performance enhancement \cite{cheng2022RISrotate, chen2024rotatable,feng20256DIRS,lu2024wireless,han2024movableSuper}. Despite the aforementioned promising directions, there are still several unresolved issues and challenges. First, the effectiveness of MA in enhancing IRS-aided systems awaits further investigation. For instance, it is still unknown whether integrating static IRS with MA can achieve performance comparable to that of using dynamic IRS alone. Second, there is a lack of system-level design in MA-IRS-aided networks, e.g., the joint optimization of the numbers of IRSs and IRS elements, deployment/optimization for MA and IRS, and MA design for multiple IRSs. Moreover, the theoretical performance limits of MA and IRS synergy are still open, and how to fully leverage their complementary roles remains to be explored.

\subsection{Other Miscellaneous Topics}
\subsubsection{Spectrum Sharing}
Cognitive radio (CR) technology enables secondary users (SUs) to access the spectrum resources originally allocated to primary users (PUs), given that their imposed interference can be properly controlled. This approach has gained widespread adoption in recent years, primarily due to the escalating challenges posed by constrained spectrum resources\cite{zhang2010dynamic}. MAs can enhance the performance of CR networks by intelligently reconfiguring wireless channels between the secondary Tx and PUs/SUs, effectively suppressing co-channel interference and improving rate performance \cite{zhou2024MASymbiotic}. Particularly, the authors in \cite{wei2024joint} demonstrated that with a sufficiently large movable region at the secondary Tx, it is possible to maximize the received SNRs at the SUs while simultaneously nulling interference towards all SUs. However, to fully leverage the benefits of MAs, the secondary Tx must obtain the CSI for all PUs, which presents a significant challenge due to the lack of dedicated feedback channels from the primary to the secondary network. This is a critical issue that requires further investigation.

\subsubsection{Relay Systems}
MAs are also expected to enhance wireless relaying by improving both the reception and transmission processes between the source, relay, and destination. However, unlike conventional MA position optimization, MA-assisted relay systems introduce a new two-stage optimization problem, where the positions of MAs need to be adjusted separately for the source-relay and relay-destination transmissions. The authors in \cite{li2024movable} and \cite{li2025MArelay} focused on MA-assisted amplify-and-forward (AF) and decode-and-forward (DF) relay systems and proposed an AO algorithm to derive a locally optimal solution for the MAs' positions across both stages. Nonetheless, further exploration is needed for more relay configurations, such as full-duplex and multi-hop relays.

\subsubsection{MmWave and THz Communications}
MmWave and THz communications offer vast bandwidth and ultra-high data rates for future 6G systems, but they come with high hardware costs and energy consumption due to the need for numerous antennas and RF chains to mitigate severe path loss at such high frequencies \cite{rappaport2019wireless, zhu2019millim, Ning2023THzbeam, Akyildiz2022Terahertz}. A more efficient approach is to use sparse arrays that generate multiple high-power lobes directed toward users. However, this can lead to suboptimal performance due to fixed spatial correlations of steering vectors over different steering angles for conventional sparse arrays with fixed geometry. These challenges can be effectively addressed by utilizing MAs at mmWave/THz BSs, which allow for adaptive tuning of inter-antenna spacing to accommodate user clusters with varying distributions \cite{wang2024movable, shen2024frequencyMA}. Additionally, MAs can dynamically adjust the aperture of the antenna array, thereby enhancing near-field propagation for more efficient multi-stream transmissions, even under LoS conditions \cite{zhu2024nearfield, zhu2024suppressing}.

\subsubsection{Next-Generation Multiple Access}
The concept of next-generation multiple access seeks to enhance the connectivity and spectral efficiency of future wireless networks by leveraging advanced technologies such as non-orthogonal multiple access (NOMA) and rate-splitting multiple access (RSMA) \cite{liu2022NGMA,Clerckx2024multiple,Mao2022RSMA}. Integrating MAs into NOMA and RSMA systems offers new design DoFs in reshaping channel conditions. For single-antenna systems, the MA position optimization basically increases the channel power gains of multiple users such that the overall multiple access performance is improved for both NOMA and RSMA \cite{Li2024MAupNOMA,He2024MANOMA,gao2024MApowerNOMA,ghadi2024fluid}. For multi-antenna systems, the results become more interesting, as the MA position optimization not only improves the channel power gain but also changes the channel correlation among users \cite{xiao2024MANOMA,Zhou2024MAdownNOMA,amhaz2024optimizing,zheng2024FASNOMA,zhang2024movableRSAM}. Instead of monotonously reducing the channel correlation among all users like SDMA systems, MA-enabled NOMA and RSMA systems may benefit from an increase in the channel correlation among a selected subset of users. This facilitates parallel transmissions because the interference among these users can be effectively managed via successive interference cancellation in NOMA or common message decoding in RSMA. However, more future studies are needed to explore optimal antenna positioning and power/rate allocation strategies for MA-enabled NOMA and RSMA, providing deeper insights into such system design.

\subsection{Lessons Learned}
In this section, we have examined the integration of MAs in various wireless systems, such as space-air-ground integrated networks, AirComp, MEC, PLS, SWIPT, and their synergy with IRS, as well as other advanced topics like spectrum sharing, relaying, mmWave and THz communications, and next-generation multiple access. The wide-ranging benefits of MAs in addressing the challenges of future wireless networks have been underscored. However, the introduction of MAs also entails additional hardware costs and signal processing overheads. Therefore, future research directions should focus on the practical deployment of MAs in specific application scenarios, developing customized algorithms and protocols to achieve the best trade-off between performance and complexity/cost.

\section{Conclusions} \label{Sec_Conclusion}
This paper has provided a comprehensive tutorial on the fundamentals and advancements of MA-enabled wireless networks. We began by reviewing the historical development of MA technologies within both the antenna and communication communities, followed by an overview of promising application scenarios for MA-aided wireless systems. To characterize the continuous variation in wireless channels w.r.t. antenna movement, we then introduced the basic field-response channel model for narrowband MA-SISO systems under far-field conditions and extended it to MA-MIMO systems, wideband systems, near-field systems, and 6DMA systems incorporating flexible antenna position and orientation. We also reviewed the state-of-the-art architectures for implementing MAs and discussed their practical constraints. A general optimization framework was also formulated to fully exploit the DoFs offered by antenna movement. In particular, two key design issues were subsequently addressed, i.e., optimizing antenna movement to achieve desired performance gains in various communication/sensing systems, and developing efficient channel acquisition algorithms to reconstruct the channel mapping between Tx and Rx regions. Furthermore, we demonstrated the effectiveness of MA-aided communication systems through prototype development and experimental results. Finally, we discussed the potential extension of MA to other wireless systems and its synergy with other emerging wireless technologies. As the exploration of MA-aided wireless systems is still in its infancy, there remain significant knowledge gaps in both theoretical research and practical implementation. We hope that this tutorial will serve as a valuable resource for researchers and engineers in this burgeoning field, inspiring them to unlock the full potential of MAs in realizing more adaptive and efficient wireless networks in the future.

\bibliographystyle{IEEEtran} % use IEEEtran.bst style
\bibliography{IEEEabrv,ref_MA}

\end{document}